\documentclass[10pt]{article}
\usepackage{amssymb}

\newtheorem{Lemma}{Lemma}[section]

\newtheorem{Proposition}[Lemma]{Proposition}
\newtheorem{Corollary}[Lemma]{Corollary}
\newtheorem{Remark}[Lemma]{Remark}
\newtheorem{Definition}[Lemma]{Definition}
\newtheorem{Hypothesis}[Lemma]{Hypothesis}
\newenvironment{Proof}
{\begin{trivlist} \item[]{\bf Proof. }}%
{\hspace*{\fill}$\rule{.3\baselineskip}{.35\baselineskip}$\end{trivlist}}

\makeatletter
\@addtoreset{equation}{section}
\makeatother

\newcommand{\C}{\mathbb{C}}

\newcommand{\R}{\mathbb{R}}
\newcommand{\Z}{\mathbb{Z}}

\newfam\bifam
\font\tenbi=cmmib10 scaled \magstep1
\font\sevenbi=cmmib10 at 11pt
\font\fivebi=cmmib10 at 6pt
\textfont\bifam = \tenbi
\scriptfont\bifam= \sevenbi
\scriptscriptfont\bifam= \fivebi

\usepackage[dvips]{epsfig}
\usepackage{graphicx}

\begin{document}

\title{\bf Nonlinear Schr\"{o}dinger lattices \\ II: Persistence and stability of discrete vortices}

\author{D.E. Pelinovsky$^1$, P.G. Kevrekidis$^2$, and D.J. Frantzeskakis$^3$, \\
{\small $^{1}$ Department of Mathematics, McMaster
University, Hamilton, Ontario, Canada, L8S 4K1} \\
{\small $^{2}$ Department of Mathematics,
University of Massachusetts, Amherst, Massachusetts, 01003-4515, USA} \\
{\small $^{3}$ Department of Physics, University of Athens,
Panepistimiopolis, Zografos, Athens 15784, Greece}
}

\date{\today}
\maketitle

\begin{abstract}
We study discrete vortices in the anti-continuum limit of the
discrete two-dimensional nonlinear Schr\"{o}dinger (NLS) equations.
The discrete vortices in the anti-continuum limit represent a finite
set of excited nodes on a closed discrete contour with a non-zero
topological charge. Using the Lyapunov--Schmidt reductions, we find
sufficient conditions for continuation and termination of the
discrete vortices for a small coupling constant in the discrete NLS
lattice. An example of a closed discrete contour is considered that
includes the vortex cell (also known as the off-site vortex). We
classify the symmetric and asymmetric discrete vortices that
bifurcate from the anti-continuum limit. We predict analytically and
confirm numerically the number of unstable eigenvalues associated
with various families of such discrete vortices.
\end{abstract}

\section{Introduction}

Following the first paper of this series \cite{PaperI}, we address
discrete systems and differential-difference equations, which have
become topics of increasing physical and mathematical importance.
The variety of physical applications where such models are relevant,
and their significant differences from the mathematical theory of
partial differential equations, contribute to recent interest in
these topics. The applicability of such models extends to areas as
diverse as nonlinear optics, atomic and soft condensed-matter
physics, as well as biophysics: specific details and references can
be found in our first paper \cite{PaperI} as well as in reviews
\cite{A97,FW98,HT99,KRB01,KB04,KF04}.

The second paper is devoted to existence and stability of coherent
structures in two-dimensional lattices, which include both discrete
solitons \cite{KB04,KF04} and discrete vortices \cite{MK01}. These
two-dimensional coherent structures have emerged recently in studies
of photorefractive crystals in nonlinear optics
\cite{ESCFS02,SKEA03} and droplets of optical lattices in
Bose-Einstein condensates \cite{CBFMMTSI01,CFFFMI03}. A significant
boost to this subject was given by the experimental realization of
two-dimensional photonic crystal lattices with periodic potentials
based on the ideas of \cite{ESCFS02}. As a result, discrete solitons
were observed in \cite{FCSEC03,MECC04}, while more complex
structures such as dipoles, soliton trains and vector solitons were
observed in \cite{YMBC04,CMEXB04,CBMY04}.

Most recently, observations of discrete vortices were reported by
two independent groups \cite{NAOKMMC04,FBCMSHC04} where the
fundamental vortices with topological charge {\em one} were
experimentally created and detected in photorefractive crystals. Two
main examples of charge-one discrete vortices include a {\em vortex
cross} (an on-site centered vortex) and a {\em vortex cell} (an
off-site centered vortex). These structures were also recently
predicted in a continuous two-dimensional model with the periodic
potential \cite{YM04}.

These discoveries have stimulated further theoretical work and
numerical computations. Thus, 
%in \cite{KMCF04}, 
while in \cite{MK01}, discrete vortices of
charge {\em two} were shown to be unstable, 
recently in \cite{KMCF04} discrete vortices
of charge {\em three} were found to be stable. Based on the
theoretical predictions of \cite{KMCF04}, further experiments on
localized structures were undertaken to unveil other interesting
structures, such as the discrete soliton necklace which is more
globally stable compared to the charge three vortex
\cite{YMMKMFC04}. The discrete vortices have been also extended to
three-dimensional discrete models \cite{KMFC04}. Furthermore,
asymmetric vortices have been recently predicted in the
two-dimensional lattices in \cite{ASK04}.

The above activity clearly signals the importance and experimental
relevance of discrete solitons and vortices in two-dimensional
discrete lattices. However, most of the above mentioned works are
predominantly of experimental or numerical nature, while the
mathematical theory of existence and stability of discrete localized
structures has not been developed to a similar extent. The aim of
the present paper is to develop a categorization of discrete
solitons and vortices in the discrete two-dimensional nonlinear
Schr\"{o}dinger (NLS) equations. We start from a well-understood
limit (the so-called anti-continuum case of zero coupling between
the lattice nodes) and examining persistence of the limiting
solutions for small coupling by means of the Lyapunov-Schmidt
theory. This method allows us to discuss persistence and stability
of the localized structures by analyzing finite-dimensional linear
eigenvalue problems. The theoretical predictions agree well with
full numerical computations of the discrete two-dimensional NLS
equation.

Our main results are summarized for the simplest localized
structures in Table 1. These results corroborate and extend the
previously reported experimental and numerical findings. We 
quantify the stability of the charge-one vortex in accordance to
\cite{MK01,NAOKMMC04,FBCMSHC04,YM04}, the instability of the
charge-two vortex in accordance to \cite{MK01,KMCF04} and the
stability of the charge-three vortex in accordance to \cite{KMCF04}.
We further demonstrate the instability of all asymmetric vortices
proposed in \cite{ASK04}. Furthermore, our results can be used
to extract the spectral stability of the dipole mode considered
in  \cite{YMBC04} and of the soliton necklace of \cite{YMMKMFC04}.
% is indeed spectrally stable. 

\vspace{0.1cm}
\begin{tabular}{|p{1.4cm}|p{3cm}|p{2.5cm}|p{2.5cm}|p{3cm}|}
\hline
contour $S_M$ & vortex of charge $L$ & \# of unstable \phantom{te} eigenvalues & \# of stable \phantom{text} eigenvalues & \# of continuation \phantom{te} parameters \\
\hline
$M=1$ & symmetric $L=1$ & none & two pairs & two parameters \\
\hline
$M=2$ & symmetric $L=1$ & six complex \phantom{text} one real & none & one parameter \\
\hline
$M=2$ & symmetric $L=2$ & one real & five pairs & two parameters \\
\hline
$M=2$ & symmetric $L=3$ & none & seven pairs & one parameter \\
\hline
$M=2$ & asymmetric $L=1$ & six real & one pair & one parameter \\
\hline
$M=2$ & asymmetric $L=2$ & three real & three pairs & two parameters \\
\hline
$M=2$ & asymmetric $L=3$ & one real & six pairs & one parameter \\
\hline
\end{tabular}

\vspace{0.1cm}

{\bf Table 1:} The numbers of small unstable, stable and zero
eigenvalues, associated to vortices of the discrete two-dimensional
NLS equation with small coupling constant.

\vspace{1cm}

The paper is structured as follows. Abstract results on existence of
discrete solitons and vortices are derived in Section 2. Persistence
of localized modes for a particular square discrete contour is
considered in Section 3. Stability of the persistent solutions is
addressed in Section 4. Analytical results are compared to numerical
computations in Section 5. Section 6 concludes the paper with a
summary of our results and discussions of interesting directions for
future study.

\section{Existence of discrete vortices}

We consider the discrete nonlinear Schr\"{o}dinger (NLS) equation
in two space dimensions \cite{KRB01}:
\begin{equation}
\label{2NLS} i \dot{u}_{n,m} + \epsilon \left( u_{n+1,m} + u_{n-1,m}
+ u_{n,m+1} + u_{n,m-1} - 4 u_{n,m} \right) + |u_{n,m}|^2 u_{n,m} = 0,
\end{equation}
where $u_{n,m}(t) : \R_+ \to \C$, $(n,m) \in \Z^2$, and $\epsilon >
0$ is the inverse squared step size of the discrete two-dimensional
NLS lattice. Time-periodic localized modes of the discrete NLS
equation (\ref{2NLS}) take the form:
\begin{equation}
\label{soliton-shape-2} u_{n,m}(t) = \phi_{n,m} e^{i (\mu - 4
\epsilon) t + i \theta_0}, \qquad \phi_{n,m} \in \C, \quad (n,m) \in
\Z^2,
\end{equation}
where $\theta_0 \in \R$ and $\mu \in \R$ are parameters. Since
localized modes in the focusing NLS lattice (\ref{2NLS}) with
$\epsilon > 0$ may exist only for $\mu > 4 \epsilon$ \cite{HT99} and
the parameter $\mu$ is scaled out by the scaling transformation,
\begin{equation}
\label{scaling-transformation} \phi_{n,m} = \sqrt{\mu}
\hat{\phi}_{n,m}, \qquad \epsilon = \mu \hat{\epsilon},
\end{equation}
the parameter $\mu > 0$ will henceforth be set to $\mu = 1$. In this
case, the complex-valued $\phi_{n,m}$ solve the nonlinear
difference equations on $(n,m) \in \Z^2$:
\begin{equation}
\label{2difference} (1 - |\phi_{n,m}|^2) \phi_{n,m} = \epsilon
\left( \phi_{n+1,m} + \phi_{n-1,m} + \phi_{n,m+1} + \phi_{n,m-1}
\right).
\end{equation}
As $\epsilon = 0$, the localized modes of the difference equations
(\ref{2difference}) are given by the limiting solution:
\begin{equation}
\label{2soliton} \phi_{n,m}^{(0)} =
\left\{ \begin{array}{cc} e^{i \theta_{n,m}}, \quad (n,m) \in S, \\
0, \quad (n,m) \in \Z^2 \backslash S, \end{array} \right.
\end{equation}
where $S$ is a finite set of nodes on the lattice $(n,m) \in \Z^2$
and $\theta_{n,m}$ are parameters for $(n,m) \in S$. Since
$\theta_0$ is arbitrary in the ansatz (\ref{soliton-shape-2}), we
can set $\theta_{n_0,m_0} = 0$ for a particular node $(n_0,m_0) \in
S$. Using this convention, we define two special types of localized
modes, called discrete solitons and vortices.

\begin{Definition}
\label{definition-soliton} The localized solution of the difference
equations (\ref{2difference}) with $\epsilon > 0$, that has all
real-valued amplitudes $\phi_{n,m}$, $(n,m) \in \Z^2$ and satisfies
the limit (\ref{2soliton}) with all $\theta_{n,m} = \{0,\pi\}$,
$(n,m) \in S$, is called a discrete soliton.
\end{Definition}

\begin{Definition}
\label{definition-vortex} Let $S$ be a simply-connected discrete
contour on the plane $(n,m) \in \Z^2$. The localized solution of the
difference equations (\ref{2difference}) with $\epsilon > 0$, that
has complex-valued $\phi_{n,m}$, $(n,m) \in \Z^2$ and
satisfies the limit (\ref{2soliton}) with $\theta_{n,m} \in
[0,2\pi]$, $(n,m) \in S$, is called a discrete vortex.
\end{Definition}

\begin{Definition}
\label{definition-symmetric} Let $S$ be a simply-connected discrete
contour on the plane $(n,m) \in \Z^2$, such that each node $(n,m)
\in S$ has exactly two adjacent nodes in vertical or horizontal
directions along $S$. If the phase difference between two adjacent
$\theta_{n,m}$ for $(n,m) \in S$ is constant in $S$, the discrete
vortex is called symmetric. Otherwise, it is called asymmetric. The
total number of $2 \pi$ phase shifts across the closed contour
$S$ is called the topological charge of the discrete vortex.
\end{Definition}

In particular, we consider the ordered simply-connected discrete
contour $S = S_M$:
\begin{eqnarray}
\nonumber S_M = \left\{
(1,1),(2,1),...,(M+1,1),(M+1,2),...,(M+1,M+1), \right. \\
\label{family-3} \left. (M,M+1),...,(1,M+1),(1,M),...,(1,2)
\right\},
\end{eqnarray}
where ${\rm dim}(S_M) = 4 M$. According to Definition
\ref{definition-symmetric}, the contour $S_M$ for a fixed $M$ could
support symmetric and asymmetric vortices with some charge $L$. An
example of the simplest vortices for $M = 1$ is the symmetric
charge-one vortex cell ($\theta_{1,1}=0$, $\theta_{2,1}=
\frac{\pi}{2}$, $\theta_{2,2} = \pi$, $\theta_{1,2}
=\frac{3\pi}{2}$) \cite{MK01,YM04} and an asymmetric charge-one
vortex ($\theta_{1,1}=0$, $\theta_{2,1}= \theta$, $\theta_{2,2} =
\pi$, $\theta_{1,2} = \pi + \theta$), where $\theta \neq \{
0,\frac{\pi}{2},\pi \}$ \cite{ASK04}.

It follows from the general method \cite{A97,MA94} that the discrete
solitons of the two-dimensional NLS lattice (\ref{2difference}) (see
Definition \ref{definition-soliton}) can be continued to the domain
$0< \epsilon < \epsilon_0$ for some $\epsilon_0 > 0$. It is more
complicated to find a configuration of $\theta_{n,m}$ for $(n,m) \in
S$ that allows us to continue the discrete vortices (see Definition
\ref{definition-vortex}) for $\epsilon > 0$. The continuation of the
discrete solitons and vortices is based on the Implicit Function
Theorem and Lyapunov--Schmidt Reduction Theorem \cite{Hale,Golub}.
Abstract results on existence of such continuations are formulated
and proved below, after the introduction of some relevant notations.

Let ${\cal O}(0)$ be a small neighborhood of $\epsilon = 0$, such
that ${\cal O}(0) = (-\epsilon_0,\epsilon_0)$ for some $\epsilon_0
> 0$. Let $N = {\rm dim}(S)$ and ${\cal T}$ be the torus on
$[0,2\pi]^N$, such that $\theta_{n,m}$ for $(n,m) \in S$ form a
vector $\mbox{\boldmath $\theta$} \in {\cal T}$. Let $\Omega =
L^2(\Z^2,\C)$ be the Hilbert space of square-summable complex-valued
sequences $\{\phi_{n,m}\}_{(n,m) \in \Z^2}$, equipped with the inner
product and the norm:
\begin{equation}
\label{inner-product} ({\bf u},{\bf v})_{\Omega} =
\sum_{(n,m)\in\Z^2} \bar{u}_{n,m} v_{n,m}, \qquad \| {\bf u}
\|^2_{L^2} = \sum_{(n,m) \in \Z^2} |u_{n,m}|^2 < \infty.
\end{equation}
Let ${\bf u}$ denote an infinite-dimensional vector in $\Omega$ that
consists of components $u_{n,m}$ for all $(n,m) \in \Z^2$.

\begin{Proposition}
\label{proposition-existence} There exists a unique (discrete
soliton) solution of the difference equations (\ref{2difference}) in
the domain $\epsilon \in {\cal O}(0)$ that satisfies (i) $\phi_{n,m}
\in \R$, $(n,m) \in \Z^2$ and (ii) $\lim_{\epsilon \to 0} \phi_{n,m}
= \phi_{n,m}^{(0)}$, where $\phi_{n,m}^{(0)}$ is given by
(\ref{2soliton}) with $\theta_{n,m} = \{ 0, \pi \}$, $(n,m) \in S$.
The solution $\mbox{\boldmath $\phi$}(\epsilon)$ is analytic in
$\epsilon \in {\cal O}(0)$.
\end{Proposition}

\begin{Proof}
Assume that $\phi_{n,m} \in \R$ for all $(n,m) \in \Z^2$. The
difference equations (\ref{2difference}) are rewritten as zeros of
the nonlinear vector-valued function:
\begin{equation}
\label{1difference} f_{n,m}(\mbox{\boldmath $\phi$},\epsilon) = (1 -
\phi_{n,m}^2) \phi_{n,m} - \epsilon \left( \phi_{n+1,m} +
\phi_{n-1,m} + \phi_{n,m+1} + \phi_{n,m-1} \right) = 0.
\end{equation}
The mapping ${\bf f} : \Omega \times {\cal O}(0) \mapsto \Omega$ is
$C^1$ on $\mbox{\boldmath $\phi$} \in \Omega$ and has a bounded
continuous Fr{\'e}chet derivative, given by:
\begin{equation}
\label{eq1} {\cal L}_{n,m} = \left( 1 - 3 \phi_{n,m}^2 \right) -
\epsilon \left( s_{+1,0} + s_{-1,0} + s_{0,+1} + s_{0,-1} \right),
\end{equation}
where $s_{n',m'}$ is the shift operator, such that $s_{n',m'}
u_{n,m} = u_{n+n',m+m'}$. It is obvious that
\begin{equation}
\label{eq2} {\bf f}(\mbox{\boldmath $\phi$}^{(0)},0) = {\bf 0},
\qquad {\rm ker} ({\cal L}^{(0)}) = \emptyset,
\end{equation}
where $\mbox{\boldmath $\phi$}^{(0)}$ is the discrete soliton of
Definition \ref{definition-soliton} and ${\cal L}^{(0)}$ is the
operator  ${\cal L}$ computed at $\mbox{\boldmath $\phi$} =
\mbox{\boldmath $\phi$}^{(0)}$ and $\epsilon = 0$. It follows from
(\ref{eq1})--(\ref{eq2}) that ${\cal L}^{(0)} : \Omega \mapsto
\Omega$ has a bounded inverse. By the Implicit Function Theorem
\cite[Appendix 1]{Golub}, there exists a local $C^1$ mapping
$\mbox{\boldmath $\phi$} : {\cal O}(0) \to \Omega$, such that
$\mbox{\boldmath $\phi$}(\epsilon)$ is continuous in $\epsilon \in
{\cal O}(0)$ and $\mbox{\boldmath $\phi$}^{(0)} = \mbox{\boldmath
$\phi$}(0)$. Moreover, since ${\bf f}(\mbox{\boldmath
$\phi$},\epsilon)$ is analytic in $\epsilon \in {\cal O}(0)$, then
$\mbox{\boldmath $\phi$}(\epsilon)$ is analytic in $\epsilon \in
{\cal O}(0)$ \cite[Chapter 2.2]{Hale}.
\end{Proof}

\begin{Remark}
\label{remark-complex} Proposition \ref{proposition-existence} does
not exclude a possibility of continuation of the limiting solution
(\ref{2soliton}) with $\theta_{n,m} = \{0,\pi\}$ for all $(n,m) \in
S$ to the complex-valued solution $\mbox{\boldmath
$\phi$}(\epsilon)$ in $\epsilon \in {\cal O}(0)$.
\end{Remark}

\begin{Proposition}
\label{proposition-reductions} There exists a vector-valued function
${\bf g} : {\cal T} \times {\cal O}(0) \mapsto \R^N$, such that the
limiting solution (\ref{2soliton}) is continued to the domain
$\epsilon \in {\cal O}(0)$ if and only if $\mbox{\boldmath $\theta$}
\in {\cal T}$ is a root of ${\bf g}(\mbox{\boldmath
$\theta$},\epsilon) = {\bf 0}$ in $\epsilon \in {\cal O}(0)$.
Moreover, the function ${\bf g}(\mbox{\boldmath $\theta$},\epsilon)$
is analytic in $\epsilon \in {\cal O}(0)$ and ${\bf
g}(\mbox{\boldmath $\theta$},0) = {\bf 0}$ for any $\mbox{\boldmath
$\theta$} \in {\cal T}$.
\end{Proposition}

\begin{Proof}
When $\phi_{n,m} \in \C$ for some $(n,m) \in \Z^2$, the difference
equations (\ref{2difference}) are complemented by the complex
conjugate equations in the abstract form:
\begin{equation}
\label{nonlinear-equations} {\bf f}(\mbox{\boldmath
$\phi$},\bar{\mbox{\boldmath $\phi$}},\epsilon) = {\bf 0}, \qquad
\bar{\bf f}(\mbox{\boldmath $\phi$},\bar{\mbox{\boldmath
$\phi$}},\epsilon) = {\bf 0}.
\end{equation}
Taking the Fr{\'e}chet derivative of ${\bf f}(\mbox{\boldmath
$\phi$},\bar{\mbox{\boldmath $\phi$}},\epsilon)$ with respect to
$\mbox{\boldmath $\phi$}$ and $\bar{\mbox{\boldmath $\phi$}}$, we
compute the linearization operator ${\cal H}$ for the difference
equations (\ref{2difference}):
\begin{equation}
\label{energy} {\cal H}_{n,m} = \left( \begin{array}{cc} 1 - 2
|\phi_{n,m}|^2 & - \phi_{n,m}^2 \\ - \bar{\phi}_{n,m}^2 & 1 - 2
|\phi_{n,m}|^2 \end{array} \right) - \epsilon \left( s_{+1,0} +
s_{-1,0} + s_{0,+1} + s_{0,-1} \right) \left(
\begin{array}{cc} 1 & 0 \\ 0 & 1 \end{array} \right).
\end{equation}
Let ${\cal H}^{(0)} = {\cal H}(\mbox{\boldmath $\phi$}^{(0)},0)$. It
is clear that ${\cal H}^{(0)} : \Omega \times \Omega \mapsto \Omega
\times \Omega$ is a self-adjoint Fredholm operator of index zero
with ${\rm dim}\;{\rm ker}({\cal H}^{(0)}) = N$. Moreover,
eigenvectors of ${\rm ker}({\cal H}^{(0)})$ re-normalize the
parameters $\theta_{n,m}$ for $(n,m) \in S$ in the limiting solution
(\ref{2soliton}). By the Lyapunov Reduction Theorem \cite[Chapter
7.1]{Golub}, there exists a decomposition $\Omega = {\rm ker}({\cal
H}^{(0)}) \oplus \omega$, such that ${\bf g}(\mbox{\boldmath
$\theta$},\epsilon)$ is defined in terms of the projections to ${\rm
ker}({\cal H}^{(0)})$. Let $\{ {\bf e}_{n,m} \}_{(n,m) \in S}$ be a
set of $N$ linearly independent eigenvectors in the kernel of ${\cal
H}^{(0)}$. It follows from the representation,
\begin{equation}
\label{energy-01}
{\cal H}_{n,m}^{(0)} = -\left( \begin{array}{cc} 1 & e^{2 i \theta_{n,m}} \\
e^{-2i \theta_{n,m}} & 1 \end{array} \right), \;\; (n,m) \in S,
\end{equation}
that each eigenvector ${\bf e}_{n,m}$ in the set $\{ {\bf e}_{n,m}
\}_{(n,m) \in S}$ has the only non-zero element $(e^{i
\theta_{n,m}}, -e^{-i\theta_{n,m}})^T$ at the $(n,m)$-th position of
${\bf u} \in \Omega$. By projections of the nonlinear equations
(\ref{nonlinear-equations}) to ${\rm ker}({\cal H}^{(0)})$, we
derive an implicit representation for the functions ${\bf
g}(\mbox{\boldmath $\theta$},\epsilon)$:
\begin{eqnarray}
\nonumber 2 i g_{n,m}(\mbox{\boldmath $\theta$},\epsilon) & = & (1 -
|\phi_{n,m}|^2) \left( e^{-i\theta_{n,m}}
\phi_{n,m} - e^{i \theta_{n,m}} \bar{\phi}_{n,m} \right) \\
\nonumber & - & \epsilon e^{-i\theta_{n,m}} \left( \phi_{n+1,m} +
\phi_{n-1,m} + \phi_{n,m+1} + \phi_{n,m-1} \right) \\
& + & \epsilon e^{i \theta_{n,m}} \left( \bar{\phi}_{n+1,m} +
\bar{\phi}_{n-1,m} + \bar{\phi}_{n,m+1} + \bar{\phi}_{n,m-1}
\right),
\end{eqnarray}
for $(n,m) \in S$, where the factor $(2i)$ is introduced for
convenient notations. Let $\phi_{n,m} = e^{i \theta_{n,m}} u_{n,m}$
for $(n,m) \in S$ and $\phi_{n,m} = u_{n,m}$ for $(n,m) \in \Z^2
\backslash S$. Since eigenvectors of ${\rm ker}({\cal H}^{(0)})$ are
excluded from the solution $\mbox{\boldmath $\phi$}$ in $\omega
\subset \Omega$, we have $u_{n,m} \in \R$ for $(n,m) \in S$, such
that
\begin{eqnarray}
\nonumber -2 i g_{n,m}(\mbox{\boldmath $\theta$},\epsilon) & = &
\epsilon e^{-i\theta_{n,m}} \left( \phi_{n+1,m} + \phi_{n-1,m} +
\phi_{n,m+1} + \phi_{n,m-1} \right) \\
\label{g-function} & - & \epsilon e^{i \theta_{n,m}} \left(
\bar{\phi}_{n+1,m} + \bar{\phi}_{n-1,m} + \bar{\phi}_{n,m+1} +
\bar{\phi}_{n,m-1} \right)
\end{eqnarray}
and ${\bf g}(\mbox{\boldmath $\theta$},0) = {\bf 0}$ for any
$\mbox{\boldmath $\theta$} \in {\cal T}$. Since ${\bf
f}(\mbox{\boldmath $\phi$},\bar{\mbox{\boldmath $\phi$}},\epsilon)$
is analytic in $\epsilon \in {\cal O}(0)$, then ${\bf
g}(\mbox{\boldmath $\theta$},\epsilon)$ is analytic in $\epsilon \in
{\cal O}(0)$ \cite[Appendix 3]{Golub}.
\end{Proof}

\begin{Corollary}
The function ${\bf g}(\mbox{\boldmath $\theta$},\epsilon)$ can be
expanded into convergent Taylor series in ${\cal O}(0)$:
\begin{equation}
\label{Taylor} {\bf g}(\mbox{\boldmath $\theta$},\epsilon) =
\sum_{k=1}^{\infty} \epsilon^k {\bf g}^{(k)}(\mbox{\boldmath
$\theta$}), \qquad {\bf g}^{(k)}(\mbox{\boldmath $\theta$}) =
\frac{1}{k!}
\partial_{\epsilon}^k {\bf g}(\mbox{\boldmath $\theta$},0).
\end{equation}
If the root $\mbox{\boldmath $\theta$}(\epsilon)$ of ${\bf
g}(\mbox{\boldmath $\theta$},\epsilon) = {\bf 0}$ is analytic in
$\epsilon \in {\cal O}(0)$, then the solution $\mbox{\boldmath
$\phi$}(\epsilon)$ is analytic in $\epsilon \in {\cal O}(0)$, such
that
\begin{equation}
\label{TaylorPhi} \mbox{\boldmath $\phi$}(\epsilon) =
\mbox{\boldmath $\phi$}^{(0)} + \sum_{k=1}^{\infty} \epsilon^k
\mbox{\boldmath $\phi$}^{(k)},
\end{equation}
where $\mbox{\boldmath $\phi$}^{(0)}$ is given by (\ref{2soliton}).
\end{Corollary}

\begin{Lemma}
\label{lemma-gauge} Let $\mbox{\boldmath $\theta$}(\epsilon)$ be a
root of ${\bf g}(\mbox{\boldmath $\theta$},\epsilon)={\bf 0}$ in
$\epsilon \in {\cal O}(0)$. An arbitrary shift $\mbox{\boldmath
$\theta$}(\epsilon) + \theta_0 {\bf p}_0$, where $\theta_0 \in \R$
and ${\bf p}_0 = (1,1,...,1)^T$, gives a one-parameter family of
roots of ${\bf g}(\mbox{\boldmath $\theta$},\epsilon)={\bf 0}$ for
the same $\epsilon$.
\end{Lemma}

\begin{Proof}
The statement follows from the symmetry of the difference equations
(\ref{2difference}) with respect to gauge transformation
\cite[Chapter 7.3]{Golub}.
\end{Proof}

\begin{Proposition}
\label{proposition-sufficient} Let $\mbox{\boldmath $\theta$}_*$ be
the root of ${\bf g}^{(1)}(\mbox{\boldmath $\theta$})={\bf 0}$ and
${\cal M}_1$ be the Jacobian matrix of ${\bf
g}^{(1)}(\mbox{\boldmath $\theta$})$ at $\mbox{\boldmath $\theta$} =
\mbox{\boldmath $\theta$}_*$. If the matrix ${\cal M}_1$ has a
simple zero eigenvalue, there exists a unique (modulo gauge
transformation) analytic continuation of the limiting solution
(\ref{2soliton}) to the domain $\epsilon \in {\cal O}(0)$.
\end{Proposition}

\begin{Proof}
By Lemma \ref{lemma-gauge}, the matrix ${\cal M}_1$ has always a
non-empty kernel with the eigenvector ${\bf p}_0 = (1,1,...,1)$ due
to gauge transformation. Let $X_0$ be the constrained subspace of
$\C^N$:
\begin{equation}
\label{subspace} X_0 = \{ {\bf u} \in \C^N : \; ({\bf p}_0,{\bf u})
= 0 \}.
\end{equation}
If the matrix ${\cal M}_1$ is non-singular in the subspace $X_0$,
then there exists a unique (modulo the shift) analytic continuation
of the root $\mbox{\boldmath $\theta$}_*$ in $\epsilon \in {\cal
O}(0)$ by the Implicit Function Theorem, applied to the nonlinear
equation ${\bf g}(\mbox{\boldmath $\theta$},\epsilon) = {\bf 0}$
\cite[Appendix 1]{Golub}.
\end{Proof}

\begin{Proposition}
\label{lemma-nonexistence} Let $\mbox{\boldmath $\theta$}_*$ be a
$(1+d)$-parameter solution of ${\bf g}^{(1)}(\mbox{\boldmath
$\theta$}) = {\bf 0}$ and ${\cal M}_1$ have a zero eigenvalue of
multiplicity $(1 + d)$, where $1 \leq d \leq N-1$. The limiting
solution (\ref{2soliton}) can be continued in the domain $\epsilon
\in {\cal O}(0)$ only if ${\bf g}^{(2)}(\mbox{\boldmath
$\theta$}_*)$ is orthogonal to ${\rm ker}({\cal M}_1)$.
\end{Proposition}

\begin{Proof}
Let ${\bf p}_0$ and $\{ {\bf p}_l \}_{l = 1}^d$ be eigenvectors of
${\rm ker}({\cal M}_1)$. We define the constrained subspace of
$X_0$:
\begin{equation}
\label{subspace2} X_d = \{ {\bf u} \in X_0 : \; ({\bf p}_l,{\bf u})
= 0, \; l = 1,...,d \}.
\end{equation}
If ${\bf g}^{(2)}(\mbox{\boldmath $\theta$}_*) \notin X_d$, the
Lyapunov-Schmidt Reduction Theorem in finite dimensions
\cite[Chapter 1.3]{Golub} shows that the solution $\mbox{\boldmath
$\theta$}_*$ can not be continued in $\epsilon \in {\cal O}(0)$.
\end{Proof}

Proposition \ref{proposition-reductions} gives an abstract
formulation of the continuation problem for the limiting solution
(\ref{2soliton}) for $\epsilon \neq 0$. Proposition
\ref{proposition-sufficient} gives a sufficient condition of
existence and uniqueness (up to gauge invariance) of such
continuations. Proposition \ref{lemma-nonexistence} gives a
sufficient condition for termination of multi-parameter solutions.
Particular applications of Propositions
\ref{proposition-reductions}, \ref{proposition-sufficient} and
\ref{lemma-nonexistence} are limited by the complexity of the set
$S$ in the limiting solution (\ref{2soliton}), since computations of
the vector-valued function ${\bf g}^{(1)}(\mbox{\boldmath
$\theta$})$, ${\bf g}^{(2)}(\mbox{\boldmath $\theta$})$, and the
Jacobian matrix ${\cal M}_1$ could be technically involved. We apply
the abstract results of Propositions \ref{proposition-reductions},
\ref{proposition-sufficient} and \ref{lemma-nonexistence} to the
simply-connected discrete contour $S_M$, defined in
(\ref{family-3}).

\section{Persistence of discrete vortices}

We consider discrete solitons and vortices on the contour $S_M$
defined by (\ref{family-3}). Let the set $\theta_j$ correspond to
the ordered contour $S_M$, starting at $\theta_1 = \theta_{1,1}$,
$\theta_2 = \theta_{2,1}$ and ending at $\theta_N = \theta_{1,2}$,
where $N = 4M$. In what follows, we use the periodic boundary
conditions for $\theta_j$ on the circle from $j = 1$ to $j = N$,
such that $\theta_0 = \theta_N$, $\theta_1 = \theta_{N+1}$, and so
on.

The discrete vortex has the charge $L$ if the phase difference
changes on $2 \pi L$ along the discrete contour $S_M$. By gauge
transformation, we can always set $\theta_1 = 0$ for convenience. We
will also choose $\theta_2 = \theta$ with $0 \leq \theta \leq \pi$
for convenience, which corresponds to discrete vortices with $L \geq
0$.

\subsection{Solutions of the first-order reductions}

Substituting the limiting solution $\phi_{n,m}^{(0)}$ in the
bifurcation function (\ref{g-function}), we find that ${\bf
g}^{(1)}(\mbox{\boldmath $\theta$})$ in the Taylor series
(\ref{Taylor}) is non-zero for the contour $S_M$ and it takes the
form:
\begin{equation}
\label{sin-bifurcation} {\bf g}^{(1)}_j(\mbox{\boldmath $\theta$}) =
\sin (\theta_{j} - \theta_{j+1} ) + \sin(\theta_j - \theta_{j-1}),
\qquad 1 \leq j \leq N.
\end{equation}
The bifurcation equations ${\bf g}^{(1)}(\mbox{\boldmath $\theta$})
= {\bf 0}$ are rewritten as a system of $N$ nonlinear equations for
$N$ parameters $\theta_1$,$\theta_2$,...,$\theta_N$ as follows:
\begin{equation}
\label{sin3} \sin(\theta_2 - \theta_1) = \sin(\theta_3 - \theta_2) =
... = \sin(\theta_N - \theta_{N-1}) = \sin(\theta_1 - \theta_N).
\end{equation}
We classify all solutions of the bifurcation equations (\ref{sin3})
and give explicit examples for $M = 1$ and $M = 2$.

\begin{Proposition}
\label{lemma-asymmetric} Let $a_j = \cos(\theta_{j+1}-\theta_j)$ for
$1 \leq j \leq N$, such that $\theta_1 = 0$, $\theta_2 = \theta$,
and $\theta_{N+1} = 2 \pi L$, where $N = 4M$, $0 \leq \theta \leq
\pi$ and $L$ is the vortex charge. All solutions of the bifurcation
equations (\ref{sin3}) reduce to the four families: \\
$\phantom{text}$ (i) discrete solitons with $\theta = \{ 0,\pi \}$
and
\begin{equation}
\label{discreteSoliton} \theta_j = \{0,\pi\}, \qquad 1 \leq j \leq
N,
\end{equation}
such that the set $\{a_j\}_{j=1}^N$ includes $l$ coefficients $a_j =
1$ and $N-l$ coefficients $a_j = -1$, where $0 \leq l \leq N$.

$\phantom{text}$ (ii) symmetric vortices of charge $L$ with $\theta
= \frac{\pi L}{2 M}$, where $1 \leq L \leq 2M-1$, and
\begin{equation}
\label{symmetric-vortices} \theta_j = \frac{\pi L (j-1)}{2 M},
\qquad 1 \leq j \leq N,
\end{equation}
such that all $N$ coefficients are the same: $a_j = a = \cos
\left(\frac{\pi L}{2M}\right)$.

$\phantom{text}$ (iii) one-parameter families of asymmetric vortices
of charge $L = M$ with $0 < \theta < \pi$ and
\begin{equation}
\label{asymmetric-vortices1} \theta_{j+1} - \theta_j =
\left\{ \begin{array}{c} \theta \\
\pi - \theta \end{array} \right\} \; {\rm mod}(2\pi), \qquad 2 \leq
j \leq N,
\end{equation}
such that the set $\{a_j\}_{j=1}^N$ includes $2M$ coefficients $a_j
= \cos\theta$ and $2M$ coefficients $a_j = - \cos \theta$.

$\phantom{text}$ (iv) zero-parameter asymmetric vortices of charge
$L \neq M$ and
\begin{equation}
\label{asymmetric-vortices2} \theta = \theta_* = \frac{\pi}{2}
\left( \frac{n +2L-4M}{n - 2M} \right), \qquad 1 \leq n \leq N-1,
\;\; n \neq 2 M,
\end{equation}
such that the set $\{a_j\}_{j=1}^N$ includes $n$ coefficients $a_j =
\cos\theta_*$ and $N-n$ coefficients $a_j = - \cos \theta_*$ and the
family (iv) does not reduce to any of the families (i)--(iii).
\end{Proposition}

\begin{Proof}
All solutions of the bifurcation equations (\ref{sin3}) are given by
the binary choice (\ref{asymmetric-vortices1}) in the two roots of
the sine-function on $\theta \in [0,2\pi]$, where the first choice
gives $a_j = \cos \theta$ and the second choice gives $a_j = - \cos
\theta$. Let us assume that there are totally $n$ first choices and
$N - n$ second choices, where $1 \leq n \leq N$. Then, we have
$$
\theta_{N+1} = n \theta + (N-n) (\pi - \theta) = (2n - N) \theta +
(N-n) \pi = 2 \pi L,
$$
where $L$ is the integer charge of the discrete vortex. There are
only two solutions of the above equation. When $\theta$ is arbitrary
parameter, we have $n = \frac{N}{2} = 2M$ and $L = M$, which gives
the one-parameter family (iii). When $\theta = \theta_*$ is fixed,
we have
$$
\theta_* = \frac{\pi}{2} \left( \frac{n + 2L - 4M}{n - 2M} \right)
$$
When $n = N-2L$, we have the family (i) with $N-2L$ phases $\theta_j
= 0$ and $2L$ phases $\theta_j = \pi$. Since discrete solitons do
not have topological charge, the parameter $L$ could be
half-integer: $L = (N-l)/2$, where $0 \leq l \leq N$. When $n = 4M$,
we have the family (ii) for any $1 \leq L \leq 2M-1$. Other choices
of $n$, which are irreducible to the families (i)--(iii), produce
the family (iv).
\end{Proof}

\begin{Remark}
\label{remark-supersymmetric} The one-parameter family (iii)
connects special solutions of the families (i) and (ii). When
$\theta = 0$ and $\theta = \pi$, the family (iii) reduces to the
family (i) with $l = 2M$. When $\theta = \frac{\pi}{2}$, the family
(iii) reduces to the family (ii) with $L = M$. We shall call the
corresponding solutions of family (i) as the super-symmetric soliton
and of family (ii) as the super-symmetric vortex.
\end{Remark}

\begin{Remark}
There exist $N_1 = 2^{N-1}$ solutions of family (i), $N_2 = 2M-1$
solutions of family (ii), and $N_3$ solutions of family (iii), where
\begin{equation}
N_3 = 2^{N-1} - \sum_{k=0}^{2M-1} \frac{N!}{k! (N-k)!}.
\end{equation}
The number $N_4$ of solutions of family (iv) can not be computed in
general. We consider such solutions only in the explicit examples of
$M = 1$ and $M = 2$.
\end{Remark}

{\bf Example $M = 1$ and $N = 4$:} There are $N_1 = 8$ solutions of
family (i), $N_2 = 1$ solution of family (ii), $N_3 = 3$ solutions
of family (iii), and no solutions of family (iv). The three
one-parameter asymmetric vortices are given explicitly by
\begin{eqnarray}
\label{asymA} & \phantom{=} & {\rm (a)} \; \theta_1 = 0, \; \theta_2
= \theta, \; \theta_3 = \pi, \; \theta_4 = \pi + \theta \\
& \phantom{=} & {\rm (b)} \; \theta_1 = 0, \; \theta_2 = \theta, \;
\theta_3 = 2 \theta, \; \theta_4 = \pi + \theta \\
\label{asymC} & \phantom{=} & {\rm (c)} \; \theta_1 = 0, \; \theta_2
= \theta, \; \theta_3 = \pi, \; \theta_4 = 2 \pi - \theta.
\end{eqnarray}

{\bf Example $M = 2$ and $N = 8$:} There are $N_1 = 128$ solutions
of family (i), $N_2 = 3$ solutions of family (ii), $N_3 = 35$
solutions of family (iii), and $N_4 = 14$ solutions of family (iv).
The three symmetric vortices have topological charge $L = 1$
($\theta = \frac{\pi}{4}$), $L = 2$ ($\theta = \frac{\pi}{2}$), and
$L = 3$ ($\theta = \frac{3 \pi}{4}$). The one-parameter asymmetric
vortices include $35$ combinations of $4$ upper choices and $4$
lower choices in (\ref{asymmetric-vortices1}), starting with the
following three solutions:
\begin{eqnarray*}
\label{asymAa} & \phantom{=} & {\rm (a)} \; \theta_1 = 0, \;
\theta_2 = \theta, \; \theta_3 = 2 \theta, \; \theta_4 = 3 \theta,
\; \theta_5 = 4 \theta, \; \theta_6 = \pi + 3 \theta, \;
\theta_7 = 2 \pi + 2 \theta, \; \theta_8 = 3 \pi + \theta, \\
& \phantom{=} & {\rm (b)} \; \theta_1 = 0, \; \theta_2 = \theta, \;
\theta_3 = 2 \theta, \; \theta_4 = 3 \theta, \; \theta_5 = \pi + 2
\theta, \; \theta_6 = \pi + 3 \theta, \;
\theta_7 = 2 \pi + 2 \theta, \; \theta_8 = 3 \pi + \theta, \\
\label{asymCa} & \phantom{=} & {\rm (c)} \; \theta_1 = 0, \;
\theta_2 = \theta, \; \theta_3 = 2 \theta, \; \theta_4 = 3 \theta,
\; \theta_5 = \pi + 2 \theta, \; \theta_6 = 2 \pi + \theta, \;
\theta_7 = 2 \pi + 2 \theta, \; \theta_8 = 3 \pi + \theta,
\end{eqnarray*}
and so on. The zero-parameter asymmetric vortices include $7$
combinations of vortices with $L = 1$ for seven phase differences
$\frac{\pi}{6}$ and one phase difference $\frac{5\pi}{6}$ and $7$
combinations of vortices with $L = 3$ for one phase difference
$\frac{\pi}{6}$ and seven phase differences $\frac{5\pi}{6}$.

\subsection{Continuation of solutions of the first-order reductions}

We compute the Jacobian matrix ${\cal M}_1$ from the bifurcation
function ${\bf g}^{(1)}(\mbox{\boldmath $\theta$})$, given in
(\ref{sin-bifurcation}):
\begin{eqnarray}
\label{Melements3} ({\cal M}_1)_{i,j} = \left\{ \begin{array}{lcl}
\cos(\theta_{j+1} - \theta_j) + \cos(\theta_{j-1} - \theta_j), &
\quad & i = j, \\ - \cos(\theta_j - \theta_i), & \quad & i = j \pm 1 \\
0, & \quad & |i - j | \geq 2 \end{array} \right.
\end{eqnarray}
subject to the periodic boundary conditions. The matrix ${\cal M}_1$
is defined by the coefficients $a_j = \cos(\theta_{j+1}-\theta_j)$
for $1 \leq j \leq N$. It has the same structure as that in the
perturbation theory of continuous multi-pulse solitons in coupled
NLS equations \cite{KK04}. Three technical results establish
location of eigenvalues of the matrix ${\cal M}_1$.

\begin{Lemma}
\label{lemma-periodic} Let $n_0$, $z_0$, and $p_0$ be the numbers of
negative, zero and positive terms of $a_j = \cos(\theta_{j+1} -
\theta_j)$, $1 \leq j \leq N$, such that $n_0 + z_0 + p_0 = N$. Let
$n({\cal M}_1)$, $z({\cal M}_1)$, and $p({\cal M}_1)$ be the numbers
of negative, zero and positive eigenvalues of the matrix ${\cal
M}_1$, defined by (\ref{Melements3}). Assume that $z_0 = 0$ and
denote:
\begin{equation}
\label{case1} A_1 =  \sum_{i=1}^N \prod_{j \neq i} a_j = \left(
\prod_{i=1}^N a_i \right) \; \left( \sum_{i=1}^N \frac{1}{a_i}
\right).
\end{equation}
If $A_1 \neq 0$, then, $z({\cal M}_1) = 1$, and either $n({\cal
M}_1) = n_0-1$, $p({\cal M}_1) = p_0$ or $n({\cal M}_1) = n_0$,
$p({\cal M}_1) = p_0 - 1$. Moreover, $n({\cal M}_1)$ is even if
$A_1> 0$ and is odd if $A_1 < 0$. If $A_1 = 0$, then $z({\cal M}_1)
\geq 2$.
\end{Lemma}

\begin{Proof}
The first statement follows from Appendix A of \cite{KK04}. Let the
determinant equation be $D(\lambda) = {\rm det}({\cal M}_1 - \lambda
I) = 0$. By induction arguments in \cite{KK04,S98}, it can be found
that $D(0) = 0$ and $D'(0) = - N A_1$. On the other hand, $D'(0) = -
\lambda_1 \lambda_2 \cdots \lambda_{N-1}$, where $\lambda_N = 0$
(which exists always with the eigenvector ${\bf p}_0 =
(1,1,...,1)^T$, see Proposition \ref{proposition-sufficient}). Then,
it is clear that $(-1)^{n({\cal M}_1)} = {\rm sign}(A_1)$. When $A_1
= 0$, at least one more eigenvalue is zero, such that $z({\cal M}_1)
\geq 2$.
\end{Proof}

\begin{Lemma}
\label{lemma-constant} Let all coefficients $a_j = \cos(\theta_{j+1}
- \theta_j)$, $1 \leq j \leq N$ be the same: $a_j = a$. Eigenvalues
of the matrix ${\cal M}_1$ are computed explicitly as follows:
\begin{equation}
\label{explicit-eigenvalue} \lambda_n = 4 a \sin^2 \frac{\pi n}{N},
\qquad 1 \leq n \leq N.
\end{equation}
\end{Lemma}

\begin{Proof}
When $a_j = a$, $1 \leq j \leq N$, the eigenvalue problem for the
matrix ${\cal M}_1$ takes the form of the linear difference equation
with constant coefficients:
\begin{equation}
a \left( 2 x_j - x_{j+1} - x_{j-1} \right) = \lambda x_j, \qquad x_0
= x_N, \; x_1 = x_{N+1},
\end{equation}
The discrete Fourier mode $x_j = \exp\left(i \frac{2 \pi j
n}{N}\right)$ for $1 \leq j,n \leq N$ results in the solution
(\ref{explicit-eigenvalue}).
\end{Proof}

\begin{Lemma}
\label{lemma-alternating} Let all coefficients $a_j =
\cos(\theta_{j+1} - \theta_j)$, $1 \leq j \leq N$ alternate the
sign: $a_j = (-1)^j a$, where $N = 4M$. Eigenvalues of the matrix
${\cal M}_1$ are computed explicitly as follows:
\begin{equation}
\label{explicit-eigenvalue-alternating} \lambda_n = - \lambda_{n+2M}
= 2 a \sin \frac{\pi n}{2M}, \qquad 1 \leq n \leq 2M,
\end{equation}
such that $n({\cal M}_1) = 2M-1$, $z({\cal M}_1) = 2$, and $p({\cal
M}_1) = 2M-1$. These numbers do not change if the set $\{a_j
\}_{j=1}^N$ is obtained from the sign-alternating set by
permutations.
\end{Lemma}

\begin{Proof}
When $a_j = (-1)^j a$, $1 \leq j \leq 4M$, the eigenvalue problem
for the matrix ${\cal M}_1$ takes the form of a coupled system of
linear difference equation with constant coefficients:
\begin{equation}
a \left( y_j - y_{j-1} \right) = \lambda x_j, \qquad a \left( x_j -
x_{j+1} \right) = \lambda y_j, \qquad 1 \leq j \leq 2M,
\end{equation}
subject to the periodic boundary conditions: $x_1 = x_{2M+1}$ and
$y_0 = y_{2M}$. The discrete Fourier mode $x_j = x_0 \exp\left(i
\frac{2 \pi j n}{2M}\right)$ and $y_j = y_0 \exp\left(i \frac{2 \pi
j n}{2M}\right)$ for $1 \leq j,n \leq 2M$ results in the solution
(\ref{explicit-eigenvalue-alternating}). In this case, we have
$n({\cal M}_1) = 2M-1$, $z({\cal M}_1) = 2$, and $p({\cal M}_1) =
2M-1$, such that $D(0) = D'(0)= 0$ in the determinant equation
$D(\lambda) = {\rm det}({\cal M}_1 - \lambda I)$. In order to prove
that $z({\cal M}_1) = 2$ remains invariant with respect to
permutations of the sign-alternating set $\{a_j\}_{j=1}^N$, we find
from Mathematica that
\begin{equation}
\label{case2} D''(0) =  \left( \prod_{i=1}^N a_i \right) \; \left(
\alpha_N \left( \sum_{i=1}^{N-1} \frac{1}{a_i a_{i+1}} +
\frac{1}{a_1 a_N} \right) + \beta_N \left( \sum_{i = 1}^{N-2}
\sum_{l = i + 2}^N \frac{1}{a_i a_l} - \frac{1}{a_1 a_N} \right)
\right),
\end{equation}
where $0 < \alpha_N < \beta_N$ are numerical coefficients. Let $A_*$
denote the sign-alternating set $\{a_j\}_{j=1}^N$, such that $a_j =
(-1)^j a$, and $A$ denote a set obtained from $A_*$ by permutations.
It is clear that
$$
\left( \sum_{i=1}^{N-1} \frac{1}{a_i a_{i+1}} + \frac{1}{a_1 a_N}
\right)_{A_*} \leq \left( \sum_{i=1}^{N-1} \frac{1}{a_i a_{i+1}} +
\frac{1}{a_1 a_N} \right)_{A}
$$
and
$$
\left( \sum_{i = 1}^{N-1} \sum_{l = i + 1}^N \frac{1}{a_i a_l}
\right)_{A_*} = \left( \sum_{i = 1}^{N-1} \sum_{l = i + 1}^N
\frac{1}{a_i a_l} \right)_{A}.
$$
Therefore, the expression in brackets in (\ref{case2}) can be
estimated as follows:
\begin{eqnarray*}
(\alpha_N - \beta_N)  \left( \sum_{i=1}^{N-1} \frac{1}{a_i a_{i+1}}
+ \frac{1}{a_1 a_N} \right)_{A} + \beta_N  \left( \sum_{i = 1}^{N-1}
\sum_{l = i + 1}^N \frac{1}{a_i a_l} \right)_{A} & \leq & \\
(\alpha_N - \beta_N)  \left( \sum_{i=1}^{N-1} \frac{1}{a_i a_{i+1}}
+ \frac{1}{a_1 a_N} \right)_{A_*} + \beta_N  \left( \sum_{i =
1}^{N-1} \sum_{l = i + 1}^N \frac{1}{a_i a_l} \right)_{A_*} & < & 0,
\end{eqnarray*}
where the last inequality follows from the fact that $D''(0) < 0$
for $A_*$. Therefore, $z({\cal M}_1) = 2$ for $A$. Combining it with
estimates from Appendix A in \cite{KK04}, we have $n({\cal M}_1) =
p({\cal M}_1) = 2 M - 1$.
\end{Proof}

Using Lemmas \ref{lemma-periodic}, \ref{lemma-constant}, and
\ref{lemma-alternating}, we classify the continuation of solutions
of the first-order reductions, which are described in the families
(i)--(iv) of Proposition \ref{lemma-asymmetric}.

For family (i), excluding the case of super-symmetric solitons (see
Remark \ref{remark-supersymmetric}), the numbers of positive and
negative signs of $a_j$ are different, such that the conditions $z_0
= 0$ and $A_1 \neq 0$ are satisfied in Lemma \ref{lemma-periodic},
and hence $z({\cal M}_1) = 1$. By Proposition
\ref{proposition-sufficient}, the family (i) has a unique
continuation to discrete solitons (see Definition
\ref{definition-soliton}). Continuations described in Remark
\ref{remark-complex} are only possible for super-symmetric solitons.

For family (ii), all coefficients $a_j$ are the same: $a_j = a =
\cos\left( \frac{\pi L}{2M} \right)$, $1 \leq j \leq N$. By Lemma
\ref{lemma-constant}, there is always a zero eigenvalue ($\lambda_N
= 0$), while remaining ($N-1$) eigenvalues are all positive for $a
> 0$ (when $1 \leq L \leq M-1$), negative for $a < 0$ (when $M+1
\leq L \leq 2 M-1$), and zero for $a = 0$ (when $L = M$). By
Proposition \ref{proposition-sufficient}, the family (ii) has a
unique continuation to symmetric vortices with charge $L$, where $1
\leq L \leq 2M-1$ and $L \neq M$ (see Definitions
\ref{definition-vortex} and \ref{definition-symmetric}).

For family (iii), there are $2M$ coefficients $a_j = \cos \theta$
and $2M$ coefficients $a_j = -\cos\theta$, which are non-zero for
$\theta \neq \frac{\pi}{2}$. By Lemma \ref{lemma-alternating}, we
have $n({\cal M}_1) = 2M - 1$, $z({\cal M}_1) = 2$, and $p({\cal
M}_1) = 2M - 1$. The additional zero eigenvalue is related to the
derivative of the family of the asymmetric discrete vortices
(\ref{asymmetric-vortices1}) with respect to the parameter $\theta$.
Therefore, continuations of family (iii) of asymmetric vortices,
including the particular cases of super-symmetric solitons of family
(i) and super-symmetric vortices of family (ii), must be considered
beyond the first-order reductions.

For family (iv), since $n \neq 2M$, the conditions  $z_0 = 0$ and
$A_1 \neq 0$ are satisfied in Lemma \ref{lemma-periodic}, and hence
$z({\cal M}_1) = 1$. By Proposition \ref{proposition-sufficient},
the family (iv) has a unique continuation to asymmetric vortices for
$\epsilon \neq 0$.

\subsection{Continuation of solutions to the second-order reductions}

Results of the first-order reductions are insufficient to conclude
persistence of the asymmetric vortices of family (iii), including
the super-symmetric soliton of family (i) and the super-symmetric
vortex of family (ii). Therefore, we continue the bifurcation
function ${\bf g}(\mbox{\boldmath $\theta$},\epsilon)$ to the second
order of $\epsilon$ in the Taylor series (\ref{Taylor}). It follows
from (\ref{2difference}) that the first-order correction of the
Taylor series (\ref{TaylorPhi}) satisfies the inhomogeneous problem:
\begin{equation}
\label{2problem} (1 - 2 |\phi_{n,m}^{(0)}|^2) \phi_{n,m}^{(1)} -
\phi_{n,m}^{(0) 2} \bar{\phi}_{n,m}^{(1)} = \phi_{n+1,m}^{(0)} +
\phi_{n-1,m}^{(0)} + \phi_{n,m+1}^{(0)} + \phi_{n,m-1}^{(0)}.
\end{equation}
We define solution of the inhomogeneous problem (\ref{2problem}) in
$\omega \subset \Omega$, such that the homogeneous solutions in
${\rm ker}({\cal H}^{(0)})$ are removed from the solution
$\mbox{\boldmath $\phi$}^{(1)}$. This is equivalent to the
constraint: $\phi_{n,m} = u_{n,m} e^{i \theta_{n,m}}$, $u_{n,m} \in
\mathbb{R}$ for all $(n,m) \in S_M$. We develop computations for
three distinct cases: $M = 1$, $M = 2$ and $M \geq 3$. This
separation is due to the special structure of the discrete contours
$S_M$.

{\bf Case $M = 1$:} The inhomogeneous problem (\ref{2problem}) has a
unique solution $\mbox{\boldmath $\phi$}^{(1)} \in \omega \subset
\Omega$:
\begin{eqnarray}
\label{solution31} \phi_{n,m}^{(1)} = - \frac{1}{2} \left[ \cos
(\theta_{j-1}-\theta_j) + \cos(\theta_{j+1}-\theta_j) \right] e^{i
\theta_j},
\end{eqnarray}
where the index $j$ enumerates the node $(n,m)$ on the contour
$S_M$,
\begin{equation}
\label{solution32} \phi_{n,m}^{(1)} = e^{i \theta_j},
\end{equation}
where the node $(n,m)$ is adjacent to the $j$-th node on the contour
$S_M$, while $\phi_{n,m}^{(1)}$ is empty for all remaining nodes. By
substituting the first-order correction term $\phi_{n,m}^{(1)}$ into
the bifurcation function (\ref{g-function}), we find the correction
term ${\bf g}^{(2)}(\mbox{\boldmath $\theta$})$ in the Taylor series
(\ref{Taylor}):
\begin{eqnarray}
\nonumber {\bf g}^{(2)}_j(\mbox{\boldmath $\theta$}) & = &
\frac{1}{2} \sin (\theta_{j+1} - \theta_{j} ) \left[ \cos(\theta_j -
\theta_{j+1}) + \cos(\theta_{j+2} -\theta_{j+1}) \right] \\
\label{sin-bifurcation2} & + & \frac{1}{2} \sin(\theta_{j-1} -
\theta_{j}) \left[ \cos(\theta_j - \theta_{j-1}) + \cos(\theta_{j-2}
-\theta_{j-1}) \right], \qquad 1 \leq j \leq N.
\end{eqnarray}
We compute the vector ${\bf g}_2 = {\bf g}^{(2)}(\mbox{\boldmath
$\theta$})$ at the asymmetric vortex solutions
(\ref{asymA})--(\ref{asymC}):
$$
\mbox{(a)} \; {\bf g}_2 = \left( \begin{array}{c} 0 \\ 0 \\ 0 \\ 0
\end{array} \right), \qquad \mbox{(b)} \;
{\bf g}_2 = \left( \begin{array}{c} 2  \\ 0 \\ -2 \\ 0
\end{array} \right) \; \sin \theta \cos \theta, \qquad
\mbox{(c)} \; {\bf g}_2 = \left( \begin{array}{c} 0 \\ -2 \\ 0 \\ 2
\end{array} \right) \sin \theta \cos \theta.
$$
The kernel of ${\cal M}_1$ is two-dimensional with the eigenvectors
${\bf p}_0$ and ${\bf p}_1$. The second eigenvector ${\bf p}_1$ is
related to derivatives of the solutions (\ref{asymA})--(\ref{asymC})
in $\theta$:
$$
\mbox{(a)} \; {\bf p}_1 = \left( \begin{array}{c} 0 \\ 1 \\ 0 \\ 1
\end{array} \right), \qquad \mbox{(b)} \;
{\bf p}_1 = \left( \begin{array}{c} 0 \\ 1 \\ 2 \\ 1
\end{array} \right), \qquad \mbox{(c)} \; {\bf p}_1 =
\left( \begin{array}{c} 0 \\ 1 \\ 0 \\ -1 \end{array} \right).
$$
The Fredholm alternative $({\bf p}_1,{\bf g}_2) = 0$ is satisfied
for the solution (a) but fails for the solutions (b) and (c), unless
$\theta = \{ 0, \frac{\pi}{2},\pi \}$. The latter cases are included
in the definitions of super-symmetric discrete solitons and vortices
(see Remark \ref{remark-supersymmetric}). By Proposition
\ref{lemma-nonexistence}, the solutions (b) and (c) can not be
continued in $\epsilon \neq 0$, while the solution (a) can be
continued up to the second-order reductions.

{\bf Case $M = 2$:} The solution $\mbox{\boldmath $\phi$}^{(1)} \in
\omega \subset \Omega$ of the inhomogeneous problem (\ref{2problem})
is given by (\ref{solution31}) and (\ref{solution32}), except for
the center node $(2,2)$, where
\begin{equation}
\label{solution32a} \phi_{2,2}^{(1)} = e^{i \theta_2} + e^{i
\theta_4} + e^{i \theta_6} + e^{i \theta_8}.
\end{equation}
The correction term ${\bf g}^{(2)}(\mbox{\boldmath $\theta$})$ is
given by (\ref{sin-bifurcation2}) but the even entries are modified
as follows:
\begin{equation}
\label{sin-bifurcation2a} {\bf g}^{(2)}_j(\mbox{\boldmath $\theta$})
\to {\bf g}^{(2)}_j(\mbox{\boldmath $\theta$}) + \sin(\theta_j -
\theta_{j-2}) + \sin(\theta_j - \theta_{j+2}) + \sin(\theta_j -
\theta_{j+4}), \qquad j = 2,4,6,8.
\end{equation}
The vector ${\bf g}_2 = {\bf g}^{(2)}(\mbox{\boldmath $\theta$})$
can be computed for each of 35 one-parameter asymmetric vortex
solutions, starting with the first three solutions:
$$
\mbox{(a)} \; {\bf g}_2 = \left( \begin{array}{c} 2 \\ 1 \\ 0 \\ -1
\\ -2 \\ -1 \\0 \\ 1 \end{array} \right) \; \sin \theta \cos \theta, \qquad
\mbox{(b)} \; {\bf g}_2 = \left( \begin{array}{c} 2  \\ 1 \\ -1 \\
-1 \\ 0 \\-1 \\-1 \\1 \end{array} \right) \; \sin \theta \cos
\theta, \qquad \mbox{(c)} \; {\bf g}_2 = \left( \begin{array}{c} 2 \\ 1 \\
-1 \\ -2 \\ 0 \\ 1 \\ -1 \\ 0 \end{array} \right) \sin \theta \cos
\theta.
$$
The second eigenvector ${\bf p}_1$ of the kernel of ${\cal M}_1$ is
related to derivatives of the family in $\theta$, e.g.
$$
\mbox{(a)} \; {\bf p}_1 = \left( \begin{array}{c} 0 \\ 1 \\ 2 \\ 3
\\ 4 \\ 3 \\2 \\ 1 \end{array} \right), \qquad \mbox{(b)} \;
{\bf p}_1 = \left( \begin{array}{c} 0 \\ 1 \\ 2 \\ 3 \\2 \\ 3 \\2 \\
1 \end{array} \right), \qquad \mbox{(c)} \; {\bf p}_1 = \left(
\begin{array}{c} 0 \\ 1 \\ 2 \\ 3 \\ 2 \\1 \\2 \\ 1 \end{array} \right).
$$
The Fredholm alternative condition $({\bf p}_1,{\bf g}_2) = 0$ fails
for all solutions of family (iii) but one, excluding the special
values $\theta = \{ 0, \frac{\pi}{2},\pi \}$. The only solution of
family (iii), where ${\bf g}_2 = {\bf 0}$, is characterized by the
alternating signs of coefficients $a_j = \cos(\theta_{j+1} -
\theta_j)$ for $1 \leq j \leq N$.

{\bf Case $M \geq 3$:} The solution $\mbox{\boldmath $\phi$}^{(1)}
\in \omega \subset \Omega$ of the inhomogeneous problem
(\ref{2problem}) is given by (\ref{solution31}) and
(\ref{solution32}), except for the four interior corner nodes
$(2,2)$,$(M,2)$,$(M,M)$, and $(2,M)$, where
\begin{equation}
\label{solution33} \phi_{n,m}^{(1)} = e^{i \theta_{j-1}} + e^{i
\theta_{j+1}}, \qquad j = 1, M+1, 2M+1, 3M+1.
\end{equation}
The correction term ${\bf g}^{(2)}(\mbox{\boldmath $\theta$})$ is
given by (\ref{sin-bifurcation2}), except for the adjacent entries
to the four corner nodes on the contour $S_M$: $(1,1)$, $(1,M+1)$,
$(M+1,M+1)$, and $(M+1,1$), which are modified by:
\begin{eqnarray}
\nonumber {\bf g}^{(2)}_j(\mbox{\boldmath $\theta$}) & \to & {\bf
g}^{(2)}_j(\mbox{\boldmath $\theta$}) + \sin(\theta_j -
\theta_{j-2}), \qquad j = 2, M+2, 2M+2, 3M+2, \\
\label{sin-bifurcation2b} {\bf g}^{(2)}_j(\mbox{\boldmath $\theta$})
& \to & {\bf g}^{(2)}_j(\mbox{\boldmath $\theta$}) + \sin(\theta_j -
\theta_{j+2}), \qquad j = M, 2M, 3M, 4M.
\end{eqnarray}
Again, there is only one solution of family (iii), where ${\bf g}_2
= {\bf 0}$, which is characterized by the alternating signs of
coefficients $a_j = \cos(\theta_{j+1} - \theta_j)$ for $1 \leq j
\leq N$. All other solutions of family (iii) do not satisfy the
Fredholm alternative condition $({\bf p}_1,{\bf g}_2) = 0$.

By using results of these computations, we classify continuations of
solutions of the super-symmetric solitons of family (i) and
asymmetric vortices of family (iii). Let ${\cal M}_2$ be the
Jacobian matrix computed from the bifurcation function ${\bf
g}^{(2)}(\mbox{\boldmath $\theta$})$, given in
(\ref{sin-bifurcation2}), (\ref{sin-bifurcation2a}) and
(\ref{sin-bifurcation2b}). Since $({\bf p}_1,{\bf g}_2) \neq 0$ for
$\theta \neq \{ 0, \frac{\pi}{2},\pi\}$, except for the case of
sign-alternating set $\{a_j\}_{j=1}^N$ with $a_j = (-1)^j a$, it
follows from regular perturbation theory that $({\bf p}_1, {\cal
M}_2 {\bf p}_1) \neq 0$. Therefore, the second zero eigenvalue of
${\cal M}_1$ bifurcates off zero for the matrix ${\cal M}_1 +
\epsilon {\cal M}_2$. By Proposition \ref{proposition-sufficient}
(which needs to be modified for the Jacobian matrix ${\cal M}_1 +
\epsilon {\cal M}_2$), the super-symmetric solutions of family (i),
which are different from sign-alternating sets $a_j = (-1)^j a$, are
uniquely continued to discrete solitons (see Definition
\ref{definition-soliton}).

By Proposition \ref{lemma-nonexistence}, all asymmetric vortices of
family (iii), except for the sign-alternating set $a_j =
\cos(\theta_{j+1} - \theta_j) = (-1)^{j+1} \cos \theta$, $1\leq j
\leq N$, can not be continued to $\epsilon \neq 0$. The only
solution which can be continued up to the second-order reductions
has the explicit form:
\begin{equation}
\label{asymmetric-vortices} \theta_{4j-3} = 2 \pi (j-1), \quad
\theta_{4j-2} = \theta_{4j-3} + \theta, \qquad \theta_{4j-1} =
\theta_{4j-3} + \pi, \quad \theta_{4j} = \theta_{4j-3} + \pi +
\theta,
\end{equation}
where $1 \leq j \leq M$ and $0 \leq \theta \leq \pi$. This solution
includes two particular cases of super-symmetric solitons of family
(i) for $\theta = 0$ and $\theta = \pi$ and super-symmetric vortices
of family (ii) for $\theta = \frac{\pi}{2}$. Continuation of the
solution (\ref{asymmetric-vortices}) must be considered beyond the
second-order reductions.

\subsection{Jacobian matrix of the second-order reductions}

The Jacobian matrix ${\cal M}_1$ of the first-order reductions is
empty for super-symmetric vortices of family (ii) with $L = M$. In
order to study stability of super-symmetric vortices, we need to
compute the Jacobian matrix ${\cal M}_2$ from the second-order
bifurcation function ${\bf g}^{(2)}(\mbox{\boldmath $\theta$})$,
given in (\ref{sin-bifurcation2}), (\ref{sin-bifurcation2a}), and
(\ref{sin-bifurcation2b}). These computations are developed
separately for three cases $M = 1$, $M = 2$, and $M \geq 3$.

{\bf Case $M = 1$:} Non-zero elements of ${\cal M}_2$ are given by:
\begin{eqnarray}
\label{Melements4} ({\cal M}_2)_{i,j} = \left\{ \begin{array}{lcl}
+1, & \quad & i = j, \\ -\frac{1}{2}, & \quad & i = j \pm 2 \\
0, & \quad & |i - j | \neq 0,2 \end{array} \right.
\end{eqnarray}
or explicitly for $N = 4$:
\begin{equation}
\label{M2a} {\cal M}_2 = \left( \begin{array}{ccccc} 1 & 0 & -1 & 0 \\
0 & 1 & 0 & -1 \\ -1 & 0 & 1 & 0 \\ 0 & -1 & 0 & 1 \end{array}
\right).
\end{equation}
The matrix ${\cal M}_2$ has four eigenvalues: $\lambda_1 = \lambda_2
= 2$ and $\lambda_3 = \lambda_4 = 0$. The two eigenvectors for the
zero eigenvalue are ${\bf p}_3 = (1,0,1,0)^T$ and ${\bf p}_4 =
(0,1,0,1)^T$. The eigenvector ${\bf p}_4$ corresponds to the
derivative of the asymmetric vortex (\ref{asymA}) with respect to
parameter $\theta$, while the eigenvector ${\bf p}_0 = {\bf p}_3 +
{\bf p}_4$ corresponds to the shift due to gauge transformation.

{\bf Case $M = 2$:} The Jacobian matrix ${\cal M}_2$ is given in
(\ref{Melements4}) except for the even entries which are modified as
follows:
\begin{eqnarray}
\label{Melements4b} ({\cal M}_2)_{i,j} \to ({\cal M}_2)_{i,j} +
\left\{ \begin{array}{lcl}
-1, & \quad & i = j, \\ + 1, & \quad & i = j \pm 2 \\
-1, & \quad & i = j \pm 4 \end{array} \right. \qquad j = 2,4,6,8.
\end{eqnarray}
The matrix ${\cal M}_2$ for $N = 8$ takes the explicit form:
\begin{equation}
{\cal M}_2 = \left( \begin{array}{ccccccccc}
1 & 0 & -\frac{1}{2} & 0 & 0 & 0 & -\frac{1}{2} & 0 \\
0 & 0 & 0 & \frac{1}{2} & 0 & -1 & 0 & \frac{1}{2}  \\
-\frac{1}{2} & 0 & 1 & 0 & -\frac{1}{2} & 0 & 0 & 0 \\
0 & \frac{1}{2} & 0 & 0 & 0 & \frac{1}{2} & 0 & -1 \\
0 & 0 & -\frac{1}{2} & 0 & 1 & 0 & -\frac{1}{2} & 0 \\
0 & -1 & 0 & \frac{1}{2} & 0 & 0 & 0 & \frac{1}{2} \\
-\frac{1}{2} & 0 & 0 & 0 & -\frac{1}{2} & 0 & 1 & 0 \\
0 & \frac{1}{2} & 0 & -1 & 0 & \frac{1}{2} & 0 & 0
\end{array} \right).
\end{equation}
The eigenvalue problem for ${\cal M}_2$ decouples into two linear
difference equations with constant coefficients:
\begin{eqnarray*}
2 x_j - x_{j+1} - x_{j-1} = 2 \lambda x_j, \qquad j = 1,2,3,4 \\
- 2 y_{j+2} + y_{j+1} + y_{j-1} = 2 \lambda y_j, \qquad j = 1,2,3,4,
\end{eqnarray*}
subject to the periodic boundary conditions for $x_j$ and $y_j$. By
the discrete Fourier transform, see the proof of Lemma
\ref{lemma-alternating}, the first problem has eigenvalues:
$\lambda_1 = 1$, $\lambda_2 = 2$, $\lambda_3 = 1$, and $\lambda_4 =
0$, while the second problem has eigenvalues: $\lambda_5 = 1$,
$\lambda_6 = -2$, $\lambda_7 = 1$, and $\lambda_8 = 0$. The two
eigenvectors for the zero eigenvalue are ${\bf p}_4 =
(1,0,1,0,1,0,1,0)^T$ and ${\bf p}_8 = (0,1,0,1,0,1,0,1)^T$, where
the eigenvector ${\bf p}_8$ corresponds to the derivative of the
asymmetric vortex (\ref{asymmetric-vortices}) with respect to
parameter $\theta$ and the eigenvector ${\bf p}_0 = {\bf p}_4 + {\bf
p}_8$ corresponds to the shift due to gauge transformation.

{\bf Case $M \geq 3$:} The Jacobian matrix ${\cal M}_2$ is given in
(\ref{Melements4}), except for the adjacent entries to the four
corner nodes on the contours $S_M$: $(1,1)$, $(1,M+1)$, $(M+1,M+1)$,
and $(M+1,1$), which are modified by
\begin{eqnarray}
\label{Melements4c} ({\cal M}_2)_{i,j} \to ({\cal M}_2)_{i,j} +
\left\{ \begin{array}{lcl}
-1, & \quad & i = j = 2,M,M+2,2M,2M+2,3M,3M+2,4M, \\ + 1, & \quad & i = j - 2 = M,2M,3M,4M \\
+1, & \quad & i = j + 2 = 2,M+2,2M+2,3M+2 \end{array} \right. \qquad
\end{eqnarray}
The matrix ${\cal M}_2$ for $N = 12$ takes the explicit form:
\begin{equation}
{\cal M}_2 = \left( \begin{array}{ccccccccccccc}
1 & 0 & -\frac{1}{2} & 0 & 0 & 0 & 0 & 0 & 0 & 0 & -\frac{1}{2} & 0 \\
0 & 0 & 0 & -\frac{1}{2} & 0 & 0 & 0 & 0 & 0 & 0 & 0 & \frac{1}{2}  \\
-\frac{1}{2} & 0 & 0 & 0 & \frac{1}{2} & 0 & 0 & 0 & 0 & 0 & 0 & 0 \\
0 & -\frac{1}{2} & 0 & 1 & 0 & -\frac{1}{2} & 0 & 0 & 0 & 0 & 0 & 0 \\
0 & 0 & \frac{1}{2} & 0 & 0 & 0 & -\frac{1}{2} & 0 & 0 & 0 & 0 & 0 \\
0 & 0 & 0 & -\frac{1}{2} & 0 & 0 & 0 & \frac{1}{2} & 0 & 0 & 0 & 0 \\
0 & 0 & 0 & 0 & -\frac{1}{2} & 0 & 1 & 0 & -\frac{1}{2} & 0 & 0 & 0 \\
0 & 0 & 0 & 0 & 0 & \frac{1}{2} & 0 & 0 & 0 & -\frac{1}{2} & 0 & 0
\\ 0 & 0 & 0 & 0 & 0 & 0 & -\frac{1}{2} & 0 & 0 & 0 & \frac{1}{2} & 0
\\ 0 & 0 & 0 & 0 & 0 & 0 & 0 & -\frac{1}{2} & 0 & 1 & 0 &
-\frac{1}{2} \\ -\frac{1}{2} & 0 & 0 & 0 & 0 & 0 & 0 & 0 & \frac{1}{2} & 0 & 0 & 0 \\
0 & \frac{1}{2} & 0 & 0 & 0 & 0 & 0 & 0 & 0 & -\frac{1}{2} & 0 & 0
\end{array} \right).
\end{equation}
The eigenvalue problem for ${\cal M}_2$ decouples into eigenvalue
problems for two $6$-by-$6$ matrix, which are related by the
Toeplitz transformation. As a result, the spectra of these two
matrices are identical with the eigenvalues, obtained with the use
of MATLAB:
\begin{eqnarray*}
\lambda_1 = \lambda_7 = -0.780776, \;\; \lambda_2 = \lambda_8 =
-0.5, \;\;\lambda_3 = \lambda_9 = 0, \\ \lambda_4 = \lambda_{10} =
0.5, \;\;\lambda_5 = \lambda_{11} = 1.28078, \;\;\lambda_6 =
\lambda_{12} = 1.5.
\end{eqnarray*}
We confirm that the matrix ${\cal M}_2$ has exactly two zero
eigenvalues, one of which is related to the derivative of the
asymmetric vortex (\ref{asymmetric-vortices}) in $\theta$ and the
other one is related to the shift due to gauge transformation.

Computations of the matrix ${\cal M}_2$ for super-symmetric vortices
of family (ii) confirm the results of the second-order reductions
for asymmetric vortices of family (iii). Although all $N_3$
solutions of family (iii) reduce to the super-symmetric vortex of
family (ii) in the first-order reductions, it is the only family
(\ref{asymmetric-vortices}) that survives in the second-order
reductions, such that the super-symmetric vortex of family (ii) with
$L = M$ and $\theta = \frac{\pi}{2}$ can be deformed and continued
up to the second-order reductions to the asymmetric vortex
(\ref{asymmetric-vortices}).

All individual results on persistence of localized modes on the
discrete contour $S_M$ are summarized as follows.

\begin{Proposition}
\label{proposition-persistence} Consider the discrete soliton and
vortices of the nonlinear equations (\ref{2difference}) that
bifurcate from the limiting solution $\phi_{n,m}^{(0)}$ in
(\ref{2soliton}) on the discrete contour $S_M$ in (\ref{family-3}).
There exists a unique (modulo gauge transformation) continuation to
the domain $\epsilon \in {\cal O}(0)$ of discrete solitons of family
(i) in (\ref{discreteSoliton}), except for the case $l = 2M$ and
$a_j = (-1)^j a$, of symmetric vortices of family (ii) in
(\ref{symmetric-vortices}), except for the case $L = M$, and of
zero-parameter asymmetric vortices of family (iv) in
(\ref{asymmetric-vortices2}). Asymmetric vortices of family (iii) in
(\ref{asymmetric-vortices1}) can not be continued to the domain
$\epsilon \in {\cal O}(0)$, except for the only solution
(\ref{asymmetric-vortices}).
\end{Proposition}

\begin{Hypothesis}
\label{hypothesis-persistence} There exists a one-parameter (modulo
gauge transformation) continuation to the domain $\epsilon \in {\cal
O}(0)$ of the family of asymmetric vortices
(\ref{asymmetric-vortices}) with $0 \leq \theta \leq \pi$, which
includes the one-parameter continuation of the two exceptions of
Proposition \ref{proposition-persistence}.
\end{Hypothesis}

We will not be proving Hypothesis \ref{hypothesis-persistence}
beyond the second-order reductions. Instead, we use Proposition
\ref{proposition-persistence} to study stability of persistent
localized modes of the discrete NLS equation (\ref{2difference}).

\section{Stability of discrete vortices}
\label{section-stability}

The spectral stability of discrete vortices is studied with the
standard linearization:
\begin{equation}
\label{linearization} u_{n,m}(t) = e^{i (1 - 4 \epsilon) t + i
\theta_0} \left( \phi_{n,m} + a_{n,m} e^{\lambda t} + \bar{b}_{n,m}
e^{\bar{\lambda} t} \right), \qquad (n,m) \in \Z^2,
\end{equation}
where $\lambda \in \C$ and $(a_{n,m},b_{n,m}) \in \C^2$ solve the
linear eigenvalue problem on $(n,m) \in \Z^2$:
\begin{eqnarray*}
\nonumber \left(1 - 2 |\phi_{n,m}|^2\right) a_{n,m} - \phi_{n,m}^2
b_{n,m} - \epsilon \left( a_{n+1,m} + a_{n-1,m} + a_{n,m+1} + a_{n,m-1}
\right) & = & i \lambda a_{n,m}, \\
- \bar{\phi}_{n,m}^2 a_{n,m} + \left(1 - 2 |\phi_{n,m}|^2\right)
b_{n,m} - \epsilon \left( b_{n+1,m} + b_{n-1,m} + b_{n,m+1} +
b_{n,m-1} \right) & = & -i \lambda b_{n,m}.
\end{eqnarray*}
The stability problem (\ref{eigenvalue}) can be formulated in the
matrix-vector form:
\begin{equation}
\label{eigenvalue} {\cal H} \mbox{\boldmath $\psi$} = i \lambda
{\cal \sigma} \mbox{\boldmath $\psi$},
\end{equation}
where $\mbox{\boldmath $\psi$} \in \Omega \times \Omega$ consists of
2-blocks of $(a_{n,m},b_{n,m})^T$, ${\cal H}$ is defined by the
linearization operator (\ref{energy}), and ${\cal \sigma}$ consists
of $2$-by-$2$ blocks of
$$
\left( \begin{array}{cc} 1 & 0 \\ 0 & -1 \end{array} \right).
$$
The discrete vortex is called spectrally unstable if there exists
$\lambda$ and $\mbox{\boldmath $\psi$} \in \Omega \times \Omega$ in
the problem (\ref{eigenvalue}), such that ${\rm Re}(\lambda) > 0$.
Otherwise, the discrete vortex is called weakly spectrally stable.
By using the Taylor series (\ref{TaylorPhi}), the linearized
operator ${\cal H}$ is expanded as follows:
\begin{equation}
\label{series-H} {\cal H} = {\cal H}^{(0)} + \epsilon {\cal H}^{(1)}
+ \epsilon^2 {\cal H}^{(2)} + {\rm O}(\epsilon^3),
\end{equation}
where ${\cal H}^{(0)}$ is defined in (\ref{energy-01}), while the
first-order and second-order corrections are given by {\small
\begin{eqnarray*}
{\cal H}_{n,m}^{(1)} = - 2  \left(
\begin{array}{cc} \bar{\phi}_{n,m}^{(0)} \phi_{n,m}^{(1)} + \phi_{n,m}^{(0)}
\bar{\phi}_{n,m}^{(1)} & \phi_{n,m}^{(0)} \phi_{n,m}^{(1)} \\
\bar{\phi}_{n,m}^{(0)} \bar{\phi}_{n,m}^{(1)} &
\bar{\phi}_{n,m}^{(0)} \phi_{n,m}^{(1)} + \phi_{n,m}^{(0)}
\bar{\phi}_{n,m}^{(1)}
\end{array} \right) - \left( \delta_{+1,0} + \delta_{-1,0} + \delta_{0,+1} +
\delta_{0,-1} \right) \left(
\begin{array}{cc} 1 & 0 \\ 0 & 1 \end{array} \right)
\end{eqnarray*}
and
\begin{eqnarray*}
{\cal H}_{n,m}^{(2)} = - 2  \left(
\begin{array}{cc} \bar{\phi}_{n,m}^{(0)} \phi_{n,m}^{(2)} + \phi_{n,m}^{(0)}
\bar{\phi}_{n,m}^{(2)} & \phi_{n,m}^{(0)} \phi_{n,m}^{(2)} \\
\bar{\phi}_{n,m}^{(0)} \bar{\phi}_{n,m}^{(2)} &
\bar{\phi}_{n,m}^{(0)} \phi_{n,m}^{(2)} + \phi_{n,m}^{(0)}
\bar{\phi}_{n,m}^{(2)}
\end{array} \right) - \left(
\begin{array}{cc} 2 |\phi_{n,m}^{(1)}|^2 & \phi_{n,m}^{(1) 2} \\
\bar{\phi}_{n,m}^{(1) 2} & 2 |\phi_{n,m}^{(1)}|^2 \end{array}
\right).
\end{eqnarray*}
}

It is clear from the explicit form (\ref{energy-01}) that the
spectrum of ${\cal H}^{(0)} \mbox{\boldmath $\varphi$} = \gamma
\mbox{\boldmath $\varphi$}$ has exactly $N$ negative eigenvalues
$\gamma = -2$, $N$ zero eigenvalues $\gamma = 0$ and infinitely many
positive eigenvalues $\gamma = +1$. The negative and zero
eigenvalues $\gamma = -2$ and $\gamma = 0$ map to $N$ double zero
eigenvalues $\lambda = 0$ in the eigenvalue problem ${\cal \sigma}
{\cal H}^{(0)} \mbox{\boldmath $\psi$} = i \lambda \mbox{\boldmath
$\psi$}$. The positive eigenvalues $\gamma = +1$ map to the
infinitely many eigenvalues $\lambda = \pm i$. Since zero
eigenvalues of ${\cal \sigma} {\cal H}^{(0)}$ are isolated from the
rest of the spectrum  of ${\cal \sigma} {\cal H}^{(0)}$, their
splitting can be studied through regular perturbation theory
\cite{HJ85}. On the other hand, if the discrete vortex solutions
$\phi_{n,m}$ for $(n,m) \in \Z^2$ decays sufficiently fast as $|n| +
|m| \to \infty$, the continuous spectral bands of ${\cal \sigma H}$
are located on the imaginary axis of $\lambda$ near the points
$\lambda = \pm i$, similarly to the case $\phi_{n,m} = 0$ for $(n,m)
\in \Z^2$ \cite{LL92}. Therefore, the infinite-dimensional part of
the spectrum does not produce any unstable eigenvalues ${\rm
Re}(\lambda) > 0$ in the stability problem (\ref{eigenvalue}) with
small $\epsilon \in {\cal O}(0)$. We shall consider how zero
eigenvalues of ${\cal H}^{(0)}$ and $\sigma {\cal H}^{(0)}$ split as
$\epsilon \neq 0$ for solutions of the nonlinear equations
(\ref{2difference}), which are categorized by Propositions
\ref{lemma-asymmetric} and \ref{proposition-persistence}.

\subsection{Splitting of zero eigenvalues in the first-order reductions}

The splitting of zero eigenvalues of ${\cal H}$ is related to the
Lyapunov--Schmidt reductions of the nonlinear equations
(\ref{nonlinear-equations}). We show that the same matrix ${\cal
M}_1$, which gives the Jacobian of the bifurcation functions ${\bf
g}^{(1)}(\mbox{\boldmath $\theta$})$, defines also small eigenvalues
of ${\cal H}$ that bifurcate from zero eigenvalues of ${\cal
H}^{(0)}$ in the first-order reductions.

\begin{Lemma}
\label{lemma-splitting} Let the Jacobian matrix ${\cal M}_1$ have
eigenvalues $\mu_j^{(1)}$, $1 \leq j \leq N$. The eigenvalue problem
${\cal H} \mbox{\boldmath $\varphi$} = \gamma \mbox{\boldmath
$\varphi$}$ has $N$ eigenvalues $\gamma_j$ in $\epsilon \in {\cal
O}(0)$, such that
\begin{equation}
\label{limiting-1} \lim_{\epsilon \to 0} \frac{\gamma_j}{\epsilon} =
\mu_j^{(1)}, \qquad 1 \leq j \leq N.
\end{equation}
\end{Lemma}

\begin{Proof}
We assume that there exists an analytical solution $\mbox{\boldmath
$\phi$}(\epsilon)$ of the nonlinear equations
(\ref{nonlinear-equations}). The Taylor series of $\mbox{\boldmath
$\phi$}(\epsilon)$ is defined by (\ref{TaylorPhi}). By taking the
derivative in $\epsilon$, we rewrite the problem
(\ref{nonlinear-equations}) in the form:
\begin{equation}
\label{derivative-H} {\cal H}_p \mbox{\boldmath $\psi$}(\epsilon) +
\epsilon {\cal H}_s \mbox{\boldmath $\psi$}(\epsilon) + {\cal H}_s
\mbox{\boldmath $\phi$}(\epsilon) = {\bf 0}, \qquad \mbox{\boldmath
$\psi$}(\epsilon) = \mbox{\boldmath $\phi$}'(\epsilon),
\end{equation}
where the linearization operator (\ref{energy}) is represented as
${\cal H} = {\cal H}_p + \epsilon {\cal H}_s$. Using the series
(\ref{TaylorPhi}) and (\ref{series-H}), we have the linear
inhomogeneous equation:
\begin{equation}
\label{inhomog-equation-1} {\cal H}^{(0)} \mbox{\boldmath
$\phi$}^{(1)} + {\cal H}_s \mbox{\boldmath $\phi$}^{(0)} = {\bf 0}.
\end{equation}
Let ${\bf e}_j(\mbox{\boldmath $\theta$})$, $j = 1,...,N$ be
eigenvectors of the kernel of ${\cal H}^{(0)}$. Each eigenvector
${\bf e}_j(\mbox{\boldmath $\theta$})$ contains the only non-zero
block $i (e^{i \theta_j},-e^{-i\theta_j})^T$ at the $j$-th position,
which corresponds to the node $(n,m)$ on the contour $S_M$. It is
clear that the eigenvectors are orthogonal as follows:
\begin{equation}
\label{orthogonality-relations-1} ({\bf e}_i(\mbox{\boldmath
$\theta$}),{\bf e}_j(\mbox{\boldmath $\theta$})) = 2 \delta_{i,j},
\qquad 1 \leq i,j \leq N.
\end{equation}
Let $\hat{\bf e}_j(\mbox{\boldmath $\theta$})$, $j = 1,...,N$ be
generalized eigenvectors, such that each eigenvector $\hat{\bf
e}_j(\mbox{\boldmath $\theta$})$  contains the only non-zero block $
(e^{i \theta_j},e^{-i\theta_j})^T$ at the $j$-th position. Direct
computations show that
\begin{equation}
\label{orthogonality-relations-2} {\cal H}^{(0)} \hat{\bf
e}_j(\mbox{\boldmath $\theta$}) = 2 i \sigma {\bf
e}_j(\mbox{\boldmath $\theta$}),\qquad 1 \leq j \leq N.
\end{equation}
The limiting solution (\ref{2soliton}) can be represented as
follows:
$$
\mbox{\boldmath $\phi$}^{(0)}(\mbox{\boldmath $\theta$}) =
\sum_{j=1}^N \hat{\bf e}_j(\mbox{\boldmath $\theta$}).
$$
By comparing the inhomogeneous equation (\ref{inhomog-equation-1})
with the definition (\ref{g-function}) of the bifurcation function
${\bf g}(\mbox{\boldmath $\theta$})$ and its Taylor series
(\ref{Taylor}), we have the correspondence:
$$
g_j^{(1)}(\mbox{\boldmath $\theta$}) = \frac{1}{2} \left( {\bf
e}_j(\mbox{\boldmath $\theta$}), {\cal H}_s \mbox{\boldmath
$\phi$}^{(0)}(\mbox{\boldmath $\theta$}) \right).
$$
Consider a perturbation to a fixed point of ${\bf
g}^{(1)}(\mbox{\boldmath $\theta$}_*) = {\bf 0}$ in the form
$\mbox{\boldmath $\theta$} = \mbox{\boldmath $\theta$}_* + \epsilon
{\bf c},$ where ${\bf c} = (c_1,c_2,...,c_N)^T \in \mathbb{R}^N$. It
is clear that
$$
\mbox{\boldmath $\phi$}^{(0)}(\mbox{\boldmath $\theta$}) =
\mbox{\boldmath $\phi$}^{(0)} + \epsilon \sum_{i=1}^N c_i {\bf e}_i
+ {\rm O}(\epsilon^2), \qquad {\bf e}_j(\mbox{\boldmath $\theta$}) =
{\bf e}_j - \epsilon c_j \hat{\bf e}_j + {\rm O}(\epsilon^2),
$$
where $\mbox{\boldmath $\phi$}^{(0)} = \mbox{\boldmath
$\phi$}^{(0)}(\mbox{\boldmath $\theta$}_*)$, ${\bf e}_j = {\bf
e}_j(\mbox{\boldmath $\theta$}_*)$, and $\hat{\bf e}_j = \hat{\bf
e}_j(\mbox{\boldmath $\theta$}_*)$. By expanding the bifurcation
function ${\bf g}^{(1)}(\mbox{\boldmath $\theta$})$ near
$\mbox{\boldmath $\theta$} = \mbox{\boldmath $\theta$}_*$, we define
the Jacobian matrix ${\cal M}_1$:
$$
g_j^{(1)}(\mbox{\boldmath $\theta$}) = g_j^{(1)} + \epsilon \left(
{\cal M}_1 {\bf c} \right)_j + {\rm O}(\epsilon^2),
$$
where
\begin{equation}
\label{correction-term1} \left( {\cal M}_1 {\bf c} \right)_j =
\frac{1}{2} \sum_{i=1}^n \left( {\bf e}_j, {\cal H}_s {\bf e}_i
\right) c_i - \frac{1}{2} c_j \sum_{i=1}^N \left( \hat{\bf e}_j,
{\cal H}_s \hat{\bf e}_i \right).
\end{equation}
On the other hand, the regular perturbation series for small
eigenvalues of the problem ${\cal H} \mbox{\boldmath $\varphi$} =
\gamma \mbox{\boldmath $\varphi$}$ are defined as follows:
\begin{equation}
\label{regular-series1} \mbox{\boldmath $\varphi$} = \mbox{\boldmath
$\varphi$}^{(0)} + \epsilon \mbox{\boldmath $\varphi$}^{(1)} +
\epsilon^2 \mbox{\boldmath $\varphi$}^{(2)} + {\rm O}(\epsilon^3),
\qquad \gamma = \epsilon \gamma_1 + \epsilon^2 \gamma_2 + {\rm
O}(\epsilon^3),
\end{equation}
where $\mbox{\boldmath $\varphi$}^{(0)} = \sum_{j=1}^N c_j {\bf
e}_j$, according to the kernel of ${\cal H}^{(0)}$. The first-order
correction term $\mbox{\boldmath $\varphi$}^{(1)}$ satisfies the
inhomogeneous equation:
\begin{equation}
\label{inhomogeneous-equation1} {\cal H}^{(0)} \mbox{\boldmath
$\varphi$}^{(1)} + {\cal H}^{(1)} \mbox{\boldmath $\varphi$}^{(0)} =
\gamma_1 \mbox{\boldmath $\varphi$}^{(0)}.
\end{equation}
Projection to the kernel of ${\cal H}^{(0)}$ gives the eigenvalue
problem for $\gamma_1$:
\begin{equation}
\label{eigenvalue-problem1} \frac{1}{2} \sum_{i=1}^N \left( {\bf
e}_j, {\cal H}^{(1)} {\bf e}_i \right) c_i = \gamma_1 c_j.
\end{equation}
By direct computations:
$$
- \frac{1}{2} \sum_{i=1}^N \left( \hat{\bf e}_j, {\cal H}_s \hat{\bf
e}_i \right) = \cos(\theta_j - \theta_{j+1}) + \cos(\theta_j -
\theta_{j-1}) = \sum_{i=1}^N \left( {\bf e}_j, {\cal H}_p^{(1)} {\bf
e}_i \right),
$$
such that the limiting relation (\ref{limiting-1}) follows from
(\ref{correction-term1}) and (\ref{eigenvalue-problem1}) with ${\cal
H}^{(1)} = {\cal H}^{(1)}_p + {\cal H}_s$.
\end{Proof}

We apply results of Lemma \ref{lemma-splitting} to the solutions of
the first-order reductions, which are described in families
(i)--(iv) of Propositions \ref{lemma-asymmetric} and
\ref{proposition-persistence}. The numbers of negative, zero and
positive eigenvalues of ${\cal M}_1$ are denoted as $n({\cal M}_1)$,
$z({\cal M}_1)$ and $p({\cal M}_1)$ respectively. These numbers
determine the numbers of small negative and positive eigenvalues of
${\cal H}$ for small positive $\epsilon$. They are predicted from
Lemmas \ref{lemma-periodic}, \ref{lemma-constant} and
\ref{lemma-alternating}.

For family (i), we compute the parameter $A_1$ in Lemma
\ref{lemma-periodic} as $A_1 = (-1)^{N-l} (2l-N)$, where $l$ is
defined in Proposition \ref{lemma-asymmetric}. By Lemma
\ref{lemma-periodic}, we have $n({\cal M}_1) = N-l-1$, $z({\cal
M}_1) = 1$, and $p({\cal M}_1) = l$ for $0 \leq l \leq 2M-1$ and
$n({\cal M}_1) = N-l$, $z({\cal M}_1) = 1$, and $p({\cal M}_1) =
l-1$ for $2M+1 \leq l \leq 4M$. In the case of super-symmetric
solitons for $l = 2M$, by Lemma \ref{lemma-alternating}, we have
$n({\cal M}_1) = 2M-1$, $z({\cal M}_1) = 2$, and $p({\cal M}_1) =
2M-1$.

For family (ii), by Lemma \ref{lemma-constant}, we have $n({\cal
M}_1) = 0$, $z({\cal M}_1) = 1$, and $p({\cal M}_1) = N-1$ for $1
\leq L \leq M-1$ and $n({\cal M}_1) = N-1$, $z({\cal M}_1) = 1$, and
$p({\cal M}_1) = 0$ for $M+1 \leq L \leq 2M - 1$, where $L$ is
defined in Proposition \ref{lemma-asymmetric}. The case of
super-symmetric vortices $L = 2M$ can only be studied in the
second-order reductions, since ${\cal M}_1 = 0$.

The family (iii) is represented by the only one-parameter solution
(\ref{asymmetric-vortices}) for each $L = M$. By Lemma
\ref{lemma-alternating}, we have $n({\cal M}_1) = 2M-1$, $z({\cal
M}_1) = 2$, and $p({\cal M}_1) = 2M-1$ for $0 \leq \theta \leq \pi$,
excluding the case of super-symmetric vortices $\theta =
\frac{\pi}{2}$ but including the case of super-symmetric solitons
$\theta = 0$ and $\theta = \pi$.

The family (iv) is characterized by the value of $\cos \theta_* \neq
0$, $L \neq M$, and $1 \leq n \leq N-1$, $n \neq 2M$, specified in
Proposition \ref{lemma-asymmetric}. The parameter $A_1$ in Lemma
\ref{lemma-periodic} is computed as $A_1 = (-1)^{N-n} (\cos
\theta_*)^{N-1} (2n-N)$, such that $z({\cal M}_1) = 1$ in all cases.
In the case $\cos \theta_* > 0$, by Lemma \ref{lemma-periodic}, we
have $n({\cal M}_1) = N-n-1$ and $p({\cal M}_1) = n$ for $1 \leq n
\leq 2M-1$ and $n({\cal M}_1) = N-n$ and $p({\cal M}_1) = n-1$ for
$2M+1 \leq n \leq N-1$. In the opposite case of $\cos \theta_* < 0$,
we have $n({\cal M}_1) = n$ and $p({\cal M}_1) = N-n-1$ for $1 \leq
n \leq 2M-1$ and $n({\cal M}_1) = n-1$ and $p({\cal M}_1) = N-n$ for
$2M+1 \leq n \leq N-1$.

The splitting of zero eigenvalue of ${\cal H}$ is related to splitting
of zero eigenvalues of $\sigma {\cal H}$ in the stability problem
(\ref{eigenvalue}).

\begin{Lemma}
\label{lemma-splitting000} Let the Jacobian matrix ${\cal M}_1$ have
eigenvalues $\mu_j^{(1)}$, $1 \leq j \leq N$. The eigenvalue problem
${\cal H} \mbox{\boldmath $\psi$} = i \lambda \sigma \mbox{\boldmath
$\psi$}$ has $N$ pairs of eigenvalues $\lambda_j$ in $\epsilon \in
{\cal O}(0)$, such that
\begin{equation}
\label{limiting-000} \lim_{\epsilon \to 0}
\frac{\lambda^2_j}{\epsilon} = 2 \mu_j^{(1)}, \qquad 1 \leq j \leq
N.
\end{equation}
\end{Lemma}

\begin{Proof}
The regular perturbation series for small eigenvalues of $\sigma
{\cal H}$ are defined as follows:
\begin{eqnarray}
\label{Lyapunoveigenvector1} \mbox{\boldmath $\psi$} =
\mbox{\boldmath $\psi$}^{(0)} + \sqrt{\epsilon} \mbox{\boldmath
$\psi$}^{(1)} + \epsilon \mbox{\boldmath $\psi$}^{(2)} + \epsilon
\sqrt{\epsilon} \mbox{\boldmath $\psi$}^{(3)} + {\rm O}(\epsilon^2),
\qquad \lambda = \sqrt{\epsilon} \lambda_1 + \epsilon \lambda_2 +
\epsilon \sqrt{\epsilon} \lambda_3 + {\rm O}(\epsilon^2),
\end{eqnarray}
where, due to the relations (\ref{orthogonality-relations-1}) and
(\ref{orthogonality-relations-2}), we have:
\begin{equation}
\label{zeroeigenvector1} \mbox{\boldmath $\psi$}^{(0)} = \sum_{j =
1}^N c_j {\bf e}_j, \qquad \mbox{\boldmath $\psi$}^{(1)} =
\frac{\lambda_1}{2}\sum_{j = 1}^N c_j \hat{\bf e}_j,
\end{equation}
according to the kernel and generalized kernel of $\sigma {\cal
H}^{(0)}$. The second-order correction term $\mbox{\boldmath
$\psi$}^{(2)}$ satisfies the inhomogeneous equation:
\begin{equation}
\label{Lyapunovlinearsystem1} {\cal H}^{(0)} \mbox{\boldmath
$\psi$}^{(2)} + {\cal H}^{(1)} \mbox{\boldmath $\psi$}^{(0)} = i
\lambda_1 {\cal \sigma} \mbox{\boldmath $\psi$}^{(1)} + i \lambda_2
{\cal \sigma} \mbox{\boldmath $\psi$}^{(0)}.
\end{equation}
Projection to the kernel of ${\cal H}^{(0)}$ gives the eigenvalue
problem for $\lambda_1$:
\begin{equation}
\label{Lyapunovreducedeigenvalue1} {\cal M}_1 {\bf c} =
\frac{\lambda_1^2}{2} {\bf c},
\end{equation}
where ${\bf c} = (c_1,c_2,...,c_N)^T$ and the matrix ${\cal M}_1$ is
the same as in the eigenvalue problem (\ref{eigenvalue-problem1}).
As a result, the relation (\ref{limiting-000}) is proved.
\end{Proof}

The numbers of negative, zero and positive eigenvalues of ${\cal
M}_1$, denoted as $n({\cal M}_1)$, $z({\cal M}_1)$ and $p({\cal
M}_1)$, are computed above. Let $r_1$, $z_1$, and $i_1$ be the
numbers of pairs of real, zero and imaginary eigenvalues of the
reduced eigenvalue problem (\ref{Lyapunovreducedeigenvalue1}). We
consider stability of families (i)--(iv), described in Propositions
\ref{lemma-asymmetric} and \ref{proposition-persistence}.

For family (i), we have $i_1 = N-l-1$, $z_1 = 1$, and $r_1 = l$ for
$0 \leq l \leq 2M-1$; $i_1 = N-l-1$, $z_1 = 2$, and $r_1 = l-1$ for
$l = 2M$; and $i_1 = N-l$, $z_1 = 1$, and $r_1 = l-1$ for $2M+1 \leq
l \leq N$, where $l$ is defined in Proposition
\ref{lemma-asymmetric}.

For family (ii), we have $i_1 = 0$, $z_1 = 1$, and $r_1 = N-1$ for
$1 \leq L \leq M-1$; $i_1 = 0$, $z_1 = N$, and $r_1 = 0$ for $L =
M$; and $i_1 = N-1$, $z_1 = 1$, and $p_1 = 0$ for $M+1 \leq L \leq
2M - 1$, where $L$ is defined in Proposition \ref{lemma-asymmetric}.

For the only one-parameter solution (\ref{asymmetric-vortices}) of
family (iii) for each $L = M$, we have $i_1 = 2M-1$, $z_1 = 2$, and
$r_1 = 2M-1$ for $0 \leq \theta \leq \pi$ and $\theta \neq
\frac{\pi}{2}$.

For family (iv) with $\cos \theta_* > 0$, we have $i_1 = N-n-1$,
$z_1 = 1$, and $r_1 = n$ for $1 \leq n \leq 2M-1$ and $i_1 = N-n$,
$z_1 = 1$, and $r_1 = n-1$ for $2M+1 \leq n \leq N-1$. In the
opposite case of $\cos \theta_* < 0$, we have $i_1 = n$, $z_1 = 1$,
and $r_1 = N-n-1$ for $1 \leq n \leq 2M-1$ and $i_1 = n-1$, $z_1 =
1$, and $r_1 = N-n$ for $2M+1 \leq n \leq N-1$.

There are several features which are not captured in the first-order
reductions. For super-symmetric solitons of family (i), when $l =
2M$ but $a_j \neq (-1)^j a$, the additional zero eigenvalue splits
at the second-order reductions, which leads to an additional
non-zero eigenvalues of the stability problem (\ref{eigenvalue}).
For super-symmetric vortices of family (ii), when $L = M$, the
matrix ${\cal M}_1 = 0$, such that non-zero eigenvalues occur only
in the second-order reductions. Finally, for symmetric vortices of
family (ii), multiple real non-zero eigenvalues of the first-order
reductions, according to the roots of $\sin^2 \frac{\pi n}{N}$ in
the explicit solution (\ref{explicit-eigenvalue}), split into the
complex domain in the second-order reductions. These questions are
studied next in the reverse order.

\subsection{Splitting of non-zero eigenvalues in the second-order reductions}

We continue the regular perturbation series
(\ref{Lyapunoveigenvector1}) to the second-order reductions. By
using the explicit first-order correction term (\ref{solution31}),
we compute the explicit solution:
\begin{equation}
\label{inhomog-solution-1a} \mbox{\boldmath $\psi$}^{(2)} =
\frac{\lambda_2}{2}\sum_{j = 1}^N c_j \hat{\bf e}_j + \frac{1}{2}
\sum_{j=1}^N (\sin(\theta_{j+1} - \theta_j ) c_{j+1} +
\sin(\theta_{j-1} - \theta_j) c_{j-1}) \hat{\bf e}_j + \sum_{j =
1}^N c_j \left( {\bf e}_{j + 1} + {\bf e}_{j-1} \right),
\end{equation}
where the vectors ${\bf e}_{j\pm 1}$ are obtained from ${\bf e}_j$
by shifts of non-zero elements of ${\bf e}_j$ from the node $(n,m) \in
S_M$ to the adjacent nodes $(n,m) \in \Z^2 \backslash S_M$. The
third-order correction term $\mbox{\boldmath $\psi$}^{(3)}$
satisfies the inhomogeneous equation:
\begin{equation}
\label{Lyapunovlinearsystem1a} {\cal H}^{(0)} \mbox{\boldmath
$\psi$}^{(3)} + {\cal H}^{(1)} \mbox{\boldmath $\psi$}^{(1)} = i
\lambda_1 {\cal \sigma} \mbox{\boldmath $\psi$}^{(2)} + i \lambda_2
{\cal \sigma} \mbox{\boldmath $\psi$}^{(1)} + + i \lambda_3 {\cal
\sigma} \mbox{\boldmath $\psi$}^{(0)}.
\end{equation}
Projection to the kernel of ${\cal H}^{(0)}$ gives the extended
eigenvalue problem for $\lambda_1$ and $\lambda_2$:
\begin{equation}
\label{Lyapunovreducedeigenvalue1a} {\cal M}_1 {\bf c} =
\frac{\lambda_1^2}{2} {\bf c} + \sqrt{\epsilon} \left( \lambda_1
\lambda_2 {\bf c} + \lambda_1 {\cal L}_1 {\bf c} \right),
\end{equation}
where the matrix ${\cal L}_1$ is defined by
\begin{eqnarray}
\label{LelementsA} ({\cal L}_1)_{i,j} = \left\{ \begin{array}{lcl}
\sin(\theta_j-\theta_i), & \quad & i = j \pm 1, \\
0, & \quad & |i - j | \neq 1 \end{array} \right.
\end{eqnarray}
subject to the periodic boundary conditions. Let $\gamma_j$ be an
eigenvalue of the symmetric matrix ${\cal M}_1$ with the linearly
independent eigenvector ${\bf c}_j$. Then,
\begin{equation}
\lambda_1 = \pm \sqrt{2 \gamma_j}, \qquad \lambda_2 = - \frac{({\bf
c}_j, {\cal L}_1 {\bf c}_j)}{({\bf c}_j,{\bf c}_j)}.
\end{equation}
Since the matrix ${\cal L}_1$ is skew-symmetric, the second-order
correction term $\lambda_2$ is purely imaginary, unless $({\bf c}_j,
{\cal L}_1 {\bf c}_j) = 0$. For discrete solitons of family (i), we
have $\sin(\theta_{j+1} - \theta_j) = 0$, such that ${\cal L}_1 = 0$
and $\lambda_2 = 0$.

For symmetric vortices of family (ii) with $L \neq M$, the matrix
${\cal M}_1$ has double eigenvalues, according to the roots of
$\sin^2 \frac{\pi n}{N}$ in the explicit solution
(\ref{explicit-eigenvalue}). Using the same discrete Fourier
transform as in the proof of Lemma \ref{lemma-constant}, one can
find the values of $\lambda_1$ and $\lambda_2$ in this case.

\begin{Lemma}
\label{lemma-constantA} Let all coefficients $a_j =
\cos(\theta_{j+1} - \theta_j)$ and $b_j = \sin(\theta_{j+1} -
\theta_j)$, $1 \leq j \leq N$ be the same: $a_j = a$ and $b_j = b$.
Eigenvalues of the reduced problem
(\ref{Lyapunovreducedeigenvalue1a}) are given explicitly:
\begin{equation}
\label{explicit-eigenvalueA} \lambda_1 = \pm \sqrt{8 a} \sin
\frac{\pi n}{N}, \quad \lambda_2 = - 2 i b \sin \frac{2\pi n}{N},
\quad 1 \leq n \leq N.
\end{equation}
\end{Lemma}

According to Lemma \ref{lemma-constantA}, all double roots of
$\lambda_1$  for $n \neq \frac{N}{2}$ and $n \neq N$ split along the
imaginary axis in $\lambda_2$. When $a > 0$, the splitting occurs in
the transverse directions to the real values of $\lambda_1$. When $a
< 0$, the splitting occurs in the longitudinal directions to the
imaginary values of $\lambda_1$. The simple roots at $n =
\frac{N}{2}$ and $n = N$ are not affected, since $\lambda_2 = 0$ for
$n = \frac{N}{2}$ and $n = N$.

For asymmetric vortices of family (iii), it follows from the
explicit solution (\ref{explicit-eigenvalue-alternating}) that
$\lambda_2 = 0$ for all roots, so that the splitting of eigenvalues
does not occur in the second-order reductions.

For asymmetric vortices of family (iv), the value of $\lambda_2$ can
not be computed in general.

\subsection{Splitting of zero eigenvalues in the second-order reductions}

We extend results of the regular perturbation series
(\ref{regular-series1}) and (\ref{Lyapunoveigenvector1}) to the case
${\cal M}_1 = 0$, which occurs for super-symmetric vortices of
family (ii) with charge $L = M$. It follows from the problem
(\ref{inhomogeneous-equation1}) with ${\cal M}_1 = 0$ that $\gamma_1
= 0$ and the first-order correction term $\mbox{\boldmath
$\varphi$}^{(1)}$ has the explicit form:
\begin{equation}
\label{correction-term2} \mbox{\boldmath $\varphi$}^{(1)} =
\frac{1}{2} \sum_{j=1}^N (c_{j+1} - c_{j-1}) \hat{\bf e}_j + \sum_{j
= 1}^N c_j \left( {\bf e}_{j + 1} + {\bf e}_{j-1} \right),
\end{equation}
where the vectors ${\bf e}_{j\pm 1}$ are the same as in the formula
(\ref{inhomog-solution-1a}). The second-order correction term
$\mbox{\boldmath $\varphi$}^{(2)}$ satisfies the inhomogeneous
equation:
\begin{equation}
\label{inhomogeneous-equation2} {\cal H}^{(0)} \mbox{\boldmath
$\varphi$}^{(2)} + {\cal H}^{(1)} \mbox{\boldmath $\varphi$}^{(1)} +
{\cal H}^{(2)} \mbox{\boldmath $\varphi$}^{(0)} = \gamma_2
\mbox{\boldmath $\varphi$}^{(0)}.
\end{equation}
Projection to the kernel of ${\cal H}^{(0)}$ gives the eigenvalue
problem for $\gamma_2$:
\begin{equation}
\label{eigenvalue-problem2} \frac{1}{2} \left( {\bf e}_j, {\cal
H}^{(1)} \mbox{\boldmath $\varphi$}^{(1)} \right) + \frac{1}{2}
\sum_{i=1}^N \left( {\bf e}_j, {\cal H}^{(2)} {\bf e}_i \right) c_i
= \gamma_2 c_j.
\end{equation}
By direct computations in the three separate cases, one can show
that the matrix on the left-hand-side of the eigenvalue problem
(\ref{eigenvalue-problem2}) is nothing but the matrix ${\cal M}_2$,
which is the Jacobian of the nonlinear function ${\bf
g}^{(2)}(\mbox{\boldmath $\theta$})$, computed for the
super-symmetric vortex of family (ii). Let the number of negative,
zero and positive eigenvalues of ${\cal M}_2$ be denoted as $n({\cal
M}_2)$, $z({\cal M}_2)$ and $p({\cal M}_2)$ respectively. For
super-symmetric vortices of family (ii), we have $n({\cal M}_2) =
0$, $z({\cal M}_2) = 2$ and $p({\cal M}_2) = 2$ for $M = 1$;
$n({\cal M}_2) = 1$, $z({\cal M}_2) = 2$ and $p({\cal M}_2) = 5$ for
$M = 2$; and $n({\cal M}_2) = 4$, $z({\cal M}_2) = 2$ and $p({\cal
M}_2) = 6$ for $M = 3$.

Splitting of zero eigenvalues of $\sigma {\cal H}$ is studied with
the regular perturbation series (\ref{Lyapunoveigenvector1}). When
${\cal M}_1 = 0$, it follows from the problem
(\ref{Lyapunovlinearsystem1}) that $\lambda_1 = 0$, such that the
regular perturbation series (\ref{Lyapunoveigenvector1}) can be
re-ordered as follows:
\begin{eqnarray}
\label{Lyapunoveigenvector2} \mbox{\boldmath $\psi$} =
\mbox{\boldmath $\psi$}^{(0)} + \epsilon \mbox{\boldmath
$\psi$}^{(1)} + \epsilon^2 \mbox{\boldmath $\psi$}^{(2)} + {\rm
O}(\epsilon^3), \qquad \lambda = \epsilon \lambda_1 + \epsilon^2
\lambda_2 + {\rm O}(\epsilon^3),
\end{eqnarray}
where
\begin{equation}
\label{zeroeigenvector2} \mbox{\boldmath $\psi$}^{(0)} = \sum_{j =
1}^N c_j {\bf e}_j, \qquad \mbox{\boldmath $\psi$}^{(1)} =
\mbox{\boldmath $\varphi$}^{(1)} + \frac{\lambda_1}{2}\sum_{j = 1}^N
c_j \hat{\bf e}_j,
\end{equation}
and $\mbox{\boldmath $\varphi$}^{(1)}$ is given by
(\ref{correction-term2}). The second-order correction term
$\mbox{\boldmath $\psi$}^{(2)}$ is found from the inhomogeneous
equation:
\begin{equation}
\label{Lyapunovlinearsystem2} {\cal H}^{(0)} \mbox{\boldmath
$\psi$}^{(2)} + {\cal H}^{(1)} \mbox{\boldmath $\psi$}^{(1)} + {\cal
H}^{(2)} \mbox{\boldmath $\psi$}^{(0)} = i \lambda_1 {\cal \sigma}
\mbox{\boldmath $\psi$}^{(1)} + i \lambda_2 {\cal \sigma}
\mbox{\boldmath $\psi$}^{(0)}.
\end{equation}
Projection to the kernel of ${\cal H}^{(0)}$ gives the eigenvalue
problem for $\lambda_1$:
\begin{equation}
\label{Lyapunovreducedeigenvalue2} {\cal M}_2 {\bf c} = \lambda_1
{\cal L}_2 {\bf c} + \frac{\lambda_1^2}{2} {\bf c},
\end{equation}
where ${\bf c} = (c_1,c_2,...,c_N)^T$, the matrix ${\cal M}_2$ is
the same as in the eigenvalue problem (\ref{eigenvalue-problem2}),
and the matrix ${\cal L}_2$ follows from the matrix ${\cal L}_1$ in
the form (\ref{LelementsA}) with $\sin(\theta_{j+1}-\theta_j) = 1$,
or explicitly:
\begin{eqnarray}
\label{Lelements} ({\cal L}_2)_{i,j} = \left\{ \begin{array}{lcl}
+1, & \quad & i = j - 1, \\ -1, & \quad & i = j + 1 \\
0, & \quad & |i - j | \neq 1 \end{array} \right.
\end{eqnarray}
subject to the periodic boundary conditions. Since ${\cal M}_2$ is
symmetric and ${\cal L}_2$ is skew-symmetric, the eigenvalues of the
problem (\ref{Lyapunovreducedeigenvalue2}) occur in pairs
$(\lambda_1,-\lambda_1)$. Computations of eigenvalues of the reduced
eigenvalue problem (\ref{Lyapunovreducedeigenvalue2}) are reported
in the three distinct cases: $M = 1$, $M = 2$ and $M \geq 3$.

{\bf Case $M = 1$:} The reduced eigenvalue problem
(\ref{Lyapunovreducedeigenvalue2}) takes the form of the difference
equation with constant coefficients:
$$
- c_{j+2} + 2 c_j - c_{j-2} = \lambda_1^2 c_j + 2 \lambda_1 \left(
c_{j+1} - c_{j-1} \right), \qquad j = 1,2,3,4,
$$
subject to the periodic boundary conditions. By the discrete Fourier
transform, the difference equation reduces to the characteristic
equation:
$$
\left( \lambda_1 + 2 i \sin \frac{\pi n}{2} \right)^2 = 0, \qquad n
= 1,2,3,4.
$$
The reduced eigenvalue problem (\ref{Lyapunovreducedeigenvalue2})
has two eigenvalues of algebraic multiplicity two at $\lambda_1 =
-2i$ and $\lambda_1 = 2i$ and zero eigenvalue of algebraic
multiplicity four.

{\bf Case $M = 2$:} The reduced eigenvalue problem
(\ref{Lyapunovreducedeigenvalue2}) takes the form of the system of
difference equations with constant coefficients:
\begin{eqnarray*}
- x_{j+1} + 2 x_j - x_{j-1} = \lambda_1^2 x_j + 2 \lambda_1 \left(
y_j - y_{j-1} \right),  \quad j = 1,2,3,4,\\
y_{j+1} - 2 y_{j+2} + y_{j-1} = \lambda_1^2 y_j + 2 \lambda_1 \left(
x_{j+1} - x_j \right), \quad j = 1,2,3,4,
\end{eqnarray*}
where $x_j = c_{2j-1}$ and $y_j = c_{2j}$ subject to the periodic
boundary conditions. The characteristic equation for the linear
system takes the explicit form:
$$
\lambda_1^4 - 2 \lambda_1^2 \left( 1 - (-1)^n - 8 \sin^2 \frac{\pi
n}{4} \right) + 8 \sin^2 \frac{\pi n}{4}  \left( 1 - (-1)^n - 2
\sin^2 \frac{\pi n}{4} \right) = 0, \qquad n = 1,2,3,4.
$$
The reduced eigenvalue problem (\ref{Lyapunovreducedeigenvalue2})
has three eigenvalues of algebraic multiplicity four at $\lambda_1 =
-\sqrt{2}i$, $\lambda_1 = \sqrt{2} i$, and $\lambda_1 = 0$, and four
simple eigenvalues at $\lambda_1 = \pm i \sqrt{\sqrt{80}+8}$ and
$\lambda_1 = \pm \sqrt{\sqrt{80} - 8}$.

{\bf Case $M \geq 3$:} Eigenvalues of the reduced eigenvalue problem
(\ref{Lyapunovreducedeigenvalue2}) with $M = 3$ are computed
numerically by using Mathematica. The results are as follows:
\begin{eqnarray*}
\lambda_{1,2} = \pm 3.68497 i, \;\; \lambda_{3,4} = \lambda_{5,6} =
\pm 3.20804 i, \;\;\lambda_{7,8} = \pm 2.25068 i, \;\;
\lambda_{9,10} = \lambda_{11,12} = \pm i,  \\
\lambda_{13,14} = \lambda_{15,16} = \pm 0.53991,
\;\;\lambda_{17,18,19,20} = \pm 0.634263 \pm 0.282851 i, \;\;
\lambda_{21,22,23,24} = 0.
\end{eqnarray*}

Using these computations of eigenvalues, we summarize that the
second-order reduced eigenvalue problem
(\ref{Lyapunovreducedeigenvalue2}) has no unstable eigenvalues
$\lambda$ when $L = M = 1$; a simple unstable (positive) eigenvalue
when $L = M = 2$; two unstable real and two unstable complex
eigenvalues when $L = M = 3$. We note that destabilization of the
super-symmetric vortex with $M = 2$ occurs due to the center node
$(2,2)$, which couples the four even-numbered nodes of the contour
$S_2$ in the second-order reductions. Due to this coupling, there
exists a simple negative eigenvalue of the Jacobian matrix ${\cal
M}_2$ and a simple positive eigenvalue in the reduced eigenvalue
problem (\ref{Lyapunovreducedeigenvalue2}). Similarly,
destabilization of the super-symmetric vortex with $M = 3$ occurs
due to the coupling of eight nodes of the contour $S_3$ with four
interior corner points $(2,2)$, $(2,M)$, $(M,M)$, and $(M,2)$, which
result in the four negative eigenvalues of the matrix ${\cal M}_2$
and four unstable eigenvalues in the reduced eigenvalue problem
(\ref{Lyapunovreducedeigenvalue2}).

We note that if the matrix ${\cal M}_2$ would be defined by the
formula (\ref{Melements4}) for any $M \geq 1$ (i.e. if all nodes
inside the contour $S_M$ would be removed by drilling a hole), the
eigenvalues of ${\cal M}_2$ would be all positive and the
eigenvalues of the reduced problem
(\ref{Lyapunovreducedeigenvalue2}) would be all purely imaginary,
similarly to the case $M = 1$.

\subsection{Additional splitting in the second-order reduction}

For super-symmetric solitons of family (i), when $l = 2M$ but $a_j
\neq (-1)^j a$, the Jacobian matrix ${\cal M}_1$ has two zero
eigenvalues with eigenvectors ${\bf p}_0$ and ${\bf p}_1$, but the
Jacobian matrix ${\cal M}_1 + \epsilon {\cal M}_2$ has only one zero
eigenvalue with eigenvector ${\bf p}_0$. Therefore, the splitting of
zero eigenvalue occurs in the second-order reduction. Extending the
regular perturbation series (\ref{regular-series1}) to the next
order, we find that $\gamma_1 = 0$ for ${\bf c} = {\bf p}_1$, and
$$
\gamma_2 = \frac{({\bf p}_1, {\cal M}_2 {\bf p}_1)}{({\bf p}_1,{\bf
p}_1)}.
$$
Extending the regular perturbation series
(\ref{Lyapunoveigenvector1}) to the second order, we have find that
$\lambda_1^2 = 0$ for ${\bf c} = {\bf p}_1$, ${\cal L}_1 = 0$, and
$$
\lambda_2^2 = 2 \frac{({\bf p}_1, {\cal M}_2 {\bf p}_1)}{({\bf
p}_1,{\bf p}_1)} = 2\gamma_2.
$$
Thus, the splitting of zero eigenvalue in the second-order reduction
is the same as the splitting of zero eigenvalues in the first-order
reductions. A positive eigenvalue $\gamma_2$ results in a pair of
real eigenvalues $\lambda_2$, while a negative eigenvalue $\gamma_2$
results in a pair of purely imaginary eigenvalues $\lambda_2$.

All individual results on stability of localized modes on the
discrete contour $S_M$ are summarized as follows.

\begin{Proposition}
\label{proposition-stability} Consider the stability problem
(\ref{eigenvalue}) in the domain $\epsilon \in {\cal O}(0)$,
associated to the families of discrete solitons and vortices of
Propositions \ref{lemma-asymmetric} and
\ref{proposition-persistence}. The following solutions are
spectrally stable in the domain $\epsilon \in {\cal O}(0)$: discrete
solitons of family (i) with $l = 0$; symmetric vortices of family
(ii) with the charge $M+1 \leq L \leq 2M-1$; and symmetric vortex of
family (ii) with the charge $L = M = 1$. All other solutions have at
least one unstable eigenvalue with ${\rm Re}(\lambda) > 0$.
\end{Proposition}

We note that stability of discrete solitons of family (i) with $l =
0$ is equivalent to stability of discrete solitons in the
one-dimensional NLS lattice, which is proved in the first paper
\cite{PaperI}. When $l = 0$, the limiting solution (\ref{2soliton})
consists of alternating up and down pulses along the contour $S_M$,
similar to Theorem 3.6 in \cite{PaperI}.

We also note that the purely imaginary eigenvalue of $\lambda$ have
negative Krein signature, such that the Hamiltonian Hopf
bifurcations may occur for larger values of $\epsilon$ beyond the
neighborhood ${\cal O}(0)$. The number of eigenvalues of negative
Krein signature is related to the closure relation for the number of
negative eigenvalues of the linearized Hamiltonian ${\cal H}$ (see
\cite{PaperI}). We can see from computations of families (i)-(iv)
that the closure relation is satisfied in all cases, except for the
super-symmetric vortices of family (ii) with $L = M$. Indeed, in the
case $L = M = 1$, the Hamiltonian ${\cal H}$ has four negative
eigenvalues for $\epsilon \in {\cal O}(0)$, while it has two pairs
of purely imaginary eigenvalues of negative Krein signature and two
pairs of zero eigenvalues, which exceeds the allowed negative index
of ${\cal H}$. Derivation of a modified closure relation for vortex
solutions of the discrete NLS equations is beyond the scope of this
manuscript.

\section{Numerical Results}
\label{section-numerics}

We perform direct numerical simulations of the discrete NLS equation
(\ref{2NLS}) in order to examine stability of the discrete vortices
in the simplest cases $M = 1$ and $M = 2$. The results are shown on
Figures \ref{depf1}-\ref{depf6} (see also summary in Table 1).

In computations of solutions of the problems (\ref{2difference}) and
(\ref{eigenvalue}), we will use an equivalent renormalization of the
problem with parameter:
$$
\varepsilon = \frac{\epsilon}{1 - 4 \epsilon}.
$$
This renormalization is equivalent to keeping the diagonal term $-4
\epsilon \phi_{n,m}$ in the right-hand-side of the difference
equations (\ref{2difference}). As a result, the discrete solitons
and vortices exist typically in the semi-infinite domain
$\varepsilon > 0$. The anti-continuous limit is not affected by the
renormalization since $\varepsilon \approx \epsilon$ for small
$\epsilon$.

On the figures \ref{depf1}--\ref{depf6}, the top left panel shows
the profile of the vortex solution for a specific value of
$\varepsilon$ by means of contour plots of the real (top left),
imaginary (top right) modulus (bottom left) and phase (bottom right)
two-dimensional profiles. The top right panel shows the complex
plane $\lambda = \lambda_r + i \lambda_i$ for the linear eigenvalue
problem (\ref{eigenvalue}) for the same value of $\varepsilon$. The
bottom panel shows the dependence of small eigenvalues as a function
of $\varepsilon$, obtained via continuation methods from the
anti-continuum limit of $\varepsilon=0$. The solid lines represent
numerical results, while the dashed lines show results of the
first-order and second-order reductions.

Figure \ref{depf1} show results for the super-symmetric vortex of
charge $L = 1$ on the contour $S_M$ with $M = 1$. In the
second-order reduction, the stability spectrum of the vortex
solution has a zero eigenvalue of algebraic multiplicity 4 and two
pairs of imaginary eigenvalues $\lambda \approx \pm 2 \varepsilon
i$. These pairs split along the imaginary axis beyond the
second-order reductions and undertake Hamiltonian Hopf bifurcations
for larger values of $\varepsilon$ upon collision with the
continuous spectrum (only the first bifurcation at $\varepsilon
\approx 0.38$ is shown on Fig. \ref{depf1}). It follows from Fig.
\ref{depf1} that the zero eigenvalue splits along the imaginary axis
for larger values of $\varepsilon$, beyond the second-order
reductions.

Figure \ref{depf2} show results for the symmetric vortex of charge
$L=1$ on the contour $S_M$ with $M=2$. There are three double and
one simple real unstable eigenvalues in the first-order reductions,
but all double eigenvalues split into the complex plane in the
second-order reductions. The asymptotic result
(\ref{explicit-eigenvalueA}) for eigenvalues $\lambda =
\sqrt{\varepsilon} \lambda_1 + \varepsilon \lambda_2$ with $N=8$,
$a=\cos(\pi/4)$ and $b=\sin(\pi/4)$ are shown on Fig. \ref{depf2} in
perfect agreement with numerical results.

Figure \ref{depf3} shows results for the super-symmetric vortex with
$L=M=2$. The second-order reductions have a pair of simple real
eigenvalues $\lambda \approx \pm \varepsilon \sqrt{\sqrt{80}-8}$, a
pair of simple imaginary eigenvalues $\lambda \approx \pm i
\varepsilon \sqrt{\sqrt{80}+8}$, zero eigenvalue of algebraic
multiplicity $4$ and a pair of imaginary eigenvalues of algebraic
multiplicity $4$ at $\lambda \approx \pm i \varepsilon \sqrt{2}$.
The bottom right panel of Fig. \ref{depf3} shows splitting of
multiple imaginary eigenvalues beyond the second-order reductions
and also subsequent Hamiltonian--Hopf bifurcations for larger values
of $\varepsilon$.

Figure \ref{depf4} shows results for the symmetric vortex with $L=3$
and $M=2$. The first-order reductions predict three pairs of double
imaginary eigenvalues, a pair of simple imaginary eigenvalues and a
double zero eigenvalue. The double eigenvalues split in the
second-order reductions along the imaginary axis, given by
(\ref{explicit-eigenvalueA}) with $N=8$, $a=\cos(3 \pi/4)$ and
$b=\sin(3 \pi/4)$. The seven pairs of imaginary eigenvalues lead to a
cascade of seven Hamiltonian Hopf bifurcations for larger values of
$\varepsilon$ due to their collisions with the continuous spectrum.
The first Hamiltonian--Hopf bifurcation when the symmetric vortex
becomes unstable occurs for $\varepsilon \approx 0.096$.

Zero-parameter asymmetric vortices of family (iv) on the contour
$S_M$ with $M = 2$ are shown in Fig. \ref{depf5} for $L = 1$ and in
Fig. \ref{depf6} for $L = 3$. In the case of Fig. \ref{depf5}, all
the phase differences between adjacent sites in the contour are
$\pi/6$, except for the last one which is $5 \pi/6$, completing a
phase trip of $2 \pi$ for a vortex of topological charge $L=1$.
Eigenvalues of the matrix ${\cal M}_1$ in the first-order reductions
can be computed numerically as follows: $\gamma_1 = -1.154$,
$\gamma_2 = 0$, $\gamma_3 = 0.507$, $\gamma_4 = 0.784$, $\gamma_5 =
1.732$, $\gamma_6 = 2.252$, $\gamma_7 = 2.957$ and $\gamma_8 =
3.314$. As a result, the corresponding eigenvalues $\lambda \approx
\pm \sqrt{2 \gamma \varepsilon}$ yield one pair of imaginary
eigenvalues and six pairs of real eigenvalues, in agreement with our
numerical results. The bottom panel of Fig. \ref{depf5} shows that
two pairs of real eigenvalues collide for $\varepsilon \approx
0.047$ and $\varepsilon \approx 0.057$ and lead to two quartets of
eigenvalues.

In the case of Fig. \ref{depf6}, all the phase differences in the
contour are $5 \pi/6$, except for the last one which is $\pi/6$,
resulting in a vortex of topological charge $L=3$. Eigenvalues of
the matrix ${\cal M}_1$ are found numerically as follows: $\gamma_1
= -3.314$, $\gamma_2 = -2.957$, $\gamma_3 = -2.252$, $\gamma_4 =
-1.732$, $\gamma_5 =-0.784$, $\gamma_6 =-0.507$, $\gamma_7 = 0$, and
$\gamma_8 =1.154$. Consequently, this solution has six pairs of
imaginary eigenvalues and one pair of real eigenvalues. The first
Hamiltonian--Hopf bifurcation occurs for $\varepsilon \approx
0.086$.

We note that numerical results for asymmetric vortices of family
(iii) are not shown. Both analytical and numerical analysis of such
solutions are delicate problems which are open for future studies.
However, all such solutions (if they persist) are unstable in the
first-order reductions.

\section{Conclusions}

In this series of two papers, we have developed the 
mathematical analysis
of discrete soliton and vortex solutions of the discrete NLS
equations close to the anti-continuum 
limit. These solutions are relevant to recent experimental and
numerical studies in the context of nonlinear optics, photonic
crystal lattices, soft condensed-matter physics, and Bose-Einstein
condensates.

In the present paper, we have examined the persistence of discrete
vortices starting from the anti-continuum limit of uncoupled
oscillators and continuing 
towards a finite coupling constant $\epsilon$. We have found
persistent families of such solutions that include symmetric and
asymmetric vortices. We have ruled out other non-persistent
solutions by means of the Lyapunov-Schmidt method. We have
subsequently categorized the persisting solutions to discrete
solitons, symmetric vortices and one-parameter and zero-parameter
asymmetric vortices. For persistent solutions, we have derived the
leading-order asymptotic approximations for small unstable and
neutrally stable eigenvalues of the stability problem, up to the
first-order and second-order corrections.

\begin{figure}[tbp]
\begin{center}
\epsfxsize=6.0cm %\centerline{}
\epsffile{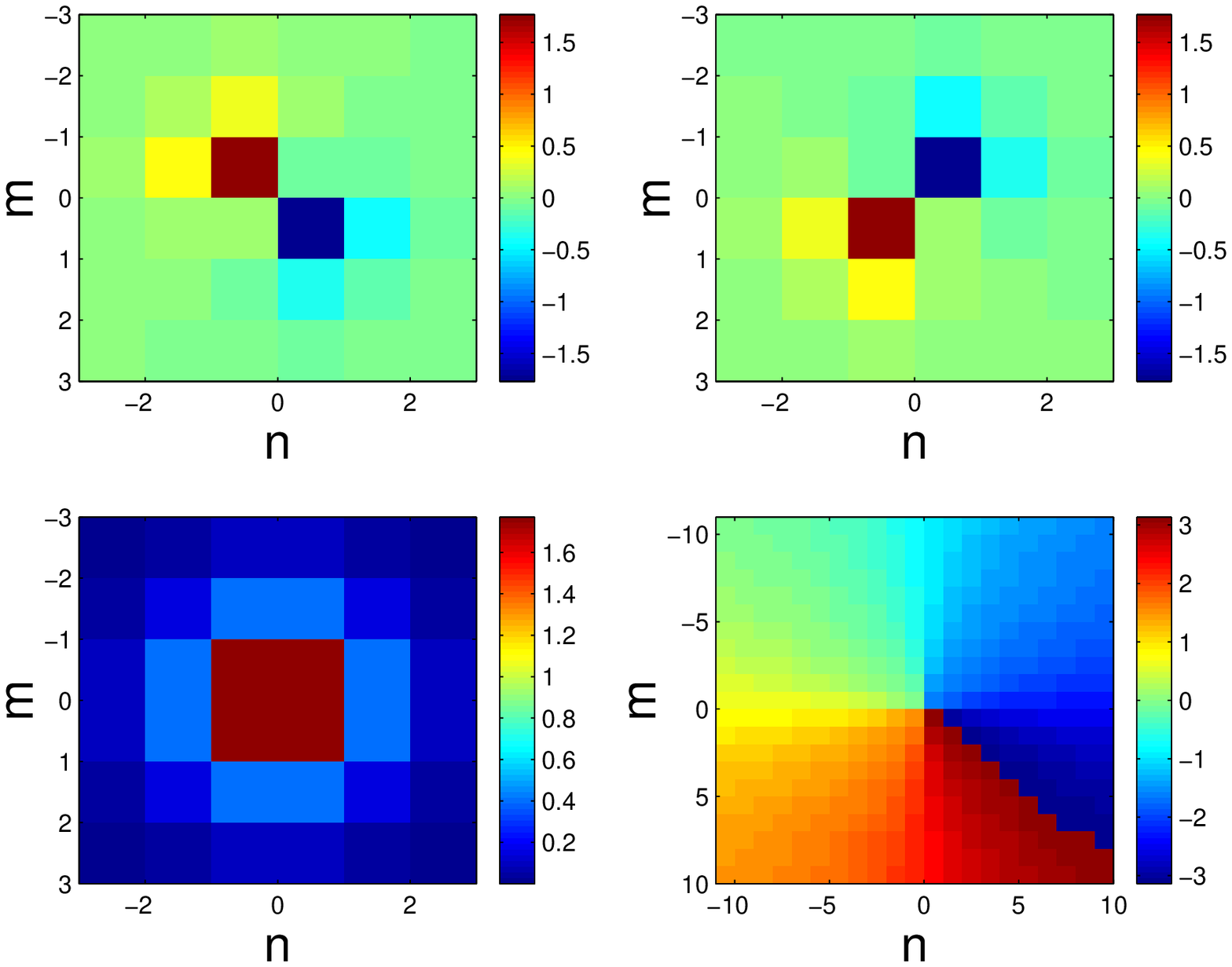}
\epsfxsize=6.0cm %\centerline{}
\epsffile{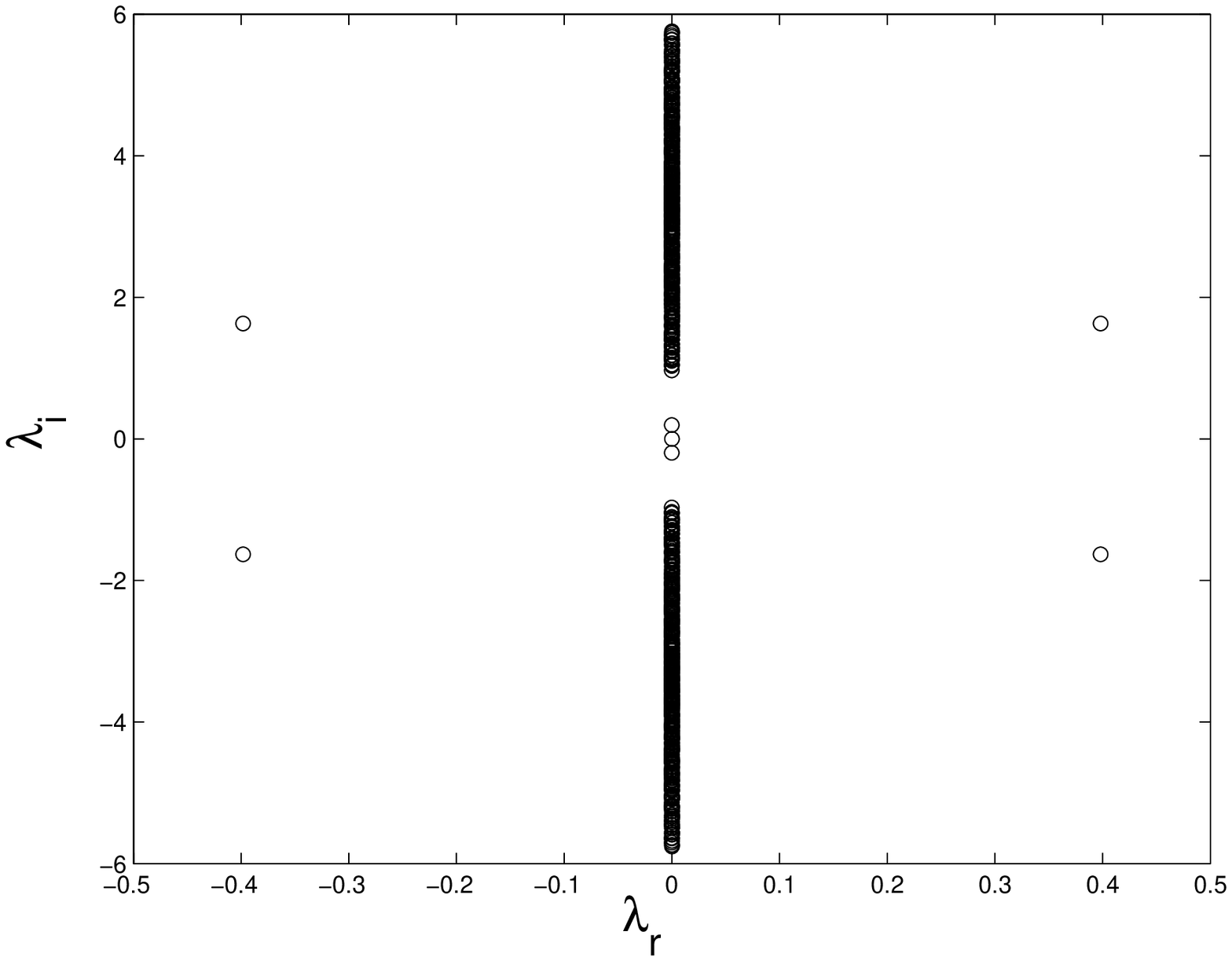} \epsfxsize=6.0cm \centerline{
\epsffile{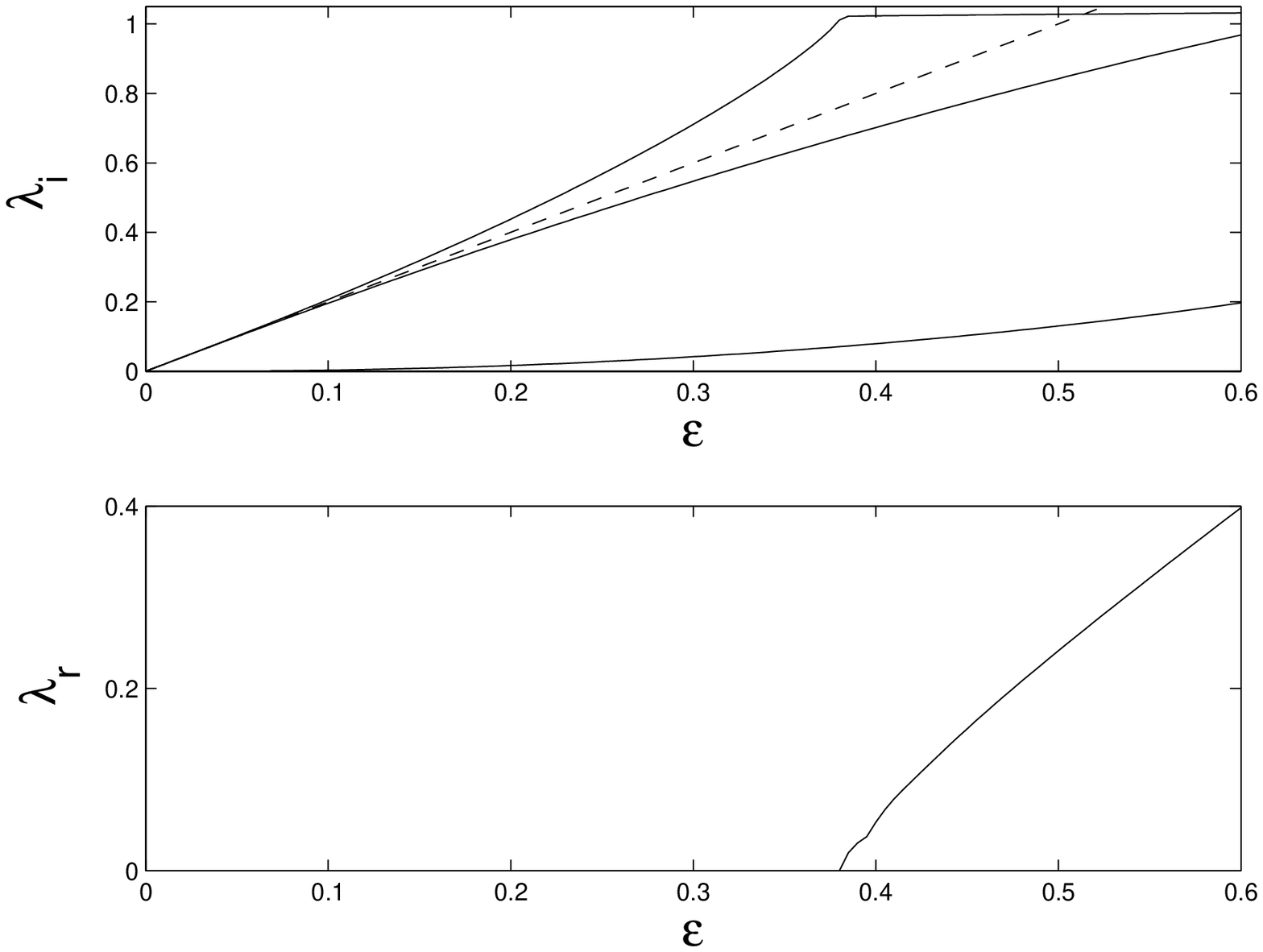} }
%\end{tabular}
\caption{The (super-symmetric) vortex cell with $L=M=1$. The top
left panel shows the profile of the solution for $\varepsilon=0.6$.
The subplots show the real (top left), imaginary (top right),
modulus (bottom left) and phase (bottom right) fields. The top right
panel shows the spectral plane $(\lambda_r,\lambda_i)$ of the linear
eigenvalue problem (\ref{eigenvalue}). The bottom panel shows the
the small eigenvalues versus $\varepsilon$ (the top subplot shows
the imaginary part, while the bottom shows the real part). The solid
lines show the numerical results, while the dashed lines show the
results of the second-order reductions.} \label{depf1}
\end{center}
\end{figure}

\begin{figure}[tbp]
\begin{center}
\epsfxsize=6.0cm %\centerline{}
\epsffile{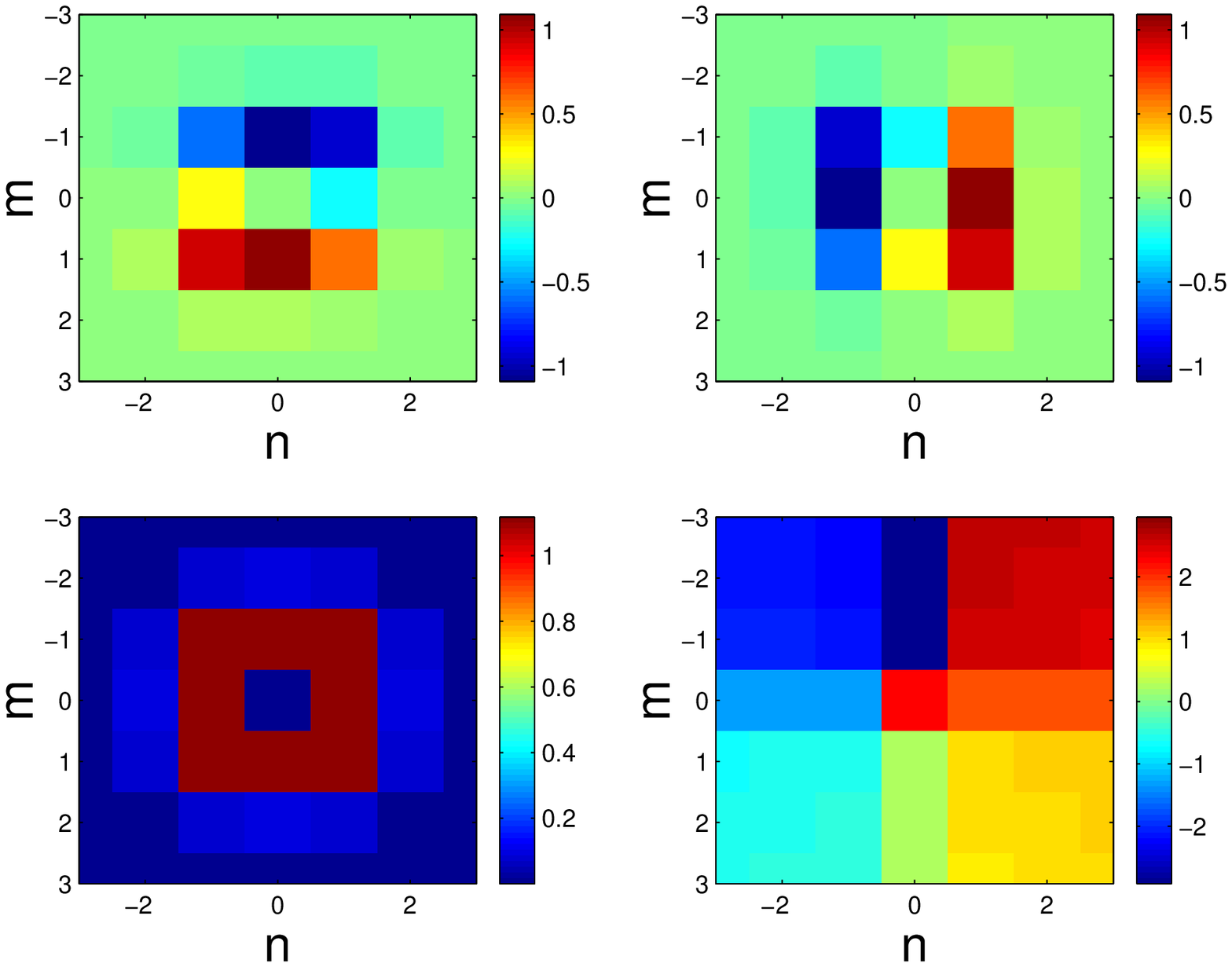}
\epsfxsize=6.0cm %\centerline{}
\epsffile{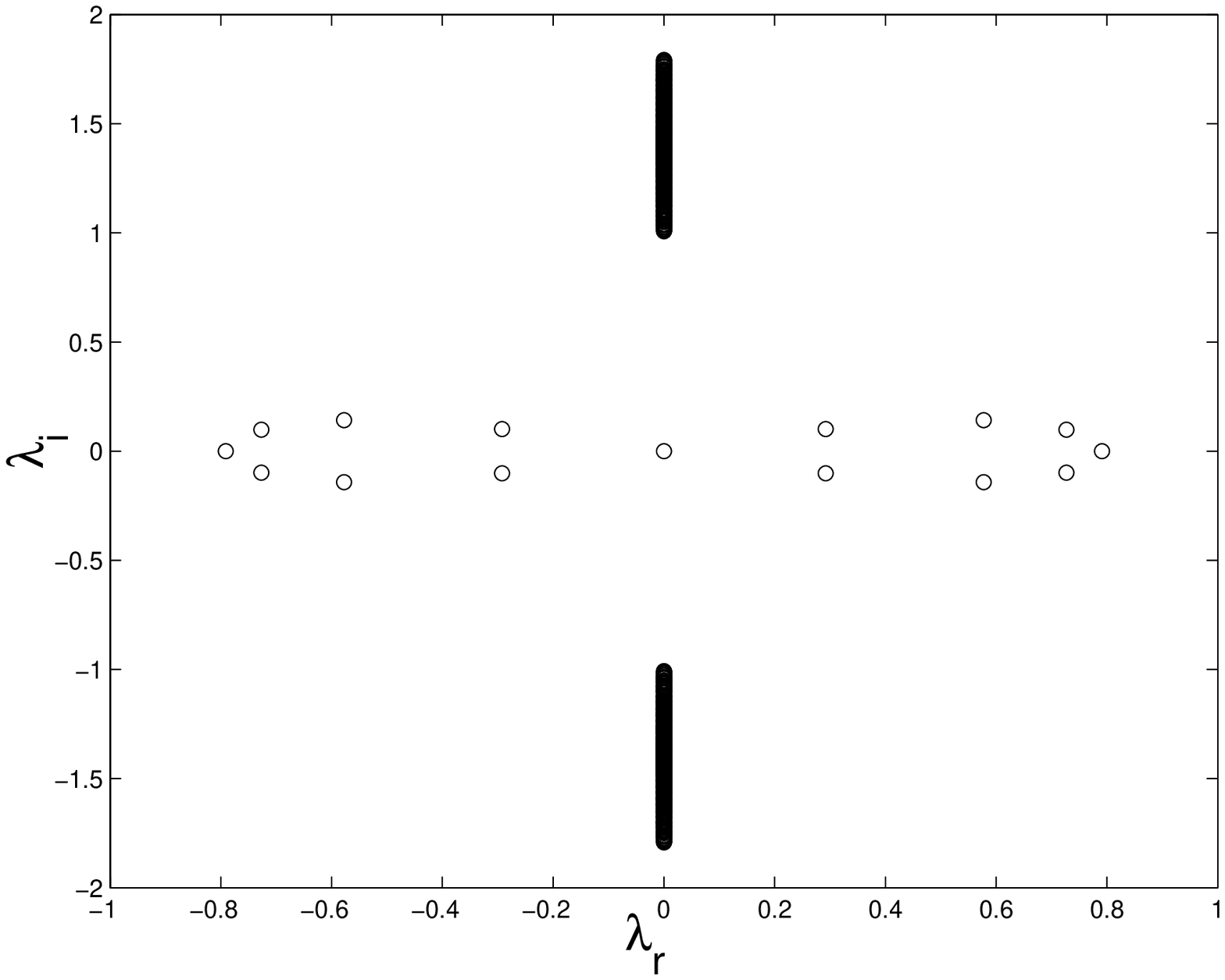}
\epsfxsize=6.0cm %\centerline{}
\centerline{ \epsffile{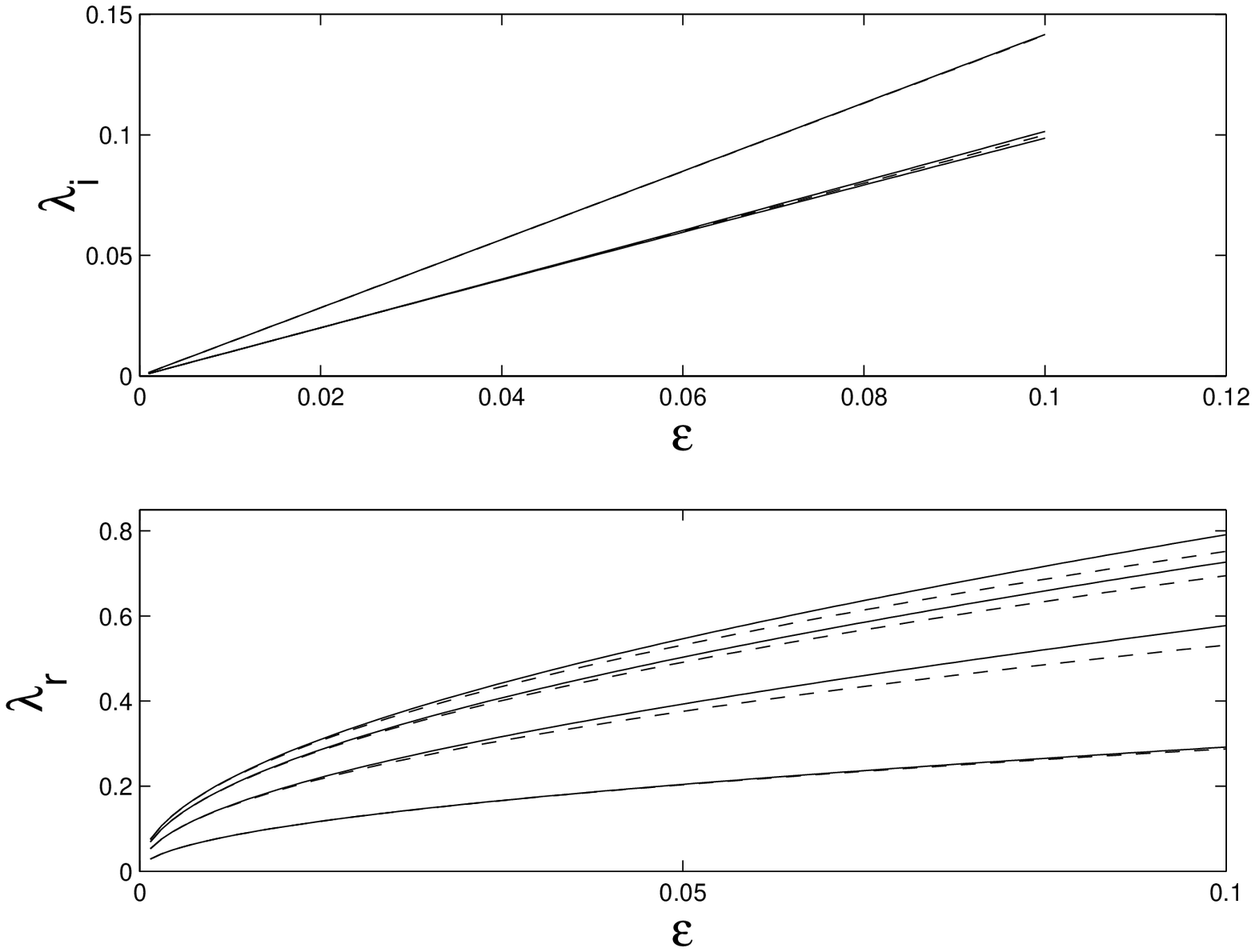} }
%\epsfxsize=7.0cm %\centerline{}
%\end{tabular}
\caption{The same features as in the previous figure are shown for
the symmetric vortex with $L=1$ and $M=2$ for
$\varepsilon=0.1$.} \label{depf2}
\end{center}
\end{figure}

\begin{figure}[tbp]
\begin{center}
\epsfxsize=7.0cm %\centerline{}
\epsffile{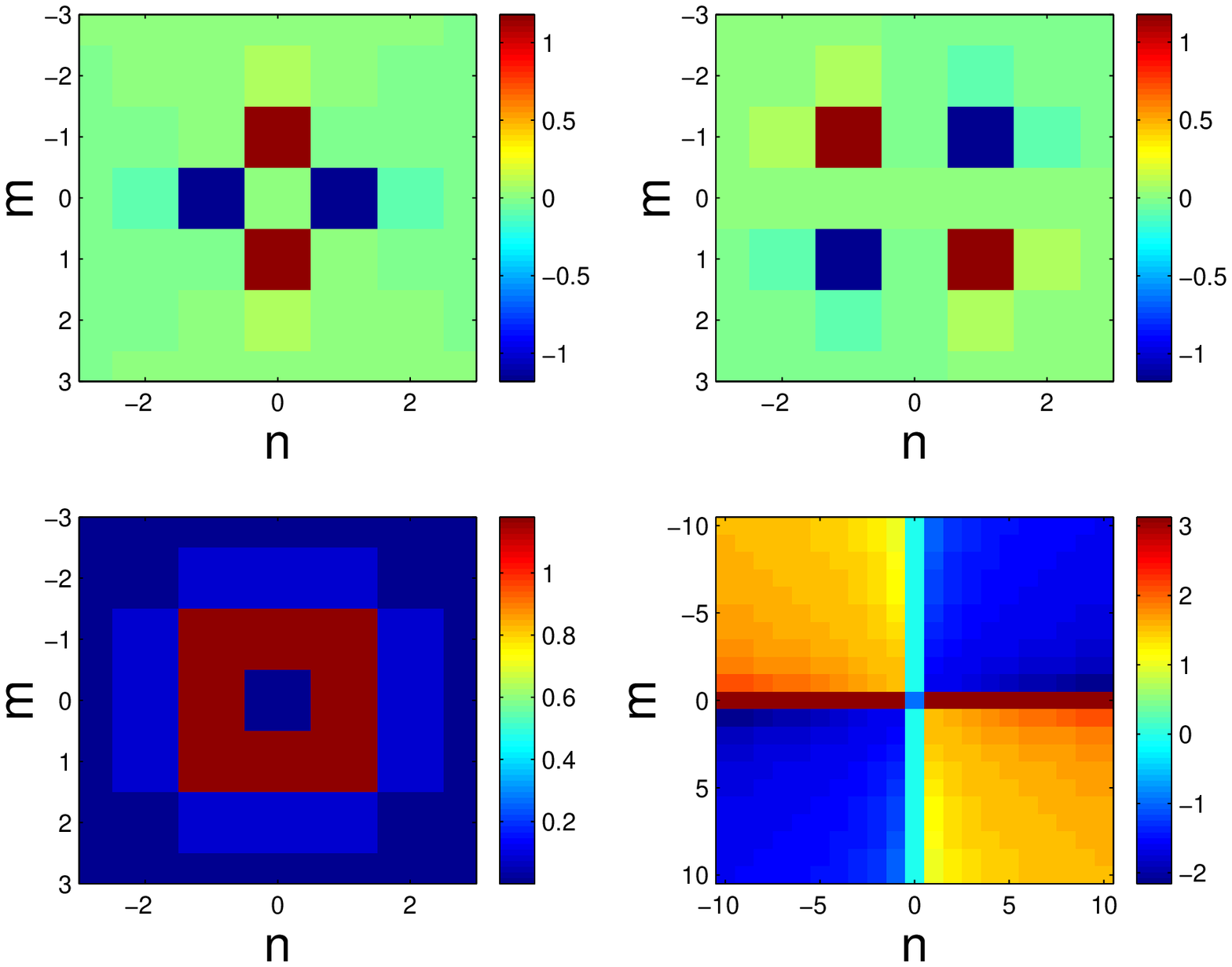}
\epsfxsize=7.0cm %\centerline{}
\epsffile{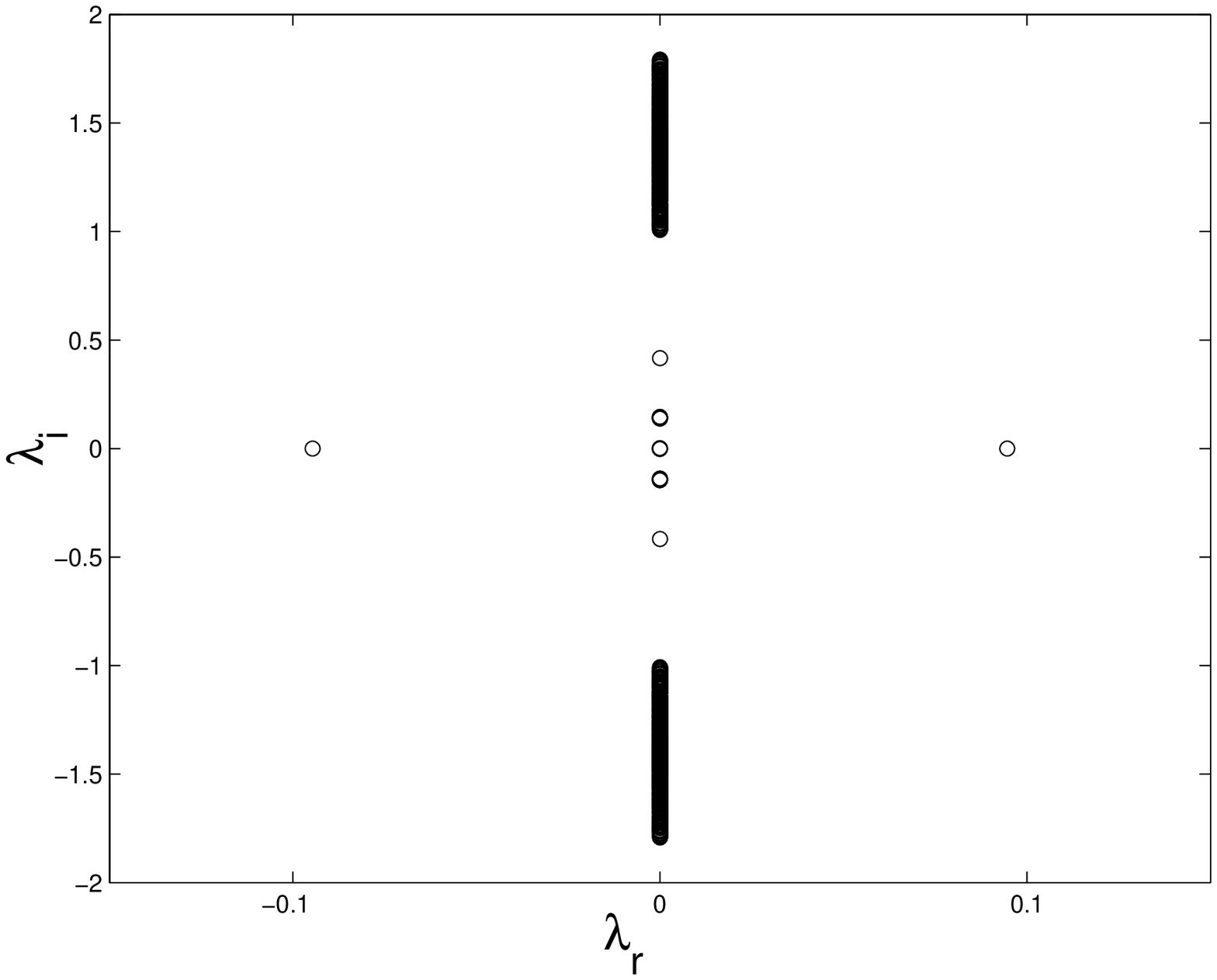}
%\centerline{
\epsfxsize=7.0cm %\centerline{}
\epsffile{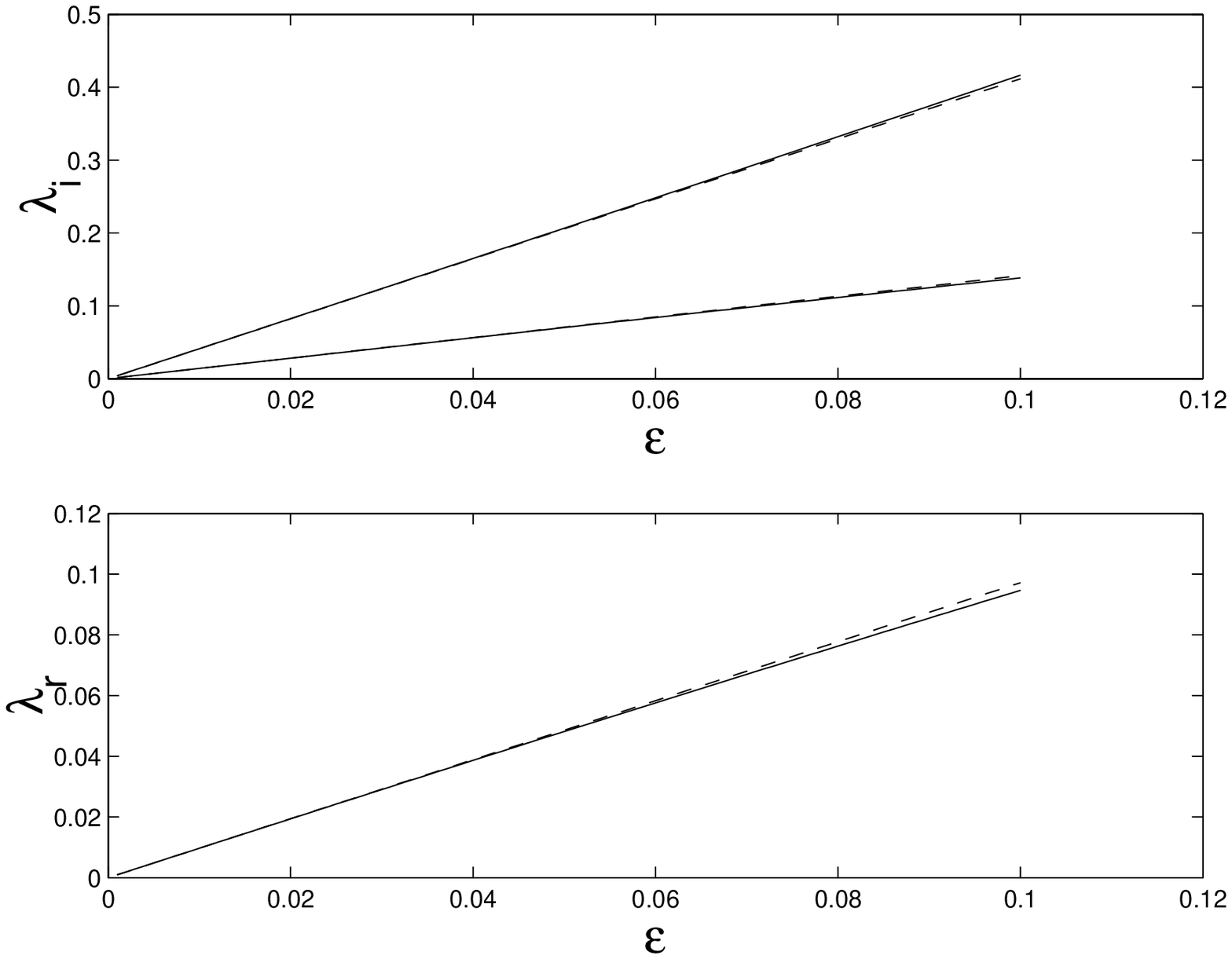}
\epsfxsize=7.0cm %\centerline{}
\epsffile{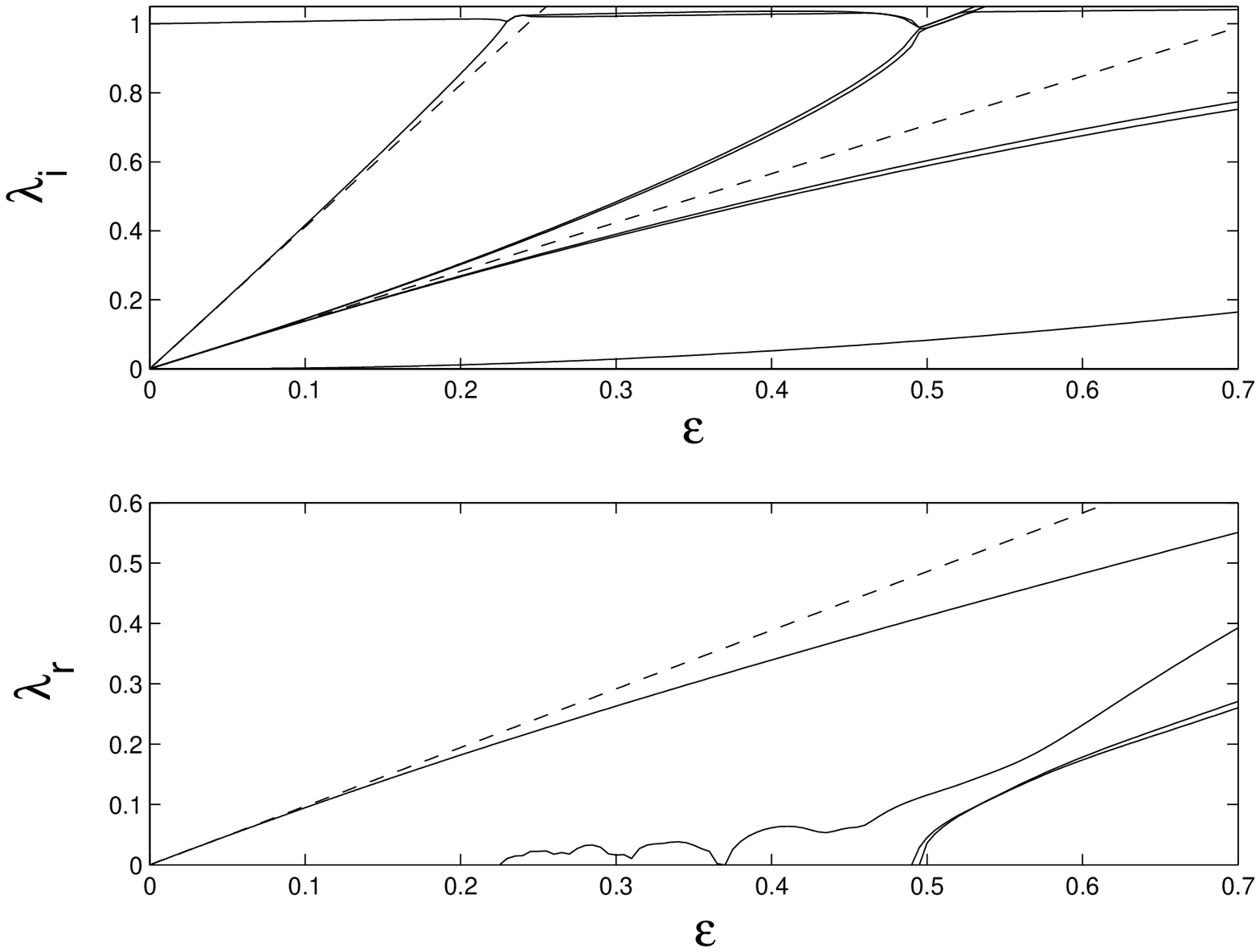}
%}
%\end{tabular}
\caption{The super-symmetric vortex with $L=M=2$ for
$\varepsilon=0.1$. The bottom right panel is an extension of the
bottom left panel to larger values of $\varepsilon$. Remarkable
agreement of the theoretical predictions (dashed lines) with the
numerical results (solid lines) can be observed for small values of
$\varepsilon$.} \label{depf3}
\end{center}
\end{figure}

\begin{figure}[tbp]
\begin{center}
\epsfxsize=7.0cm %\centerline{}
\epsffile{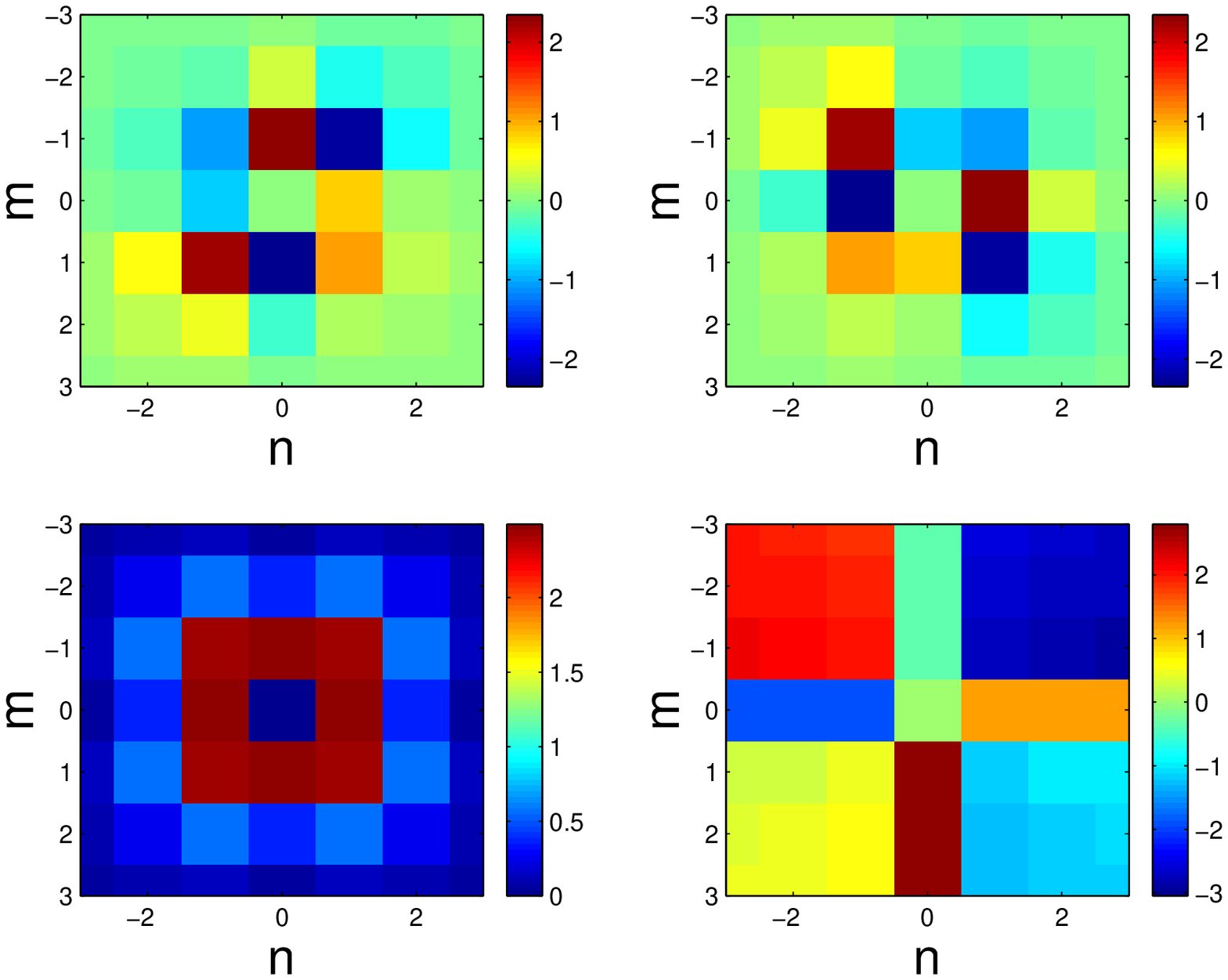}
\epsfxsize=7.0cm %\centerline{}
\epsffile{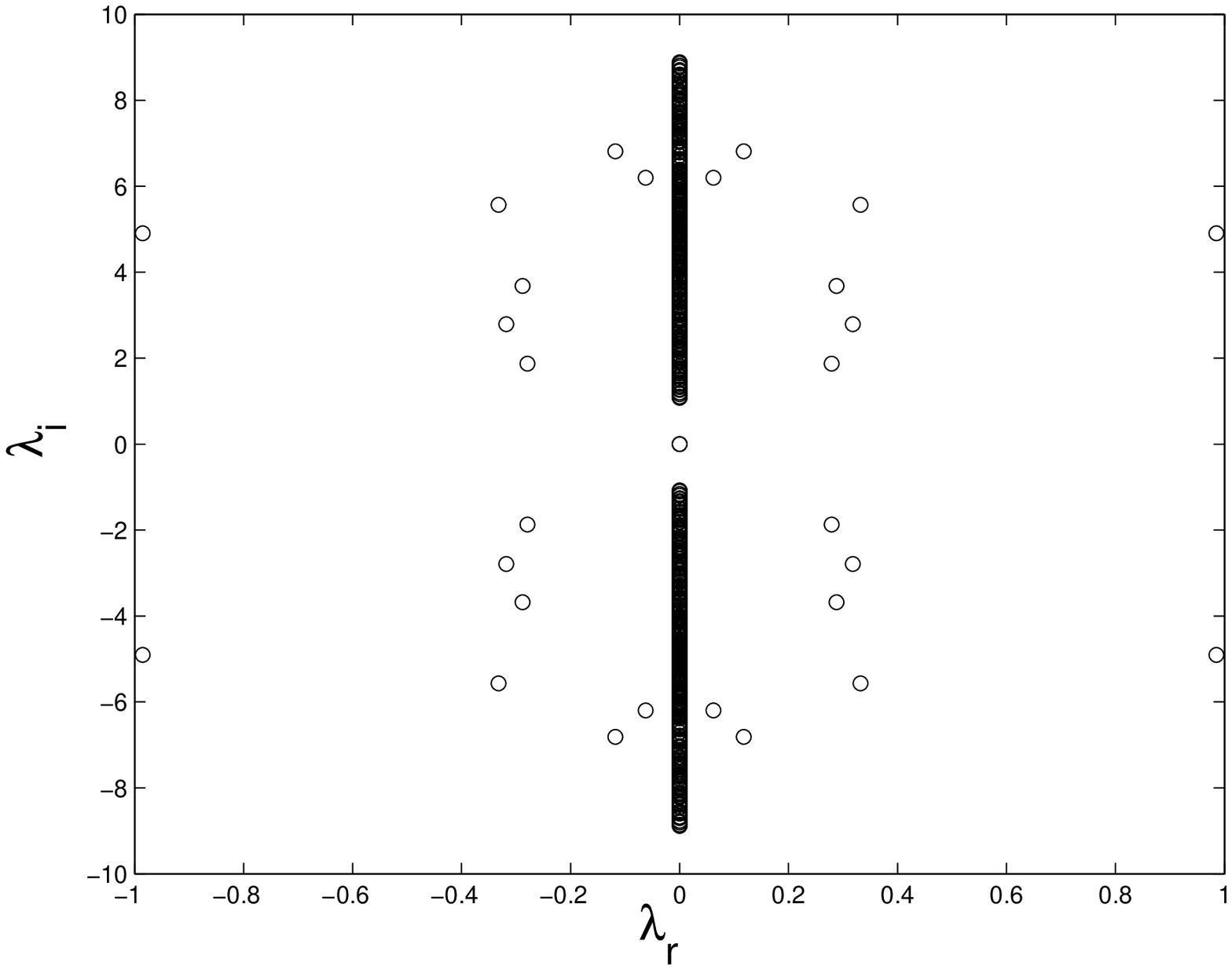}
%\centerline{
\epsfxsize=7.0cm %\centerline{}
\epsffile{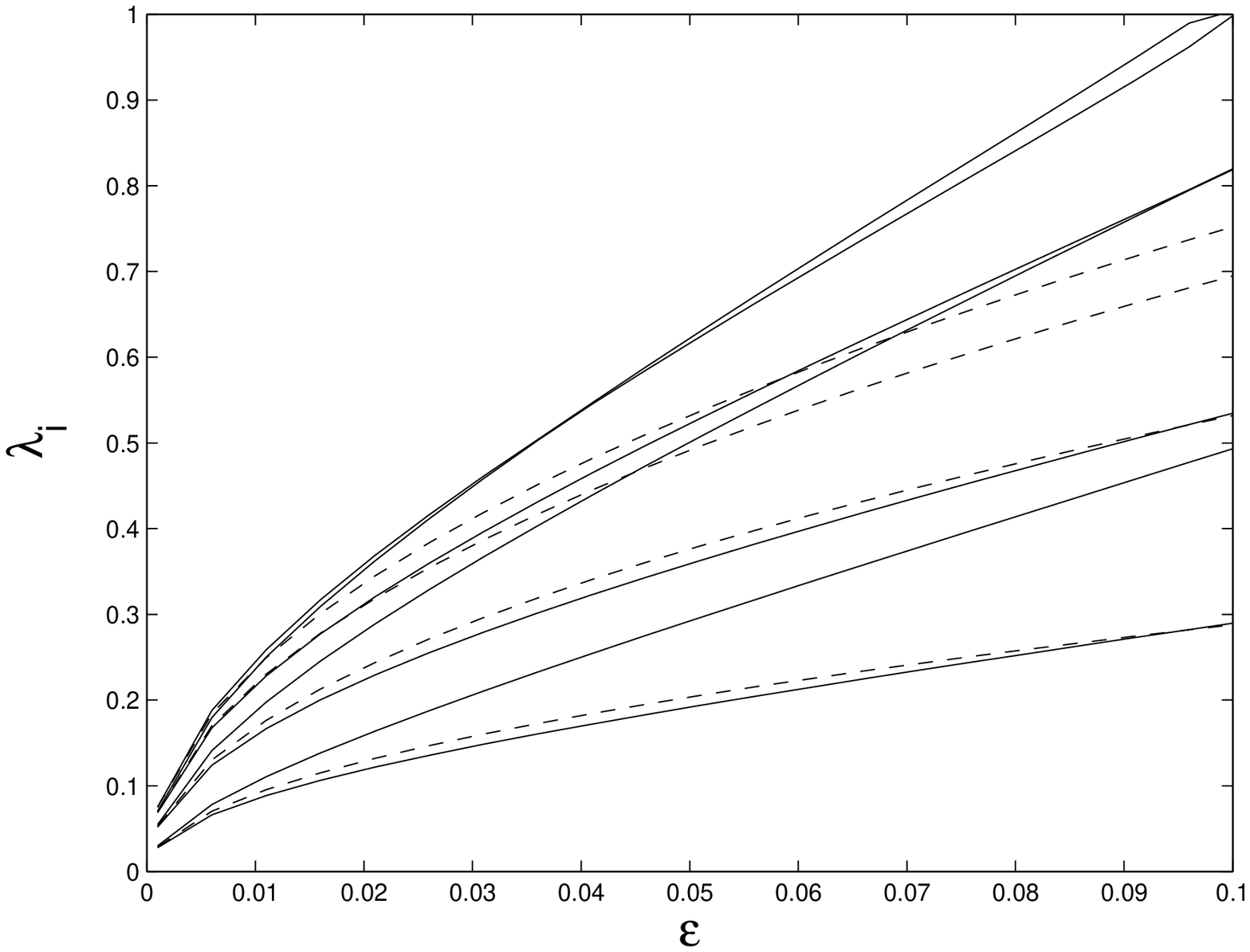}
\epsfxsize=7.0cm %\centerline{}
\epsffile{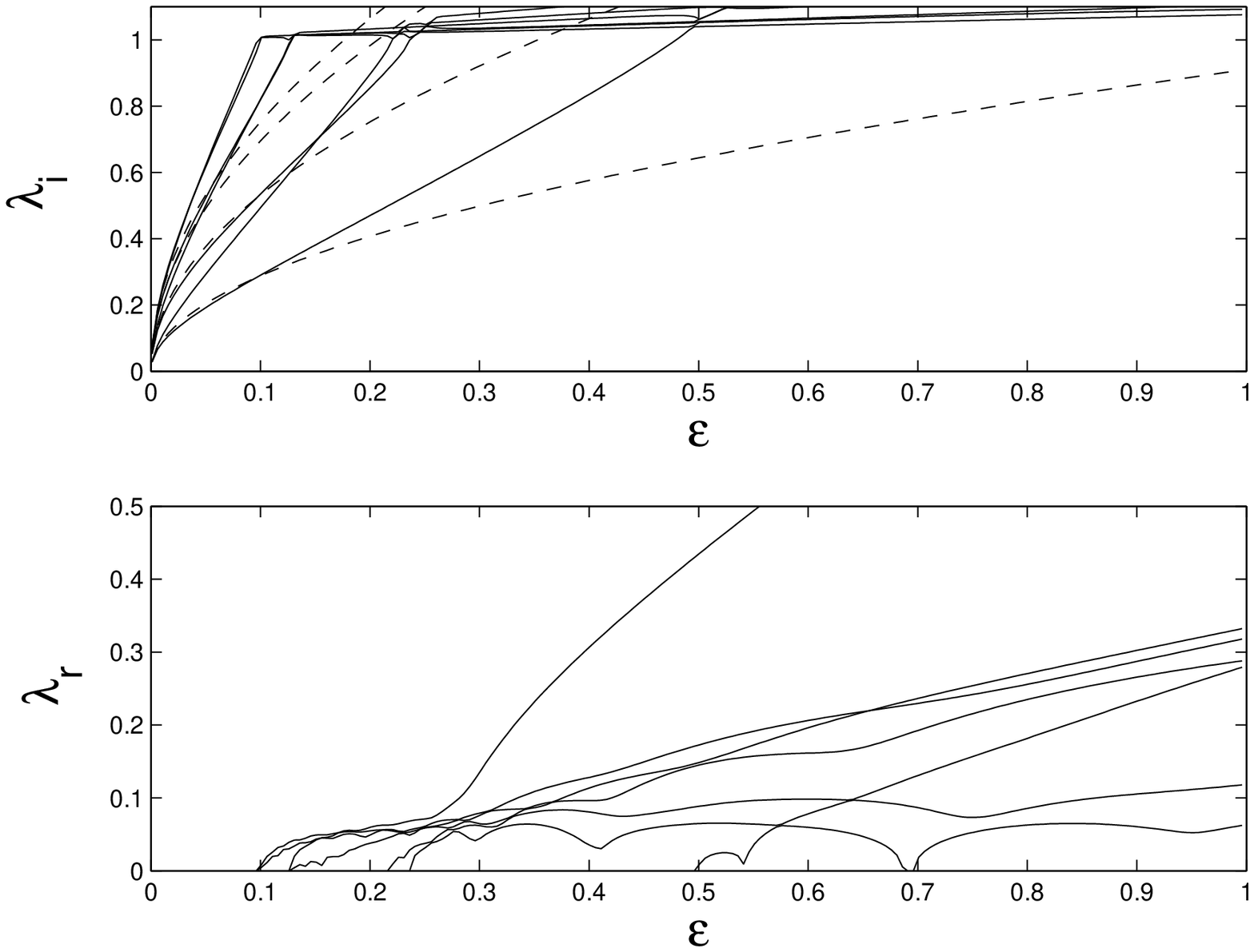}
%}
%\end{tabular}
\caption{The same features as in the previous figure are shown for
the symmetric vortex with $L=3$ and $M=2$ for
$\varepsilon=1$.} \label{depf4}
\end{center}
\end{figure}

\begin{figure}[tbp]
\begin{center}
\epsfxsize=6.0cm %\centerline{}
\epsffile{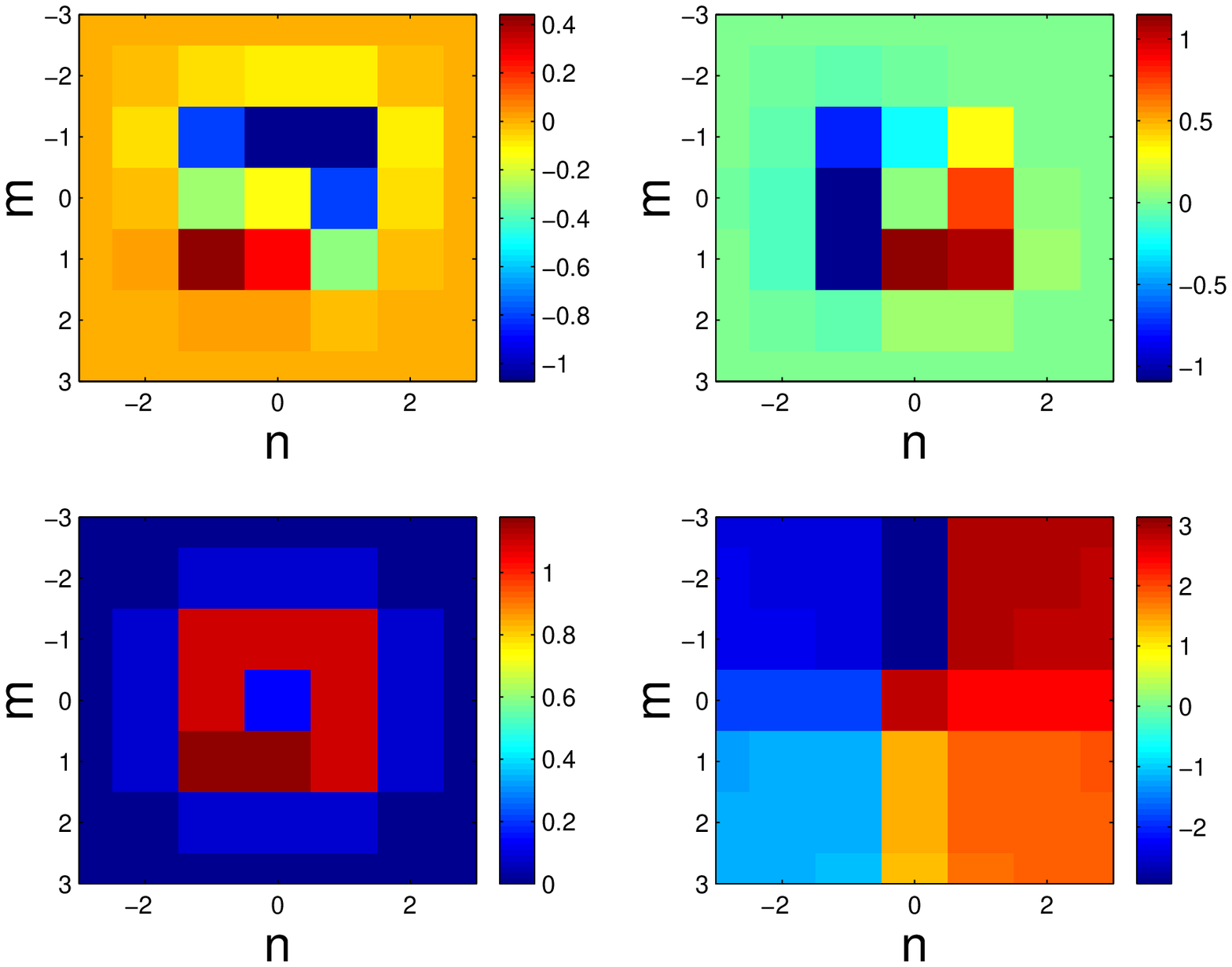}
\epsfxsize=6.0cm %\centerline{}
\epsffile{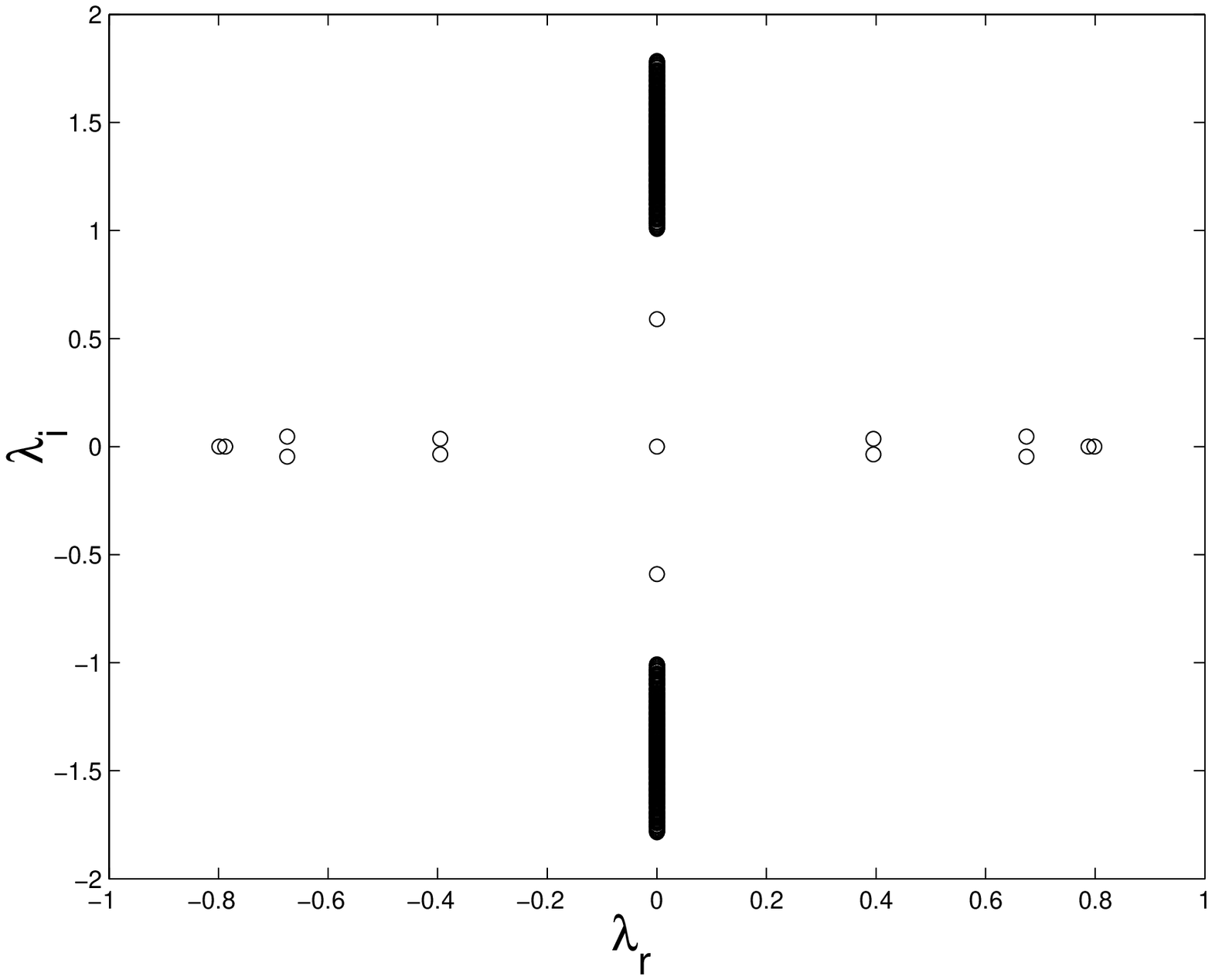}
%\centerline{
\epsfxsize=6.0cm %\centerline{}
\epsffile{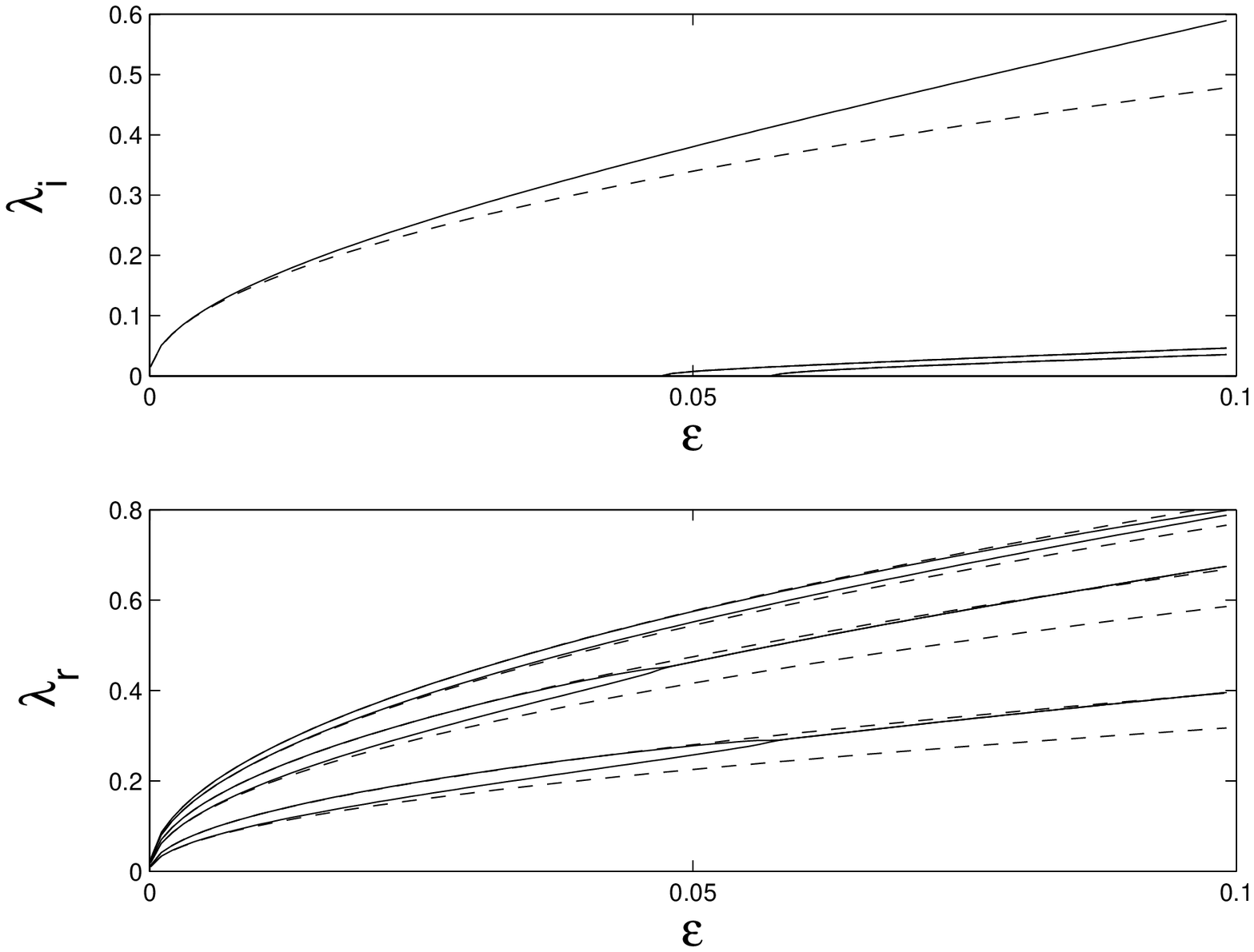}
%\epsfxsize=7.0cm %\centerline{}
%\epsffile{L3M2aa.ps}
%}
%\end{tabular}
\caption{Same as Figure 1 but for 
the asymmetric vortex with $L=1$ and $M=2$ for
$\varepsilon=0.1$.} \label{depf5}
\end{center}
\end{figure}

\begin{figure}[tbp]
\begin{center}
\epsfxsize=7.0cm %\centerline{}
\epsffile{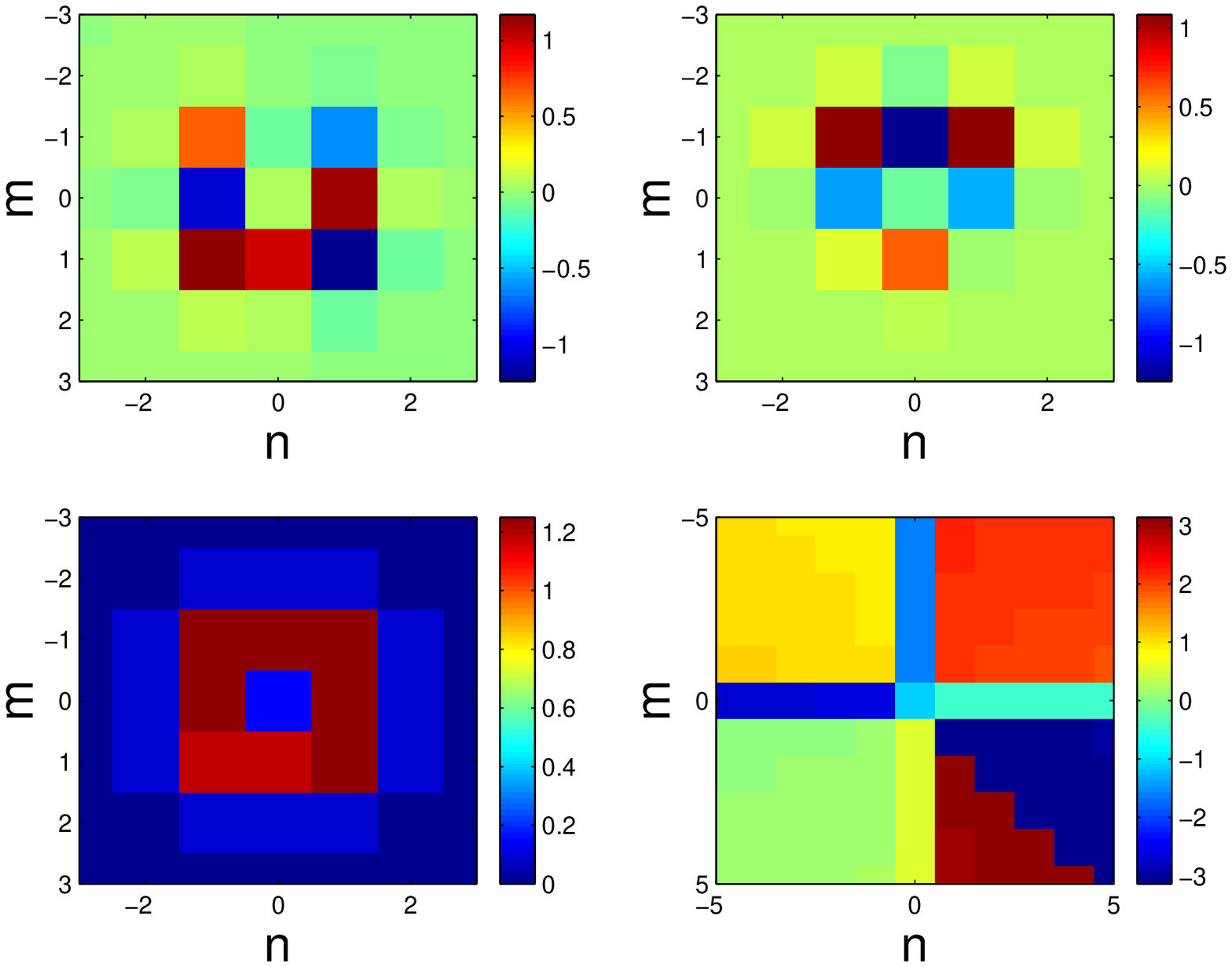}
\epsfxsize=7.0cm %\centerline{}
\epsffile{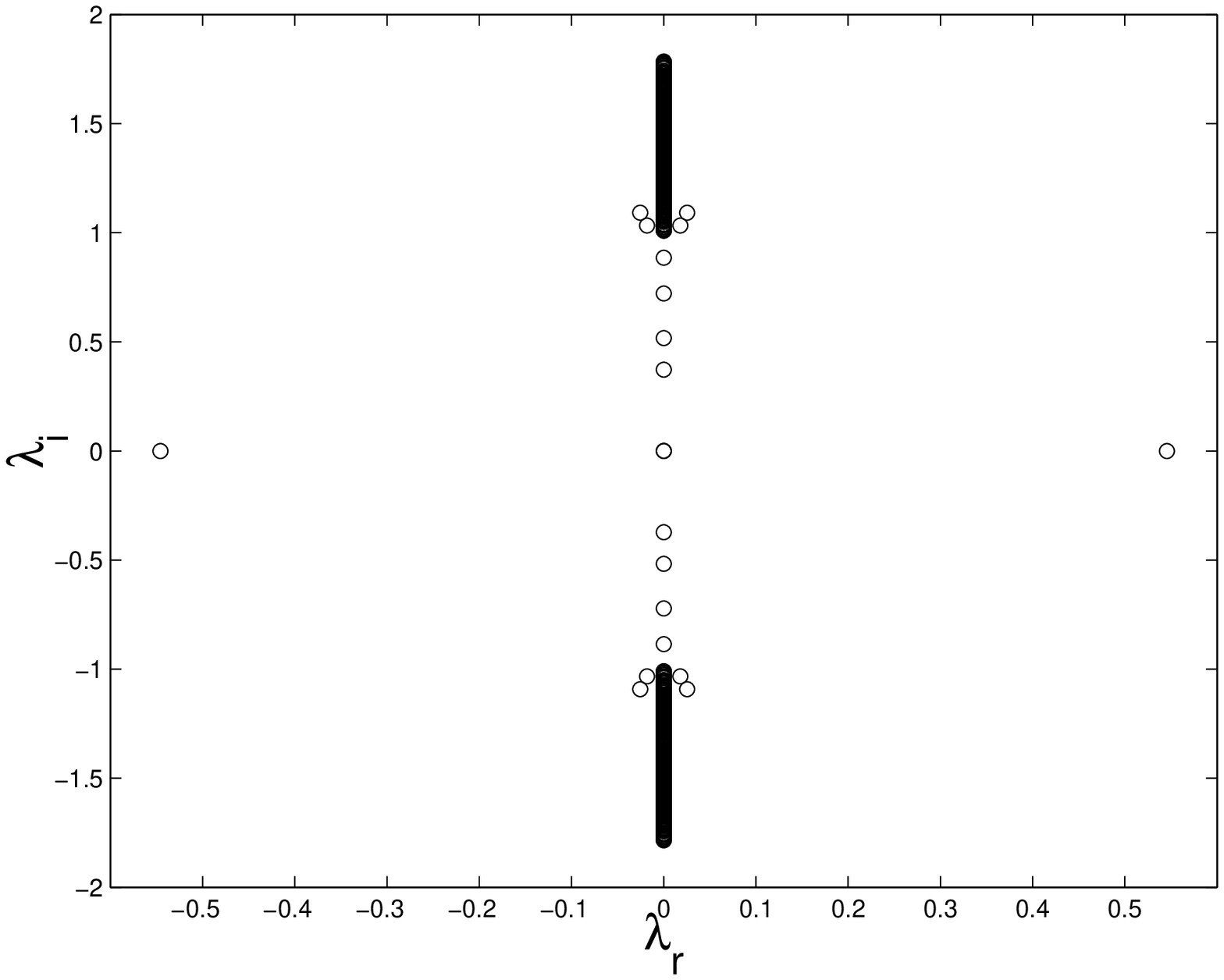}
%\centerline{
\epsfxsize=7.0cm %\centerline{}
\epsffile{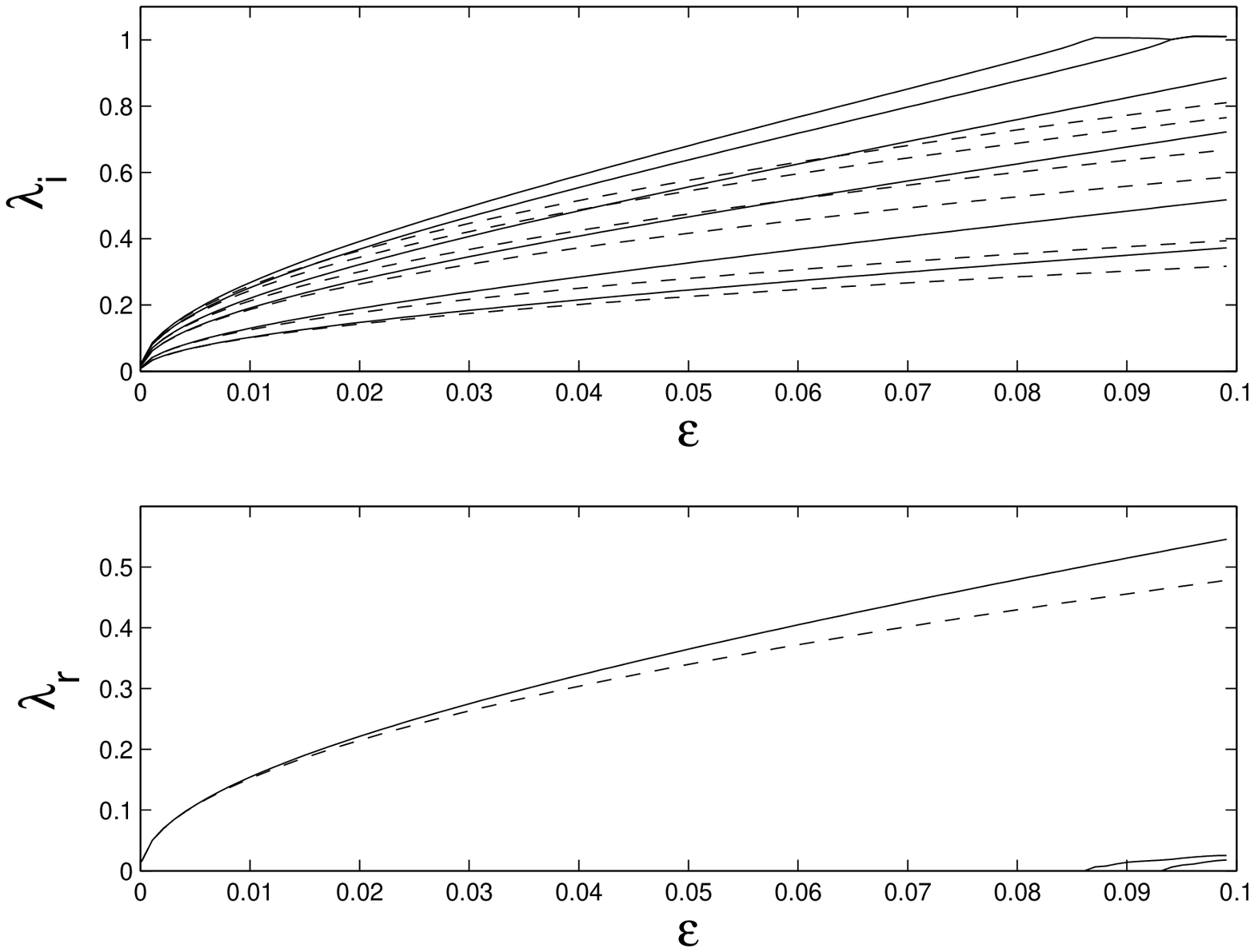}
%\epsfxsize=7.0cm %\centerline{}
%\epsffile{L3M2aa.ps}
%}
%\end{tabular}
\caption{Same as Figure 1 but for 
the asymmetric vortex with $L=3$ and $M=2$ for
$\varepsilon=0.1$.}\label{depf6}
\end{center}
\end{figure}

We have applied the results to particular computations of discrete
vortices of topological charge $L=1$, $L = 2$, and $L = 3$ on the
discrete square contours $S_M$ with $M=1$ and $M=2$. Besides
particular computations collected in Table 1 and Figs.
\ref{depf1}--\ref{depf6}, these results offer a road map on
stability predictions for larger contours, as well as predictions on
how to stabilize the discrete vortices. For example, super-symmetric
vortices of charge $L = M \geq 2$ can be stabilized by excluding the
inner nodes inside the discrete contour $S_M$.

There are remaining open problems for future analytical work. First,
it would be nice to extend this analysis to three-dimensional
structures, such as the discrete solitons, vortices, or vortex
``cubes'' (see \cite{KMFC04}). Second, other types of contours can be
studied in the two-dimensional lattice, such as the diagonal square
contours which would include the ``vortex cross'' or the octagon
(see \cite{YMMKMFC04}). Persistence of one-parameter asymmetric
vortices including the super-symmetric vortices has not yet been
proved in the present paper and it leaves space for future work.
Finally, the closure relation for negative indices of linearized
Hamiltonian associated with the discrete vortex solutions must be
derived and applied independently of the current studies. While
conceptually, the methodologies and techniques presented here can be
adapted to the problems mentioned above, actual computations of the
higher-order Lyapunov--Schmidt reductions become technically
involved.

\newpage

\begin{thebibliography}{99}

\bibitem{PaperI} D.E. Pelinovsky, P.G. Kevrekidis, and D.J.
Frantzeskakis, "Nonlinear Schr\"{o}dinger lattices, I: stability of
discrete solitons", preprint: nlin/0410005 (2004).

\bibitem{A97} S. Aubry, "Breathers in nonlinear lattices:
existence, linear stability and quantization", Physica D {\bf 103}
(1997) 201--250.

\bibitem{FW98} S. Flach and C.R. Willis, "Discrete breathers", Physics Reports
{\bf 295} (1998) 181--264.

\bibitem{HT99} D. Hennig and G. Tsironis, "Wave transmission
in nonlinear lattices", Physics Reports {\bf 307} (1999) 333--432.

\bibitem{KRB01} P.G. Kevrekidis, K.O. Rasmussen, and A.R.
Bishop, "The discrete nonlinear Schrodinger equation: A survey of
recent results", Int. J. Mod. Phys. B {\bf 15}, 2833 (2001).

\bibitem{KB04} V.V. Konotop and V.A. Brazhnyi,
"Theory of nonlinear matter waves in optical lattices", Mod. Phys.
Lett. B {\bf 18} (2004) 627-651.

\bibitem{KF04} P.G. Kevrekidis and D.J. Frantzeskakis,
"Pattern forming dynamical instabilities of Bose-Einstein
condensates", Mod. Phys. Lett. B {\bf 18}, 173-202 (2004).

\bibitem{MK01} B.A. Malomed and P.G. Kevrekidis, "Discrete
vortex solitons", Phys. Rev. E {\bf 64}, 026601 (2001).

\bibitem{ESCFS02} N.K. Efremidis, S. Sears, D.N. Christodoulides,
J.W. Fleischer, and M. Segev, "Discrete solitons in photorefractive
optically induced photonic lattices", Phys. Rev. E {\bf 66} (2002)
046602.

\bibitem{SKEA03} A.A. Sukhorukov, Yu.S. Kivshar, H.S. Eisenberg, and Y. Silberberg,
"Spatial optical solitons in waveguide arrays", IEEE J. Quantum
Elect. {\bf 39} 31-50 (2003).

\bibitem{CBFMMTSI01} F.S.\ Cataliotti, S. Burger, C. Fort, P. Maddaloni, F. Minardi,
A. Trombettoni, A. Smerzi, and M. Inguscio, "Josephson junction
arrays with Bose-Einstein condensates", Science {\bf  293} (2001)
843-846.

\bibitem{CFFFMI03} F.S. Cataliotti, L. Fallani, F. Ferlaino, C. Fort, P. Maddaloni, and
M. Inguscio, "Superfluid current disruption in a chain of weakly
coupled Bose-Einstein condensates", New. J. Phys. {\bf 5} (2003) 71.

\bibitem{FCSEC03} J.W. Fleischer, T. Carmon, M. Segev, N.K. Efremidis
and D.N. Christodoulides, "Observation of discrete solitons in
optically induced real time waveguide arrays", Phys. Rev. Lett. {\bf 90}
(2003) 023902.

\bibitem{MECC04} H. Martin, E.D. Eugenieva, Z. Chen and D.N. Christodoulides,
"Discrete Solitons and Soliton-Induced Dislocations in Partially
Coherent Photonic Lattices", Phys. Rev. Lett. {\bf 92} (2004) 123902.

\bibitem{YMBC04} J. Yang, I. Makasyuk, A. Bezryadina and Z. Chen,
"Dipole solitons in optically induced two-dimensional photonic
lattices", Opt. Lett. {\bf 29} (2004) 1662-1664.

\bibitem{CMEXB04}
Z. Chen, H. Martin, E.D. Eugenieva, J. Xu and A. Bezryadina,
"Anisotropic Enhancement of Discrete Diffraction and Formation of
Two-Dimensional Discrete-Soliton Trains", Phys. Rev. Lett. {\bf 92} (2004)
143902.

\bibitem{CBMY04} Z. Chen, A. Bezryadina, I. Makasyuk and J. Yang,
"Observation of two-dimensional lattice vector solitons", Opt. Lett.
{\bf 29} (2004) 1656-1658.

\bibitem{NAOKMMC04} D.N. Neshev, T.J. Alexander, E.A. Ostrovskaya,
Yu.S. Kivshar, H. Martin, I. Makasyuk and Z. Chen, "Observation of
Discrete Vortex Solitons in Optically Induced Photonic Lattices",
Phys. Rev. Lett. {\bf 92} (2004) 123903.

\bibitem{FBCMSHC04} J.W. Fleischer, G. Bartal, O. Cohen, O. Manela,
M. Segev, J. Hudock and D.N. Christodoulides, "Observation of
Vortex-Ring ``Discrete'' Solitons in 2D Photonic Lattices",  Phys.
Rev. Lett. {\bf 92} (2004) 123904.

\bibitem{YM04} J. Yang and Z. Musslimani,
"Fundamental and vortex solitons in a two-dimensional optical
lattice", Opt. Lett. {\bf 28} (2003) 2094-2096.

\bibitem{KMCF04} P.G. Kevrekidis, B.A. Malomed, Z. Chen
and D.J. Frantzeskakis, "Stable higher-order vortices and
quasi-vortices in the discrete nonlinear Schr{\"o}dinger equation",
Phys. Rev. E, in press (2004).

\bibitem{YMMKMFC04} J. Yang, I. Makasyuk, H. Martin, P.G. Kevrekidis,
B.A. Malomed, D.J. Frantzeskakis and Z. Chen, "Necklace-like
solitons in optically induced photonic lattices", preprint (2004).

\bibitem{KMFC04} P.G. Kevrekidis, B.A. Malomed, D.J. Frantzeskakis
and R. Carretero-Gonz{\'a}lez, "Three-dimensional solitary waves and
vortices in a discrete nonlinear Schrodinger lattice", Phys. Rev.
Lett. {\bf 93} (2004) 080403.

\bibitem{ASK04} T.J. Alexander, A.A. Sukhorukov, and Yu.S.
Kivshar, "Asymmetric vortex solitons in nonlinear periodic
lattices", Phys. Rev. Lett. {\bf 93}, 063901 (2004).

\bibitem{MA94} R.S. MacKay and S. Aubry, "Proof of
existence of breathers for time-reversible or Hamiltonian networks
of weakly coupled oscillators", Nonlinearity {\bf 7} (1994) 1623--1643.

\bibitem{Hale} S.N. Chow and J.K. Hale, {\em Methods of Bifurcation Theory}
(Springer, Verlag, 1982).

\bibitem{Golub} M. Golubitsky and D.G. Schaeffer, {\em
Singularities and Groups in Bifurcation Theory}, vol. 1,
(Springer-Verlag, New York, 1985).

\bibitem{KK04} T. Kapitula and P.G. Kevrekidis,
"Linear stability of perturbed Hamiltonian systems: theory and a
case example", J. Phys. A: Math. Gen. {\bf 37} (2004) 7509-7526.

\bibitem{S98} B. Sandstede, "Stability of multiple-pulse
solutions", Trans. Amer. Math. Soc. {\bf 350} (1998) 429--472.

\bibitem{HJ85} R. Horn and C. Johnson, {\em Matrix
Analysis}, (Cambridge University Press, 1985).

\bibitem{LL92} H. Levy and F. Lessman, {\em Finite
Difference Equations} (Dover, New York, 1992).
\end{thebibliography}
\end{document}